\documentclass[12pt,letterpaper, twoside, openany]{article}
\usepackage[top=1in, left =1in, right=1in, bottom=1in]{geometry}

\usepackage{lineno}
\usepackage{lipsum}
\usepackage{amsmath, amssymb}
\usepackage{bm}
\usepackage{amsfonts, xr, enumerate, graphicx}
\usepackage{amsthm}
\usepackage{color}
\usepackage{csquotes}
\usepackage{float}
\usepackage{rotating}
\usepackage{titlesec}
\usepackage{subfigure} 
\usepackage{bbm}
\usepackage{hyperref}
\usepackage{comment}

\DeclareMathOperator*{\Var}{Var}

\newtheorem{theorem}{Theorem}
\newtheorem{assumption}{Assumption}
\newtheorem*{algorithm}{Algorithm}

\newtheorem{corollary}{Corollary}
\newtheorem{definition}{Definition}
\newtheorem{proposition}{Proposition}
\newtheorem{lemma}{Lemma}
\newtheorem{example}{Example}

\newtheorem*{remark}{Remark}

\newcommand{\bG}{\mathbb{G}}
\newcommand{\bR}{\mathbb{R}}
\newcommand{\bW}{\mathbb{W}}
\newcommand{\bF}{\mathbb{F}}

\newcommand{\bZ}{\mathbb{Z}}
\newcommand{\bQ}{\mathbb{Q}}
\newcommand{\bP}{\mathbb{P}}

\newcommand{\cV}{\mathcal{V}}
\newcommand{\cR}{\mathcal{R}}
\newcommand{\cI}{\mathcal{I}}
\newcommand{\cH}{\mathcal{H}}
\newcommand{\cT}{\mathcal{T}}
\newcommand{\cF}{\mathcal{F}}

\newcommand{\cN}{\mathcal{N}}

\newcommand{\Exp}{\mathbb{E}}

\begin{document}
	\baselineskip=28pt 
	
	\newgeometry{top=1in,  left =1in, right=1in, bottom=1in}
	\title{\Large Distance Correlation in Multiple Biased Sampling Models}

\author{Yuwei Ke, Hok Kan Ling, Yanglei Song \\ Department of Mathematics and Statistics, Queen's University}

	
	\maketitle
\begin{abstract}
Testing the independence between random vectors is a fundamental problem in statistics. Distance correlation, a recently popular dependence measure, is universally consistent for testing independence against all distributions with finite moments. However, when data are subject to selection bias or collected from multiple sources or schemes, spurious dependence may arise. This creates a need for methods that can effectively utilize data from different sources and correct these biases. In this paper, we study the estimation of distance covariance and distance correlation under multiple biased sampling models, which provide a natural framework for addressing these issues. Theoretical properties, including the strong consistency and asymptotic null distributions of the distance covariance and correlation estimators, and the rate at which the test statistic diverges under sequences of alternatives approaching the null, are established. A weighted permutation procedure is proposed to determine the critical value of the independence test. Simulation studies demonstrate that our approach improves both the estimation of distance correlation and the power of the test.
\end{abstract}

\section{Introduction}


Multiple biased sampling models, first introduced by \cite{vardi1982nonparametric} and \cite{vardi1985empirical}, address situations where $K \geq 1$ independent samples are collected, each potentially biased relative to a reference distribution $G$ on $\mathbb{R}^m$. For $i=1,\ldots,K$, the observations in the $i$th sample follow a biased distribution of $G$ of the form 
\begin{equation*}
	F_i(A) := \frac{ \int_A w_i(z) \, dG(z)}{\int_{\mathbb{R}^{m}} w_i(z) \, dG(z) }, \quad A \in \mathcal{B},
\end{equation*}
where $\mathcal{B}$ is the Borel $\sigma$-algebra on $\mathbb{R}^m$, and $w_i(\cdot)$ is a known, nonnegative, real-valued  weight function.
Several interesting applications of the multiple biased sampling models for multivariate data have been discussed in \cite{vardi1985empirical}. One particular case is when all the weight functions represent truncation. Examples of such cases may occur when the detection limit of an equipment has improved due to technological advances so that new measuring equipment will have a larger measuring range, and data collected using new equipment would be truncated into a larger interval than data collected using old equipment. Similarly, in human perception experiments, it is often natural to associate a weight function with each participant. These examples are closely related to the enriched stratified sampling discussed in \cite{gill1988large}.

Another important special case of multiple biased sampling models arises from outcome-dependent sampling, which is a sampling scheme that depends on the outcome (see Chapter 15 of \cite{qin2017biased}). When the outcome is discrete, it corresponds to the well-known case-control sampling in epidemiology and choice-based sampling in econometrics. 
\cite{zhou2002semiparametric} illustrated that this design can offer improved statistical efficiency, compared with a simple random sample of the same size. In other words, it could provide a cost-efficient approach to studying the determinants of a continuous outcome, especially when measurements of the exposure are costly. 
For instance, \cite{zhou2007outcome} illustrated outcome-dependent sampling in the study of the association between the score on the Bayley Scale of Infant Development (outcome) and polychlorinated biphenyls (exposure), where they drew a random sample of cohort members and two additional random samples, one from each tail of the outcome distribution. For more references on the topic of biased sampling and methods for combining multiple data sources, see \cite{qin2017biased}, \cite{kedem2017statistical}, and the references therein.


Let $(X, Y)$ be a random vector from a distribution $G$ on $\mathbb{R}^{p + q}$. Let $G_X$ and $G_Y$  denote the marginal distributions of $G$ corresponding to the random vectors $X$ and $Y$, respectively. Given independent observations from $G$, a fundamental problem in statistics is testing the independence between $X$ and $Y$: $H_0: G= G_X G_Y$ versus $H_1: G \neq G_X G_Y$. Independence tests have been applied across a broad range of fields, including statistical genetics \cite{liu2010versatile}, survival analysis \cite{tsai1990testing}, public health \cite{reshef2011detecting}, linguistics \cite{nguyen2017kernel}, etc.

However, when data are subject to selection bias or are collected from multiple sources or schemes, spurious dependence may arise if these issues are not properly accounted for. In such cases, one might falsely detect a dependence between $X$ and $Y$, even though the true relationship may be driven by biases introduced by the data collection process, as illustrated in the following examples.

\begin{example}[\cite{mahfoud1982weighted}]\label{eg:1}
     Let $(X, Y)$ be the independent lifetimes of two components in parallel, forming a kit. If one records the lifetimes of the two components of the kit in use, the joint density is a weighted version of their original density with the weight function $w(x, y) = \max(x, y)$. Additionally, the observed lifetimes will show negative dependence although the lifetimes of the two components are independent.
\end{example}

\begin{example}[\cite{tenzer2022testing}]\label{eg:2}
Let $X$ be the time from admission to the intensive care units (ICU) to infection, and $Y$ be the time from infection to discharge from the ICU. Due to the sampling mechanism, the data are length biased according to the total length of stay in the ICU, resulting in the weight function $w(x, y) = x + y$. It is of interest to test whether $X$ and $Y$ are independent, but the weight function masks the true dependence between them.
\end{example}

\begin{example}\label{eg:3}
    Suppose that the joint density of the underlying variables of interest is $g(x, y) = g_X(x) g_Y(y)$, for all $x,y>0$, where $g_X$ and $g_Y$ are two densities with supports on $\mathbb{R}_+$. Now, consider a random sample with joint density proportional to $\mathbbm{1}(x > y)g_X(x)g_Y(y)$ and another random sample with joint density proportional to $\mathbbm{1}(x < 2y) g_X(x)g_Y(y)$. Clearly, $(X, Y)$ following either of the two biased densities are not independent. Hence, testing independence based on either or both of the two samples will tend to conclude that $X$ and $Y$ are dependent. 
\end{example}

This creates a need for methods that can effectively utilize data from different sources while correcting for these biases. A recent work by \cite{tenzer2022testing} focuses on testing the quasi-independence of two random variables in the presence of one biased sample. However, to the best of our knowledge, no existing methods address the challenge of combining multiple data sources subject to biased sampling in independence testing. 

To fill this gap, we study the estimation of distance covariance and distance correlation, introduced in \cite{szekely2007measuring}, under the multiple biased sampling framework presented by \cite{vardi1985empirical}. Distance correlation is a widely used and modern measure of dependence between two random vectors, $X$ and $Y$, which may have different dimensions. A key property of distance correlation is that it is equal to zero if and only if $X$ and $Y$ are independent, making it a natural choice for testing the independence of random vectors in various applications. In particular, distance correlation is universally consistent for testing independence against all distributions with finite moments.

Specifically, given biased data collected under multiple biasing mechanisms, we first estimate the unbiased distribution $G$ by the nonparametric maximum likelihood estimator (NPMLE), denoted by $\bG_n$. We then estimate the distance covariance and correlation associated with $G$ by the corresponding quantities associated with $\bG_n$. Given $\bG_n$, our estimators retain the same computation complexity as the sample distance correlation for a single, unbiased sample. 

Moreover, we establish the strong consistency of our bias-sampled distance covariance and correlation estimators, and we derive their asymptotic distributions under the null hypothesis that $X$ and $Y$ are independent. These results ensure the consistency of the independence test based on our estimators. Furthermore, we establish the rate at which the test statistic diverges under sequences of alternatives approaching the null.  
In terms of proof techniques, we build upon results from \cite{gill1988large} and \cite{clemenccon2022statistical} for the properties of the NMPLE $\bG_n$. Further, as we shall see, the bias-sampled distance covariance estimator is a rank-$4$ \emph{degenerate} $U$-statistic with estimated parameters involving multiple samples. The degeneracy and multiplicity make applying techniques from the $U$-statistics literature challenging. Instead, we argue along the lines of \cite{szekely2007measuring}, which exploits the properties of characteristic functions.


 A permutation test is proposed to determine the critical value of the independence test. Under biased sampling, however, the permutations may not be equally likely. Following  \cite{tenzer2022testing}, we propose a Metropolis-Hastings algorithm to generate the permutations. 
Simulation studies demonstrate that our approach improves both the estimation of distance correlation and the power of the test. 
We also illustrate our methodology with two real datasets, focusing on biases induced by outcome-dependent sampling, a particular instance of the multiple biased sampling considered here.

The organization of this article is as follows. In Section \ref{sect:est_def}, we first review the distance correlation and the multiple biased sampling models. Then, we introduce our proposed estimators for distance covariance and distance correlation. Section \ref{sect:testing_indep_permutation} discusses the independence test and the associated permutation test.
Theoretical properties, including the strong consistency and asymptotic null distributions of the distance covariance and correlation estimators, and the rate at which the test statistic diverges under a sequence of alternatives approaching the null, are established in Section \ref{sect:theoretical}. In Section \ref{sect:simulation}, we conduct various simulation studies to evaluate the finite-sample performance of our proposed estimator and permutation tests under different biased mechanisms and varying numbers of biased samples. Additionally, two real datasets are used in Section \ref{sect:real} to demonstrate the practical applications of our method. We conclude in Section \ref{sec:conclusion} and present all proofs in the Appendix.

\section{Estimation of distance correlation under biased sampling models}\label{sect:est_def}
Consider a random vector $(X,Y) \in \mathbb{R}^{p+q}$ from an unknown distribution function $G$.
Let $\varphi_{X,Y}$, $\varphi_X$ and $ \varphi_Y$ be the characteristic functions of $(X,Y)$, $X$ and $Y$, respectively. Suppose that $\mathbb{E}(\|X\|+\|Y\|) < \infty$. The distance covariance is defined as the square root of
\begin{equation}\label{def:dcov}
	\mathcal{V}^2(X, Y) := \int_{\mathbb{R}^{p+q}} \frac{|\varphi_{X,Y}(s, t) - \varphi_X(s)\varphi_Y(t)|^2}{c_p c_q \|s\|^{1+p} \|t\|^{1+q}}\,ds\,dt,
\end{equation}
where  
\begin{equation}\label{def:cp_const}
	c_d := \frac{\pi^{(1+d)/2}}{\Gamma((1+d)/2)} \quad \text{ for } d \in \mathbb{N}.
\end{equation}
Here, for a complex-valued number $z$, $|z|$ is the modulus of $z$. The Euclidean distance of a vector is denoted by $\|\cdot\|$ and we omit the subscript of the vector dimension to simplify the notation as it will be clear from the context. Note that in this paper, the integrals at $0$ and $\infty$ are meant in principal value sense as in \cite[Lemma 1]{szekely2007measuring}. The distance variance is defined as the square root of
\begin{equation*}
    \mathcal{V}^2(X) = \mathcal{V}^2(X, X).
\end{equation*}
The distance correlation between $X$ and $Y$  is the nonnegative number $\mathcal{R}(X,Y)$ defined by
\begin{equation}\label{def:dcorrelation}
    \mathcal{R}^2(X, Y) =  
     \frac{\mathcal{V}^2(X,Y)}{\sqrt{\mathcal{V}^2(X)\mathcal{V}^2(Y)}} \mathbbm{1}\left(  \mathcal{V}^2(X) \mathcal{V}^2(Y) > 0 \right),
\end{equation}
where $\mathbbm{1}(\cdot)$ is the indicator function.

Given a random sample $\{(X_i^{u}, Y_i^{u}), i = 1,\ldots,n\}$ from $G$, the classic nonparametric maximum likelihood estimator (NPMLE) of $G$ is simply the empirical distribution function, defined as $\mathbb{G}_n^u(x, y) := n^{-1}\sum^n_{i=1} \mathbbm{1}(X_i^{u} \leq x, Y_i^{u} \leq y)$ for $(x,y) \in \bR^{p+q}$, where $\leq$ denotes the componentwise comparison. The sample distance covariance and correlation are the distance covariance and correlation under $\bG_n^{u}$, respectively (see \cite{szekely2007measuring}). 


In a one-sample biased sampling situation, we do not observe data directly from $G$. Instead, we observe a random sample $\{(X_i, Y_i), i = 1,\ldots,n\}$ from $F$, which is a biased version of $G$ according to some known, nonnegative, real-valued biasing or weight function $w:\bR^{p+q} \to [0,\infty)$, where
\begin{equation*}
	F(A) := \frac{ \int_A w(z) \, dG(z)}{\int_{\mathbb{R}^{p+q}} w(z) \, dG(z) }, \quad A \in \mathcal{B},
\end{equation*}
with $\mathcal{B}$ denoting the Borel $\sigma$-algebra on $\mathbb{R}^{p+q}$ and $\int w(z) \, dG(z) \in (0, \infty)$. The NPMLE of $G$ (see \cite{vardi1985empirical}) is then given by the following: for  $x \in \mathbb{R}^p, y \in \mathbb{R}^q$,
\begin{equation}\label{eq:1_biased_NPMLE}
\mathbb{G}_n(x,y) := \sum^n_{i=1} \widehat{p}_i \mathbbm{1}(X_i \leq x, Y_i \leq y),\;\; \text{ where }\;\; \widehat{p}_i := \frac{w^{-1}(X_i,Y_i)}{\sum^n_{j=1}w^{-1}(X_j,Y_j)}.
\end{equation}
We assume that $w$ is positive on the support of $G$; otherwise, $G$ is not identifiable. Under this assumption, $\widehat{p}_i$ is well-defined almost surely. Further, we may estimate the distance covariance/correlation under $G$ by the corresponding quantities under $\bG_n$.

In this paper, we mainly consider data collected from \emph{multiple} potentially biased distributions, while maintaining the objective of inferring the distance covariance/correlation under the unbiased distribution $G$.

\subsection{Multiple biased samples}
\cite{vardi1982nonparametric} and \cite{vardi1985empirical} genearlized the one-sample biased sampling model to allow for $K > 1$ different biased samples. Specifically, suppose that the weight functions $w_1,\ldots,w_K:\bR^{p+q} \to [0,\infty)$ are  known, nonnegative, real-valued functions and that $G$ is an unknown distribution function on $(\mathbb{R}^{p+q}, \mathcal{B}$). The corresponding biased distributions are
\begin{equation*}
	F_i(A) := \frac{ \int_A w_i(z) \, dG(z)}{\int_{\mathbb{R}^{p+q}} w_i(z) \, dG(z) }, \quad A \in \mathcal{B}, \quad  i=1, \ldots, K,
\end{equation*}
where $W_i := \int w_i(z)\,dG(z) \in (0, \infty)$. Note that we allow the possibility that one of the weight functions is identical to $1$, corresponding to the unbiased samples. In the multiple biased sampling model, we observe $K > 1$ different independent samples: 
\begin{equation*}
	(X_{i,1},Y_{i1}),\ldots, (X_{i,n_i},Y_{i, n_i}) \stackrel{i.i.d.}{\sim} F_i, \quad i=1,\ldots, K,
\end{equation*}
where $n_i$ is the sample size in the $i$th biased samples. Let $n:= n_1 + \ldots + n_K$ denote the total sample size and write $\lambda_{n,i} := n_i /n$ for the sampling fraction from $F_i$ for $i=1,\ldots, K$. Denote the combined samples by $ ((\widetilde{X}_1,\widetilde{Y}_1),\ldots,(\widetilde{X}_n,\widetilde{Y}_n))$, which is equal to
\begin{equation*}
    ((X_{11},Y_{11}),\ldots,(X_{1,n_1},Y_{1,n_1}),\ldots,(X_{K1},Y_{K1}),\ldots,(X_{K,n_K},Y_{K,n_K})).
\end{equation*}
When $K =1$, the combined sample $(\widetilde{X}_i, \widetilde{Y}_i)$ is just the original sample. Let $w_{ij} := w_i(\widetilde{X}_j, \widetilde{Y}_j)$ and $p_j$ denote the weight that a general distribution on $\bR^{p+q}$ places on the data point $(\widetilde{X}_j, \widetilde{Y}_j)$.
Given $p := (p_1,\ldots,p_n)$, the likelihood (see \cite{vardi1985empirical}) is proportional to
\begin{equation*}
    L(p) = \prod^n_{j=1} p_j \prod^s_{i=1} \left( \sum_{k=1}^n p_k w_{ik}\right)^{-n_i}.
\end{equation*}
We denote the maximizer of $L(\cdot)$ as $\widehat{p} = (\widehat{p}_1,\ldots,\widehat{p}_n)$, and the corresponding NPMLE of $G$ as $\mathbb{G}_n$, which is defined as $\mathbb{G}_n(x,y) = \sum^n_{i=1} \widehat{p}_i \mathbbm{1}(\widetilde{X}_i \leq x, \widetilde{Y}_i \leq y), x \in \mathbb{R}^p, y \in \mathbb{R}^q$. We discuss the existence and equivalent characterization of $\bG_n$ in Subsections \ref{subsect:properties_MLE} and its computation in \ref{subsect:computation_MLE}. 

Based on the NPMLE $\bG_n$, we propose the following estimators for distance covariance and correlation under the unbiased distribution $G$.


\begin{definition}\label{def:Vn_Rn}
We define the bias-sampled distance covariance estimator between $X$ and $Y$ under $K$-sample biased sampling model, for $K \geq 1$, as the square root of
\begin{equation*}
	\widehat{\mathcal{V}}^2_n(X, Y) :=  \int_{\mathbb{R}^{p+q}} \frac{|\varphi^{(n)}_{X,Y}(s, t) - \varphi^{(n)}_X(s)\varphi^{(n)}_Y(t)|^2}{c_p c_q \|s\|^{1+p} \|t\|^{1+q}}\,ds\,dt,
\end{equation*}
where $\varphi^{(n)}_{X,Y}(s, t) := \int_{\mathbb{R}^{p+q}} e^{\sqrt{-1} (s^\top x + t^\top y) } \,d\mathbb{G}_n(x,y)$, $\varphi^{(n)}_X(s) := \varphi^{(n)}_{X,Y}(s,0)$, and $\varphi^{(n)}_Y(t) := \varphi^{(n)}_{X,Y}(0, t)$.    The bias-sampled distance variance estimator of $X$ and the bias-sampled distance correlation estimator between $X$ and $Y$ under $K$-sample biased sampling model are then defined as the square root of
of $\widehat{\mathcal{V}}_n^2(X) = \widehat{\mathcal{V}}_n^2(X,X)$ and
\begin{equation*}
    \widehat{\mathcal{R}}_n^2(X,Y):= \frac{\widehat{\mathcal{V}}^2_n(X,Y)}{\sqrt{\widehat{\mathcal{V}}^2_n(X)\widehat{\mathcal{V}}^2_n(Y)}}\mathbbm{1}(\widehat{\mathcal{V}}^2_n(X)\widehat{\mathcal{V}}^2_n(Y)>0),
\end{equation*}
respectively.
\end{definition}

For simplicity, we will write $\widehat{\mathcal{V}}^2_n$ and $\widehat{\mathcal{R}}^2_n$ for 
$\widehat{\mathcal{V}}^2_n(X, Y)$ and $\widehat{\mathcal{R}}^2_n(X, Y)$, respectively, hereafter. The following proposition provides a computational formula for $\widehat{\mathcal{V}}_n^2$ in terms of the combined samples and the NPMLE $\bG_n$ (or equivalently $\widehat{p}$). The computational complexity is the same as that for a single unbiased sample, 
which is $O(n^2)$.
\begin{proposition}\label{prop:computation}
For $K \geq 1$, we have
\begin{align*}
    \widehat{\mathcal{V}}^2_n &= \sum_{i=1}^n \sum_{j=1}^n \|\widetilde{X}_i -\widetilde{X}_j\| \cdot \|\widetilde{Y}_i - \widetilde{Y}_j\| \widehat{p}_i \widehat{p}_j \\
    & \quad \quad - 2 \sum^n_{k=1}\left\{ \widehat{p}_k \left( \sum^n_{i=1} \|\widetilde{X}_k - \widetilde{X}_i\|\widehat{p}_i\right) 
    \left( \sum^n_{i=1} \|\widetilde{Y}_k -\widetilde{Y}_i\|\widehat{p}_i\right) \right\}\\
    & \quad \quad + \left( \sum_{i=1}^n \sum_{j=1}^n \|\widetilde{X}_i -\widetilde{X}_j\|  \widehat{p}_i \widehat{p}_j \right) \left( \sum_{i=1}^n \sum_{j=1}^n \|\widetilde{Y}_i -\widetilde{Y}_j\|  \widehat{p}_i \widehat{p}_j \right).
\end{align*}
\end{proposition}

\begin{proof}
    See Appendix \ref{appendix:23}.
\end{proof}

With Proposition \ref{prop:computation}, the estimators for the distance variance and correlation can also be computed accordingly.

\subsection{Properties of NPMLE}\label{subsect:properties_MLE}

We start with the existence of the NPMLE $\bG_n$. First, we introduce additional notations.
Let $\mathcal{X}_0$ be the support of $G$ and
\begin{equation}\label{def:X_pos}
	\mathcal{X}^+ := \bigcup_{i=1}^K\{x: w_i(x) > 0 \}.
\end{equation}
Consider the graph $\bm{G}$ on the $K$ vertices $\{1,\ldots,K\}$ defined as follows: $i \leftrightarrow j$ if and only if $\int \mathbbm{1}(w_i  > 0 )\mathbbm{1}(w_j  > 0 )\,dG > 0$. To say that the graph $\bm{G}$ is connected means that every pair of $i,j$ is connected by a path.

\begin{assumption} 
\label{assumption:biased sampling}
~
\begin{enumerate}[(i)]
    \item $\mathcal{X}^+ = \mathcal{X}_0$.
    \item The graph $\bm{G}$ is connected.
\end{enumerate}
\end{assumption}

Assumption \ref{assumption:biased sampling} (i) is necessary because one cannot hope to estimate $G$ if $\mathcal{X}^+$ is a proper subset of $\mathcal{X}_0$. Assumption \ref{assumption:biased sampling} (ii)  ensures that $G$ is identifiable, and with probability $1$, for all sufficiently large $n$, the NPMLE exists and is unique; see Proposition 1.1 and Corollary 1.1 in \cite{gill1988large}.  

Given the observed data, we can determine if the NPMLE exists and is unique as follows. Define $\mathbb{F}_{nj}$ as the empirical distribution function associated with the $j$th samples. Define a directed graph $\bm{G}_n$ on the $K$ vertices by $i \rightarrow j$ if and only if $\int \mathbbm{1}(w_i > 0) \,d\mathbb{F}_{nj} > 0$. $\bm{G}_n$ is said to be strongly connected if for any two vertices $i$ and $j$, there is a directed path from $i$ to $j$ and from $j$ to $i$. If $\bm{G}_n$ is strongly connected, then the NPMLE exists and is unique; see Theorem 1.1 in \cite{gill1988large}. Likelihood-based conditions for the unique existence of the NPMLE are also provided in \cite{davidov2009existence}.

Next, we discuss equivalent characterization for the NPMLE $\bG_n$. For $x \in \bR^p$, $y \in \bR^q$ and $\theta = (\theta_1,\ldots,\theta_K) \in \bR^K$, denote by $\delta_{(x,y)}$  the Dirac measure charged at $(x,y)$ and define
\begin{align}
    \label{def:avg_weight_funcs}
\widetilde{\omega}_n(x,y; \theta) := \sum^K_{i=1} \frac{\lambda_{n,i}w_i(x, y)}{\theta_{i}}.
\end{align}
Furthermore, define the empirical measure
	\begin{equation*}
		\mathbb{F}_n :=\frac{1}{n}\sum^{n}_{j=1}\delta_{(\widetilde{X}_{j},\widetilde{Y}_{j})} = \frac{1}{n}\sum^K_{i=1}\sum^{n_i}_{j=1}\delta_{(X_{ij},Y_{ij})}.
	\end{equation*}
 Then, as discussed in \cite{gill1988large}, 
 the NPMLE $\mathbb{G}_n$ of $G$ and $\bW_n =(\mathbb{W}_{n1},\ldots,\mathbb{W}_{nK})$ of $W = (W_1,\ldots,W_K)$ are characterized by the following set of equations: for any Borel set $A \subset \bR^{p+q}$ and $i = 1,\ldots,K$,
\begin{align}
\begin{split}
    &\mathbb{G}_n(A)   = \int_A \left[ \widetilde{\omega}_n(x,y; \bW_n) \right]^{-1} d \mathbb{F}_n(x,y),  \quad \bG_n(\bR^{p+q}) = 1, \\
 & \bW_{n,i} =   \int_{\bR^{p+q} } w_i(x,y) 
	\left[\widetilde{\omega}_n(x,y; \bW_n)\right]^{-1}d \bF_n(x,y).
 \end{split}
\label{def:NPMLE}
\end{align}
The computational algorithm in Subsection \ref{subsect:computation_MLE} basically solves the above equations iteratively.

\subsection{Computation of NPMLE}\label{subsect:computation_MLE}

To compute the NPMLE, \cite{vardi1985empirical}  proposed solving a system of $K-1$ nonlinear equations in $K-1$ unknowns. In discussing Vardi's paper, \cite{mallows1985} suggests an alternative iterative scheme for solving these nonlinear equations. This algorithm is particularly simple and does not require solving any equations.
\cite{davidov2010note} shows that if the NPMLE exists and is unique, the algorithm will converge to the NPMLE. Their numerical studies demonstrate that the algorithm is significantly faster than the method proposed by \cite{vardi1985empirical} in terms of CPU time required. Thus, we adopt this algorithm in our implementation.  In the following algorithm, $\widehat{p}_j^{(t)}$ and $\widehat{W}_i^{(t)}$ denote the estimates of the NPMLE $\widehat{p}_j$ and $\bW_{n,i}$ at the $t$th iteration, respectively.


\begin{algorithm}[Computation of $\mathbb{G}_n$]
~
    \begin{enumerate}[1.]
    \item Initialize the vector $\widehat{W}^{(0)} = (\widehat{W}_1^{(0)},\ldots,\widehat{W}_K^{(0)}) > 0$ and set $t = 0$.

    \item For $j=1,\ldots,n$, set 
    \begin{equation*}
        \widetilde{p}_j^{(t)} = \left( \sum^K_{i=1}\frac{\lambda_{n,i} w_{ij}}{\widehat{W}_i^{(t)}} \right)^{-1} = 
        \left(\widetilde{\omega}_n(\widetilde{X}_{j},\widetilde{Y}_{j}; \widehat{W}^{(t)})\right)^{-1},
    \end{equation*}
    where $\widetilde{\omega}_n(\cdot)$ is defined in \eqref{def:avg_weight_funcs}. 
    Standardize them as follows: for $j = 1,\ldots,n$,
    $$
    \widehat{p}_j^{(t)} = \frac{\widetilde{p}_j^{(t)}}{\sum^n_{\ell=1}\widetilde{p}_{\ell}^{(t)} }.
    $$

    \item For $i=1,\ldots,K$, set
    \begin{equation*}
        \widehat{W}_i^{(t+1)} = \sum^n_{j=1}\widehat{p}_j^{(t)} w_{ij}.
    \end{equation*}

    \item Set $t = t + 1$.
\end{enumerate}
Iterate steps 2 to 4 until a convergence criterion is met, for example, $\|\widehat{p}^{(t)} - \widehat{p}^{(t-1)}\| < \varepsilon$ for some prespecified small $\varepsilon$, where $\widehat{p}^{(t)} := (\widehat{p}_1^{(t)}, \ldots, \widehat{p}_n^{(t)})$. The approximate NPMLE of $G$ is then given by $\mathbb{G}_n^{(T)}(x,y) = \sum^n_{i=1} \widehat{p}_i^{(T)} \mathbbm{1}(\widetilde{X}_i \leq x, \widetilde{Y}_i \leq y)$ for $, x \in \mathbb{R}^p, y \in \mathbb{R}^q$, where $T$ is the index for the final iteration.
\end{algorithm}


\section{Testing independence under multiple biased sampling models}\label{sect:testing_indep_permutation}
As mentioned in the introduction, one of the most important applications of distance covariance and distance correlation is the test of independence: $H_0: G= G_X G_Y$ versus $H_1: G \neq G_X G_Y$. In the examples in the introduction, we already see that without accounting for the biased sampling mechanism, the observed variables can become dependent when the underlying variables are independent.
In the presence of multiple biased samples, an independence test can be based on $n \widehat{\mathcal{V}}_n^2$ or $n\widehat{\mathcal{R}}_n^2$, where we reject the null hypothesis for large values of the test statistic. Note that in those examples, both Assumptions \ref{assumption:biased sampling} (i) and (ii) are satisfied.

To use $n\widehat{\mathcal{V}}^2_n$ as a test statistic, we need to find the critical value of the test. Similar to the unbiased case, the asymptotic null distribution of $n\widehat{\mathcal{V}}^2_n$ depends on the underlying distribution of $X$ and $Y$ which are generally unknown; see  Theorem \ref{thrm:null_distribution} below. Thus, we propose a permutation test to find the critical value or $p$-value of the test, which is also a standard practice in testing independence. In the unbiased case, all the permutations are equally likely under the null hypothesis. Under biased sampling, different permutations are no longer equally likely under the null hypothesis. In a one-sample biased sampling model, \cite{tenzer2022testing} has studied the distribution of permutations under quasi-independence. We now discuss the extension to multiple biased sampling models.

Let $\mathcal{S}_n := \prod^K_{i=1} S_{n_i}$, where $S_{n_i}$ is the set of all permutations of $\{1,\ldots,n_i\}$. That is, $\pi_i \in S_{n_i}$ is a bijection from $\{1,\ldots,n_i\}$ to $\{1,\ldots,n_i\}$.
Let $\mathcal{D}_n = \{ (X_{ij}, Y_{ij}): i=1,\ldots,K,j=1,\ldots,n_i\}$. With some abuse of notation, for any $\pi = (\pi_1,\ldots,\pi_K) \in \mathcal{S}_n$, let $\pi(\mathcal{D}_n)$ be the permuted sample:
\begin{equation*}
    \pi(\mathcal{D}_n) := \{ (X_{ij}, Y_{i,\pi_i(j)}):i=1,\ldots,K, j=1,\ldots,n_i\}.
\end{equation*}
As it is customary to permute either the $X_{ij}$'s or $Y_{ij}$'s, we keep the indices of the $X_{ij}$'s in the above definition. Assume that for some $\sigma$-finite measures $\nu_1$ on $\bR^{p}$ and $\nu_2$ on $\bR^{q}$, $X$ and $Y$ has a joint density $g$ relative to the product measure $\nu_1 \times \nu_2$. Denote by $g_X$ and $g_Y$ the marginal densities of $X$ and $Y$ with respect to $\nu_1$ and $\nu_2$ respectively. Further, let $f_i$ be the joint $\nu_1\times\nu_2$ density of $X$ and $Y$ in the $i$th biased sample for $i = 1,\ldots,n$.

Let $d_n$ be the realization of $\mathcal{D}_n$. 
For any $\pi = (\pi_1,\ldots,\pi_K) \in \mathcal{S}_n$, let $\mathbb{P}(\pi|d_n)$ be the probability of observing the permutation $\pi \in \mathcal{S}_n$ 
conditional on ${d}_n$, which is defined as follows:
\begin{equation}\label{eq:permut_prob}
    \mathbb{P}(\pi|d_n) = \prod^K_{i=1} \frac{\prod_{j=1}^{n_i} f_i(x_{ij}, y_{i,\pi_i(j)})}{  \sum_{\pi_i' \in S_{n_i}} \prod_{j'=1}^{n_i} f_i(x_{ij'},y_{i,\pi_i'(j')}) }.
\end{equation}
Under $H_0$, $f_i(x, y) \propto w_i(x,y)g_X(x) g_Y(y)$. Thus,
\begin{align}
\mathbb{P}_0(\pi|d_n) &= \prod^K_{i=1} \frac{\prod_{j=1}^{n_i} w_i(x_{ij}, y_{i,\pi_i(j)}) g_X(x_{ij}) g_Y(y_{i,\pi_i(j)}) }{  \sum_{\pi_i' \in S_{n_i}} \prod_{j'=1}^{n_i} w_i(x_{ij'}, y_{i,\pi_i'(j')}) g_X(x_{ij'})g_Y(y_{i,\pi_i'(j')}) } \nonumber \\
    &=\prod^K_{i=1} \left\{ \frac{1}{\text{per}(\mathcal{W}_i)} \prod^{n_i}_{j=1} \mathcal{W}_i(j, \pi_i(j)) \right\}, \label{eq:permut_prob_H0}
\end{align}
where $\mathcal{W}_i(j, j') := w_i(x_{ij}, y_{ij'})$ and $\text{per}(\mathcal{W}_i) := \sum_{\pi_i' \in S_{n_i}} \prod^{n_i}_{j'=1} \mathcal{W}_i(j',\pi_i'(j'))$. Note that the probability in \eqref{eq:permut_prob_H0} is free of the unknown density $g$ and only depends on the weight functions and the data. The subscript $0$ denotes the probability is under $H_0$.

\begin{lemma}\label{lemma:joint_dist_same}
 Conditionally on $\mathcal{D}_n$, let $\pi$ be a permutation having the conditional probability given in \eqref{eq:permut_prob}. Then, $\pi(\mathcal{D}_n) \stackrel{d}{=} \mathcal{D}_n$.         
\end{lemma}
\begin{proof}
    See Appendix \ref{appendix:23}.
\end{proof}

Note that Lemma \ref{lemma:joint_dist_same} is valid under both $H_0$ and $H_1$. Under $H_0$, the probability in \eqref{eq:permut_prob} becomes \eqref{eq:permut_prob_H0}.

Let $\mathcal{X} := \mathbb{R}^{p+q}$.
Suppose that $T_n: \mathcal{X}^n \rightarrow \mathcal{X}^n$ is a test statistic. We shall show that as a result of Lemma \ref{lemma:joint_dist_same}, the permutation test for any $T_n$ has level $\alpha$. We follow similar notations as in the supplement of \cite{pfister2018kernel} to introduce the permutation test.
First, note that the set of all possible permutations may depend on $d_n$. This occurs, for example, when one of the weight functions is $w(x, y) = \mathbbm{1}(y > x)$.
Since the  permutations are not necessarily equally likely, we define the weighted resampling distribution function for $T_n$ as
\begin{equation*}
    \widehat{R}_{T_n}(d_n; t) := \sum_{\pi \in S_n} \mathbb{P}_0( \pi | d_n) \mathbbm{1}(T_n( \pi(d_n)) \leq t).
\end{equation*}
For any $\alpha \in (0, 1)$, let $\phi_n(d_n) = \mathbbm{1}(T_n(d_n) > (\widehat{R}_{T_n}(d_n;\cdot))^{-1}(1-\alpha))$ be the resampling test, where $(\widehat{R}_{T_n}(d_n;\cdot))^{-1}$ is the quantile function of $\widehat{R}_{T_n}(d_n;\cdot)$, that is, $(\widehat{R}_{T_n}(d_n;\cdot))^{-1}(\alpha) := \inf\{ t \geq 0 : \widehat{R}_{T_n}(d_n;t) \geq \alpha\}$. 
\begin{proposition}\label{prop:permu_level}
    The test $\phi(\mathcal{D}_n)$ is a test at level $\alpha$, when testing $H_0$ against $H_1$.
\end{proposition}
\begin{proof}
    See Appendix \ref{appendix:23}.
\end{proof}

In practice, one samples a finite number, say $B$, of permutations  $\pi_{1}^*,\ldots,\pi_{B}^*$, when performing a permutation test, where the permutations are drawn with probabilities according to \eqref{eq:permut_prob_H0}.  The Monte-Carlo approximated resampling distribution function is given by
\begin{align*}
    \widehat{R}_{T_n}^B(d_n; t) &:= \frac{1}{B}\sum_{b=1}^B \mathbbm{1}(T_n(\pi_{b}^*(d_n)) \leq t) \\
    &= \sum_{\pi \in \mathcal{S}_n} \frac{\sum_{b=1}^B \mathbbm{1}(\pi_{b}^*(d_n) = \pi(d_n))}{B} \mathbbm{1}(T_n(\pi(d_n)) \leq t).
\end{align*}
By the strong law of large numbers, with probability $1$, we have
\begin{equation*}
    \lim_{B \rightarrow \infty} \widehat{R}_{T_n}^B(d_n)(t) = \widehat{R}_{T_n}(d_n)(t).
\end{equation*}
We reject $H_0$ when $T_n(d_n) > (\widehat{R}_{T_n}^B(d_n;\cdot)^{-1}(1-\alpha)$. Alternatively, one can obtain a $p$-value as in the following algorithm.

\begin{algorithm}[Weighted-permutation test]
~
    \begin{enumerate}
        \item Compute the test statistic $\widehat{\mathcal{V}}_n^2$.
        \item Generate $B$ permutations $\pi_1,\ldots,\pi_B$ according to \eqref{eq:permut_prob_H0}.
        \item Compute the test statistic for the permuted datasets and obtain $\widehat{\mathcal{V}}_n^{2,(1)},\ldots,\widehat{\mathcal{V}}_n^{2,(B)}$.
        
        \item Output the $p$-value as $\frac{1}{B+1}\left(1 + \sum_{b=1}^B\mathbbm{1}( \widehat{\mathcal{V}}_n^{2,(b)} \geq \widehat{\mathcal{V}}_n^2)\right)$.
    \end{enumerate}
\end{algorithm}

\subsection{Permutation Generation}
In the special case where all the weights are indicator functions, $\mathbb{P}_0(\pi|d_n)$ under $H_0$ is simply the uniform distribution over all valid permutations. Specifically, if all the weight functions $w_i(x, y)$ take the form of 
\begin{equation}\label{eq:separable_weight}
\prod^{p}_{i=1} \mathbbm{1}(x_i \in A_i ) \prod^q_{i=1} \mathbbm{1}(y_i \in B_i),    
\end{equation}
where $A_i, B_i \subset \mathbb{R}$, then we can simply permute the indices $(1,\ldots,n_i)$ of $Y_{i1},\ldots,Y_{in_i}$ in each sample because all $y_{i,n_i}$ values are in $\prod^q_{i=1} B_i$ and any permutation of $(1,\ldots,n_i)$ is valid.

For a general weight function, since the number of possible permutations could be very large, it may not be computationally feasible to directly draw permutations according to the probability in \eqref{eq:permut_prob_H0} exactly. One possible solution is to draw the permutations using a Markov Chain Monte Carlo algorithm.
In the one-sample biased sampling situation, \cite{tenzer2022testing} proposed a Metropolis-Hastings (MH) algorithm for drawing the permutations. In the multiple samples case, we can follow the same way by using the MH algorithm to draw permutations within each sample, which is described below.

For $i=1,\ldots,K$, let $\pi_{it} := (\pi_{it}(1),\ldots,\pi_{it}(n_i))$ be the permutation in the $i$th sample at step $t$. Define the neighbors of $\pi_{it}$, $\text{Neig}(\pi_{it})$, as all the permutations obtained from $\pi_{it}$ by a single swap of two distinct elements. That is
\begin{equation*}
\text{Neig}(\pi_{it}):= \{ \pi^{j \leftrightarrow j'}_{t} := (\pi_t(1),\ldots,\pi_t(j'),\ldots,\pi_t(j),\ldots,\pi_t(n)), \quad \forall j < j' \}.
\end{equation*}
The proposal distribution in the MH algorithm is chosen as the uniform distribution on $\text{Neig}(\pi_{it})$. The acceptance ratio is
\begin{equation*}
 \frac{P(\pi_{it}^{j \leftrightarrow j'})}{P(\pi_{it})} = \frac{ \mathcal{W}_i(j,\pi_{it}(j')) \mathcal{W}_i(j',\pi_{it}(j))}{\mathcal{W}_i(j,\pi_{it}(j))\mathcal{W}_i(j',\pi_{it}(j'))}.
\end{equation*}
Let $M_0$ be the burn-in number and suppose that we retain a permutation after every $M$ steps. The steps of this MH algorithm are as follows.

\begin{algorithm}[MH algorithm for multiple biased sampling of permutations]
~
    \begin{enumerate}
        \item Set $i=1$.
        \item Set $\pi_{i0} = (1,\ldots,n_i)$.
        \item For $ t = 0$ to $M_0 + (B-1)M$, do
        \begin{enumerate}[(i)]
            \item Sample $\pi_{it}^{j \leftrightarrow j'} \sim \text{Uniform}(\text{Neig}(\pi_{it}))$ and $U \sim U(0, 1)$.
            \item If $\displaystyle U \leq \frac{ \mathcal{W}_i(j,\pi_{it}(j')) \mathcal{W}_i(j',\pi_{it}(j))}{\mathcal{W}_i(j,\pi_{it}(j))\mathcal{W}_i(j',\pi_{it}(j'))}$, then set $\pi_{i, t + 1} \leftarrow \pi_{it}^{j \leftrightarrow j'}$; else set $\pi_{i,t+1} \leftarrow \pi_{it}$.
        \end{enumerate}
        \item The resulting $B$
        permutations for the $i$th sample are $\pi_{i,M_0}, \pi_{i,M_0+M},\ldots, \\\pi_{i,M_0 + (B-1)M}$.
        \item Repeat steps 2 to 4 for $i=2,\ldots,K$. The combined permutations are $\pi_t := (\pi_{1t},\ldots,\pi_{Kt})$, for $t= M_0, M_0+M,\ldots,M_0+(B-1)M$.
    \end{enumerate}
\end{algorithm}

\section{Theoretical Results}\label{sect:theoretical}
First, we focus on the case where the unbiased distribution $G$ is fixed, that is, $G$ does not depend on $n$. 
In this section, we focus on the case when $K \geq 2$. When $K=1$, similar results can also be established using similar arguments. We recall an assumption and some facts from \cite{gill1988large}.

\begin{assumption}\label{assumption:lambda}
There exist positive numbers $\lambda_1,\ldots,\lambda_K$ such that for $i = 1,\ldots,K$,
$$
\lim_{n \to \infty} \lambda_{n,i} = \lambda_i.
$$
\end{assumption}


\begin{lemma}[Corollary 1.1 and 2.1 from \cite{gill1988large}]\label{lemma:consistency_G}
Suppose that Assumptions \ref{assumption:biased sampling} and \eqref{assumption:lambda} hold 
As $n \to \infty$, almost surely, the solution to \eqref{def:NPMLE} has a unique solution and 
\begin{equation*}
\sup_{s \in \mathbb{R}^p, t \in \mathbb{R}^q}|	\mathbb{G}_n(s, t) - G(s, t) | \rightarrow 0.
\end{equation*}
\end{lemma}

The first statement above ensures the uniqueness of NPMLE $\bG_n$ for the unbiased distribution $G$, while the second shows strong consistency. Based on the above lemma, we can show that the estimators $\widehat{\cV}_n^2$ and  $\widehat{\cR}_n^2$ in Definition \ref{def:Vn_Rn} are strongly consistent for the distance covariance in \eqref{def:dcov} and correlation in \eqref{def:dcorrelation}. Recall that $(X,Y) \in \bR^{p+q}$ is a generic random vector from the unbiased distribution $G$.

\begin{theorem}\label{thm:consistency_dcov}
 Suppose that Assumptions \ref{assumption:biased sampling} and \ref{assumption:lambda} hold. Furthermore, assume   $\Exp(\|X\| + \|Y\|) < \infty$. Then, almost surely,
	\begin{equation*}
 		\lim_{n \rightarrow \infty}\widehat{\cV}^2_n = \mathcal{V}^2, \quad \text{ and } \quad 
		\lim_{n \rightarrow \infty}\widehat{\cR}^2_n = \mathcal{R}^2.
  \end{equation*}
\end{theorem}
\begin{proof}
    See Appendix \ref{app:consistency_dcov}.
\end{proof}

Next, we derive the limiting distribution of the test statistics $n \widehat{\cV}_n^2$ and $n\widehat{\cR}^2_n$ under the null, i.e., when $X$ and $Y$ are independent. We introduce additional notations.

First, for each $(s,t) \in \bR^{p+q}$, define a function on $\bR^{p+q}$ as follows: for each $(x,y) \in \bR^{p+q}$,
\begin{equation}
    \label{def:hst_xy}
h_{s,t}(x,y) := \left(e^{\sqrt{-1}(s^\top x)} - \varphi_X(s)\right)\left(e^{\sqrt{-1}(t^\top y )} -\varphi_Y(t) \right).
\end{equation}
For $(x,y) \in \bR^{p+q}$, define
\begin{equation}\label{def:r_underline_omega}
r(x,y) := \left[ \sum_{i=1}^{K} \lambda_i \frac{\omega_i(x,y)}{W_i}\right]^{-1}, \quad
\underline{\omega}(x,y) := \left[\frac{\omega_1(x,y)}{W_1},\ldots, \frac{\omega_K(x,y)}{W_K}\right]^\top.
\end{equation}
Furthermore, denote by $\Lambda$ the diagonal matrix with diagonal entries being $(\lambda_1,\ldots,\lambda_K)$, and define a $K\times K$ matrix as follows:
\begin{equation}
    \label{def:M_matrix}
    M := \Lambda - \int r(x,y) \underline{\omega}(x,y)\underline{\omega}(x,y)^\top \ dG(x,y).
\end{equation}
As discussed in \cite{gill1988large}, $M$ is singular, and we denote by $M^{-}$ the Moore Penrose Inverse of $M$. 

Define a covariance function $C:\bR^{p+q} \times  \bR^{p+q}$ as follows: for $(s,t),(s',t') \in \bR^{p+q}$,
\begin{align}
    \label{def:cov_function}
    \begin{split}
&C((s,t), (s',t')) := \Exp\left[r(X,Y) h_{s,t}(X,Y) \overline{h_{s',t'}(X,Y)}\right]  \\
& \qquad + \Exp\left[ r(X,Y) h_{s,t}(X,Y) \underline{\omega}^\top(X,Y) \right] M^{-1} 
\Exp\left[r(X,Y) \overline{h_{s',t'}(X,Y)} \underline{\omega}(X,Y)  \right].
\end{split}
\end{align}
Finally, for each $\delta \in (0,1)$, define
\begin{equation}\label{def:D_delta}
	D(\delta) := \{(s, t) \in \bR^{p+q}: \delta \leq \|s\| \leq 1/\delta, \delta \leq \|t\|  \leq 1/\delta \},
\end{equation}
and we denote by $\mathcal{C}(D(\delta))$ the space of (uniformly) continuous functions on $D(\delta)$. We equip $\mathcal{C}(D(\delta))$ with the supremum norm and denote by $\overset{d}{\longrightarrow}$ the weak convergence in $\mathcal{C}(D(\delta))$ or $\bR^{d}$ for some integer $d \geq 1$ \cite{billingsley2013convergence}.  

\begin{theorem}\label{thrm:convergence_of_characterists}
Assume the null holds, that is, $X$ and $Y$ are independent. Furthermore,  suppose that Assumptions \ref{assumption:biased sampling} and \ref{assumption:lambda} hold, and that
\begin{equation}
    \label{def:moment_cond1}
    \Exp[(1+r(X,Y))(1 + \|X\|^2 + \|Y\|^2)] < \infty.
\end{equation}
 Then, there exists a zero-mean Gaussian process $\bZ := \{\bZ(s,t): (s,t) \in \bR^{p+q}\}$ such that (i) it has a continuous sample path almost surely, and (ii) for any $\delta \in (0,1]$,
$$
\left\{\sqrt{n}  \left(\varphi_{X,Y}^{(n)}(s,t) - \varphi^{(n)}_X(s)\varphi^{(n)}_Y(t)\right) \ : (s,t) \in D(\delta) \right\} \overset{d}{\longrightarrow}  \left\{\bZ(s,t): (s,t) \in D(\delta)\right\}
$$
in the space $\mathcal{C}(D(\delta))$.
\end{theorem}
\begin{proof}
    See Appendix \ref{app:cf_distribution}. The proof relies on \cite[Theorem 2.2] {gill1988large} and \cite{dudley1983invariance}.
\end{proof}

\begin{theorem}\label{thrm:null_distribution}
Assume the null holds, that is, $X$ and $Y$ are independent. Furthermore,  suppose that Assumptions \ref{assumption:biased sampling} and \ref{assumption:lambda} hold, and that
\begin{equation}
    \label{def:moment_cond2} \Exp\left[(1+r(X,Y))(1 + \|X\|^2)(1 + \|Y\|^2) + r^2(X,Y) (1 + \|X\|\|Y\|)\right] < \infty. 
\end{equation}
Then we have
$$
n \widehat{\cV}_n^2 \quad \overset{d}{\longrightarrow} \quad
\bQ := \int_{\bR^{p+q}} \frac{|\bZ(s,t)|^2}{c_p c_q \|s\|^{1+p} \|t\|^{1+q}}\,ds\,dt,
$$
where $\bZ$ appears in Theorem \ref{thrm:convergence_of_characterists}. Furthermore, $\bQ$ is finite almost surely.
\end{theorem}
\begin{proof}
    See Appendix \ref{app:null_distribution}.
\end{proof}

\begin{remark}
Since $\bZ$ has continuous sample paths, the integral above is well-defined. Under the null,  
if $\cV^2(X,X) \cV^2(Y,Y)>0$,
by Theorems \ref{thm:consistency_dcov} and \ref{thrm:null_distribution}, we have as $n \to \infty$,
$$
n\widehat{\cR}_n^2 \quad \overset{d}{\longrightarrow} \quad
\frac{\bQ}{\sqrt{\cV^2(X,X) \cV^2(Y,Y)}}.
$$
\end{remark}

\begin{remark}
By Proposition \ref{prop:computation} and also as shown in the proof of the above theorem, $\widehat{\cV}_n^2$ is a rank-$4$ \emph{degenerate} $U$-statistic with estimated parameters (i.e., $\bW_n$), and with $K$ samples. 

Our strategy is to first establish functional weak convergence for characteristic functions on compact sets $\{D(\delta):\delta>0\}$ and apply the
continuous mapping theorem. Then, we use the tools from the $U$-statistics literature to control the integral outside the compact sets.
\end{remark}


For a fixed alternative, that is, for some unbiased distribution $G$ with a distance covariance $\cV^2 > 0$, by Theorem \ref{thm:consistency_dcov}, $n \widehat{\cV}_n^2$ goes to infinity almost surely. In the next subsection, we consider a sequence of alternatives that approaches the null and study the rate under which the test statistics $n \widehat{\cV}_n^2 $ would diverge.

 
\subsection{Changing unbiased distributions}\label{sec:alternative}

In this subsection, we study the asymptotic behaviour of $n\widehat{\cV}_n^2$ under alternatives that approach the null as $n \to \infty$. Specifically, we consider a parametric family of probability densities $\{g_{\rho}(\cdot): \rho \in [0,1]\}$ with respective to some $\sigma$-finite measure $\nu$ on $\bR^{p+q}$. We denote by $\cV^2(\rho)$ the distance covariance between random vectors $X \in \bR^{p}$ and $Y \in \bR^{q}$, where $(X,Y)$ has the joint  density $g_{\rho}$, that is,
\begin{align*}
    \cV^2(\rho) :=  \int_{\mathbb{R}^{p+q}} \frac{|\varphi_{X,Y}(s, t;\rho) - \varphi_X(s;\rho)\varphi_Y(t;\rho)|^2}{c_p c_q \|s\|^{1+p} \|t\|^{1+q}}\,ds\,dt, 
\end{align*}
where $\varphi_{X,Y}(\cdot;\rho)$, $ \varphi_X(\cdot;\rho)$ and $\varphi_Y(\cdot;\rho)$ are respectively the characteristics function of $(X,Y)$, $X$ and $Y$. We will assume $\cV^2(0) = 0$, that is, if $(X,Y)$ has the density $g_0$, then $X$ and $Y$ are independent.

Furthermore, we consider the following triangular array setup. Let $\rho_n \in (0,1)$ be a sequence of positive numbers. For each $n \geq 1$, assume that the unbiased distribution is $G_n$, which has the density $g_{\rho_n}$. For each $1 \leq i \leq K$, define
$$
F_{n,i}(A) := \frac{ \int_A w_i(z) \, dG_n(z)}{W_{n,i}}, \quad A \in \mathcal{B}, \quad \text{ where } \quad
W_{n,i} :=\int_{\mathbb{R}^{p+q}} w_i(z) \, dG_n(z),
$$
and we assume $0 < W_{n,i} < \infty$. Moreover, assume that
\begin{equation*}
	(X_{i,1}^{(n)},Y_{i,1}^{(n)}),\ldots, (X_{i,n_i}^{(n)},Y_{i, n_i}^{(n)}) \stackrel{i.i.d.}{\sim} F_{n,i}.
\end{equation*}
That is, the unbiased distribution changes with the sample size $n$, while the biasing mechanism, i.e., $\{w_i(\cdot):1 \leq i \leq K\}$, does not depend on $n$. Given the data $\{(X_{i,j},Y_{i,j}): 1 \leq i \leq K, 1 \leq j \leq n_i\}$, we compute the NPMLE $\bG_n$ and $\bW_n$ as in \eqref{def:NPMLE} and $\widehat{\cV}_{n}^2$ as in Definition \ref{def:Vn_Rn}.

First, let  $\bm{G}_0$  be the (undirected) graph with vertices in
$\{1,\ldots,K\}$, and edge between $k$ and $\ell$ $(k\neq \ell)$ if and only if
$$
\int \mathbbm{1}(w_{k}(x,y)  > 0 )\mathbbm{1}(w_{\ell}(x,y)  > 0 ) g_0(x,y)\,d\nu(x,y) > 0.
$$
\begin{assumption}
\label{assumption:alter_G0}
~
\begin{enumerate}[(i)]
    \item If $(X,Y)$ has the joint density $g_0$, $X$ and $Y$ are independent, i.e., $\cV^2(0) = 0$.
    
    \item $\bm{G}_0$ is connected.
    
    \item There exists some $\epsilon > 0$ such that for any $(x,y) \in \bR^{p+q}$ and $1 \leq i \leq K$,
    $$
    \epsilon \mathbbm{1}(w_i(x,y)  > 0 ) \leq w_i(x,y)  \leq 1.
    $$
\end{enumerate}
\end{assumption}
Part (iii) in Assumption \ref{assumption:alter_G0} is the same as \cite[Assumption 6]{clemenccon2022statistical}. 
It requires $w_i(\cdot)$ to be bounded above and bounded away from zero uniformly on its support.   The upper bound $1$ is unimportant, as long as $\omega_i(\cdot,\cdot)$ is bounded, since the weight functions are equivalent up to a multiplicative constant. 
 \cite{clemenccon2022statistical} provides non-asymptotic analysis for the NPMLE under multiple biased sampling and our analysis uses their results.

Second, we assume that the densities $\{g_{\rho}\}$ are smooth with respect to $\rho$ near $0$ and its $\rho$-derivatives near zero can be controlled uniformly. 


\begin{assumption}\label{assumption:dominated}
There exists some $\epsilon > 0$ such that the following holds.
    \begin{enumerate}[(i)]
\item   For $\nu$-almost everywhere $(x,y) \in \bR^{p+q}$, the function $\rho \to g_{\rho}(x,y)$ is twice differentiable on $[0,\epsilon)$.

\item There exist functions $U:\bR^{p+q} \to [1,\infty)$ and $\widetilde{U}:\bR^{p+q} \to [1,\infty)$ such that for any $\rho \in [0,\epsilon)$ and $\nu$-almost everywhere $(x,y) \in \bR^{p+q}$,
\begin{align*}
     |g_\rho(x,y)| & \leq  U(x,y) g_0(x,y), \\
     \max\left\{\left(\frac{d g_{\rho}(x,y)}{d \rho}\right)^2, \frac{d^2 g_{\rho}(x,y)}{d \rho^2} \right\} & \leq \widetilde{U}(x,y) g_{\rho}(x,y),
\end{align*}
and furthermore, for a random vector $(X,Y)$ with the density $g_0$, 
\begin{align*}
    \Exp\left[(1+\|X\|)(1+\|Y\|)  U(X,Y)\widetilde{U}(X,Y) \right] < \infty.
\end{align*}
\end{enumerate}
\end{assumption}


\begin{lemma}\label{lemma:alternative_derivatives}
 Suppose that $\cV^2(0) = 0$ and that Assumption \ref{assumption:dominated} holds. Then,  the function  $\rho \mapsto \cV^{2}(\rho)$ is twice differentiable on $[0,\epsilon)$ for some $\epsilon > 0$. Furthermore, its first derivative at zero vanishes, i.e.,
 $$
\left. \frac{d (\cV^2(\rho))}{d \rho} \right\vert_{\rho = 0} = 0.
 $$
\end{lemma}
\begin{proof}
    See Appendix \ref{app:alter_V_derivatives}.
\end{proof}

\begin{remark}
The above lemma concerns properties of the family of unbiased distributions $\{G_{\rho}\}$. It does not concern the biasing mechanism, i.e., $\{w_i(\cdot,\cdot)\}$.
\end{remark}
\begin{remark}
If $X$ and $Y$ have a joint zero-mean normal distribution with
$\text{Var}(X) = \text{Var}(Y) = 1$ and $\text{Cov}(X,Y) = \rho \in [0,1)$,  then by  \cite[Theorem 7]{szekely2007measuring},
$$
\cV^{2}(\rho) = \frac{4}{\pi}\left(
\rho \arcsin(\rho) + \sqrt{1-\rho^2}- \rho \arcsin(\rho/2) - \sqrt{4-\rho^2}+1
\right).
$$
It is elementary to see that for $\rho \in [0,1)$
\begin{align*}
\frac{d (\cV^2(\rho))}{d\rho} &=  \frac{4}{\pi}\left(\arcsin(\rho)- \arcsin(\rho/2)\right), \\
\frac{d^2 (\cV^2(\rho))}{d\rho^2} &=  \frac{4}{\pi}\left(\frac{1}{\sqrt{1-\rho^2}} - \frac{1}{\sqrt{4-\rho^2}}\right).    
\end{align*}
Thus, as expected from Lemma \ref{lemma:alternative_derivatives}, the first derivative at zero vanishes, and the second derivative at zero is $2/\pi$, i.e., a positive quantity.
\end{remark}

We will assume that the second derivative at zero is positive, that is,
\begin{equation}
    \label{def:second_derivative_positve}
\left.\frac{d^2 (\cV^2(\rho))}{d\rho^2} \right\vert_{\rho = 0} > 0.
\end{equation}
By the previous lemma, it implies that 
$\cV^2(\rho)$ is of order $\rho^2$ for $\rho$ that is close to zero. Thus,  if $\rho_n \to 0$ as $n \to \infty$, $\{G_{\rho_n}\}$ is a sequence of unbiased distributions under which $X$ and $Y$ are dependent, that is, a sequence of alternatives approaching the null. We next show that if $\sqrt{n} \rho_n \to \infty$, then $n\widehat{\cV}_{n}^2$ goes to infinity in probability, that is, for any $M > 0$, $\lim_{n \to \infty} \bP(n\widehat{\cV}_{n}^2 \geq M) = 1$. 

We impose the following assumption that strengthens Assumption \ref{assumption:lambda}.

\begin{assumption}
\label{assumption:alter_lambda}
 There exist positive numbers $\lambda_1,\ldots,\lambda_K$ such that for $i = 1,\ldots,K$,
$$
\limsup_{n \to \infty} \sqrt{n}\left|\lambda_{n,i} - \lambda_i\right| < \infty. 
$$
\end{assumption}

\begin{theorem}\label{thrm:alternative}
Suppose that Assumptions  \ref{assumption:alter_G0}, \ref{assumption:dominated} and \ref{assumption:alter_lambda} hold, and that
$\rho_n \to 0$ as $n \to \infty$. As $n \to \infty$, with probability approaching one, the NPMLE $(\bG_n,\bW_n)$ exists and is  unique. 
Furthermore, assume that \eqref{def:second_derivative_positve} holds, 
and  that $\sqrt{n} \rho_n \to \infty$ as $n \to \infty$. Then, $n\widehat{\cV}_{n}^2$ goes to infinity in probability.
\end{theorem}
\begin{proof}
    See Appendix \ref{app:alternative}.
\end{proof}

\begin{remark}
For a parametric family of distributions $\{G_{\rho}\}$,
if $\cV^2(0) = 0$, then for a sequence of alternatives $\{G_{\rho_n}\}$ that approaches the null at a rate slower than the parametric rate $n^{-1/2}$, i.e., $\rho_n \to 0$ and $\sqrt{n}\rho_n \to \infty$, the test statistic goes to infinity in probability.
\end{remark}

\section{Simulation studies}\label{sect:simulation}
We first illustrate the estimation of the distance correlation in the presence of biased samples using our bias-corrected sample distance correlation. We consider the underlying distribution to be a bivariate normal distribution with correlation $\rho$, where its squared distance correlation has a known formula as a function of $\rho$, as shown in Theorem 7 of \cite{szekely2007measuring}.
To assess the performance of our bias-sampled distance correlation estimator, we consider various weight functions listed below. 
\begin{table}[H]
  \centering
    \begin{tabular}{l|lllllll}
    \hline
          & $\mathcal{D}_{1,n}$    & $\mathcal{D}_{2,n}$   & $\mathcal{D}_{3,n}$    & $\mathcal{D}_{4,n}$    & $\mathcal{D}_{5,n}$ & $\mathcal{D}_{6,n}$ &
           \\ \hline
    $w(x, y)$ & $1$  & $\mathbbm{1}(x < 0.5)$ & $\mathbbm{1}(x < 1.5)$ & $\mathbbm{1}(x > -0.5)$ & $|xy|$ & $|x|+|y|$ \\
    \hline
    \end{tabular}%
\end{table}%
In the above table, $\mathcal{D}_{i,n}$ denotes the biased samples of size $n$ according to the corresponding weight function.
We use the notation $\widehat{R}_{N,A}$, where $A \subset \{1,\ldots,6\}$, to denote 
the bias-corrected sample distance correlation using $\cup_{a \in A} \mathcal{D}_{a,n}$ with total sample size being $N$. 
Here, we consider $\widehat{\mathcal{R}}_{2n,\{1,2\}}$, $\widehat{\mathcal{R}}_{2n,\{1,3\}}$, $\widehat{\mathcal{R}}_{2n,\{2, 4\}}$, $\widehat{\mathcal{R}}_{n,\{5\}}$, $\widehat{\mathcal{R}}_{n,\{6\}}$, $\widehat{\mathcal{R}}_{n}^u$ and $\widehat{\mathcal{R}}_{2n}^u$.
For example, $\widehat{R}_{2n,\{1,2\}}$ means we use both $\mathcal{D}_{1,n}$ and $\mathcal{D}_{2,n}$ in the estimation, and there are a total of $2n$ observations.
For comparison, we benchmark them with the sample distance correlation using unbiased data of sample sizes $n$ and $2n$. We denote the corresponding estimates as $\widehat{\mathcal{R}}_{n}^u$ and $\widehat{\mathcal{R}}_{2n}^u$, respectively. Table \ref{table:bivariate_bias_sd}  shows the bias ($\times 10^{-2})$ and the standard deviation ($\times 10^{-2}$) of the estimates with the sample size in each sample being $50$, $100$, $250$ or $500$, and the correlation value being $0.2, 0.4, 0.6$ or $0.8$.  $10,000$ repetitions are used to estimate the biases and standard deviations. The results for negative correlations are similar and therefore omitted.

From Table \ref{table:bivariate_bias_sd}, we observe that the bias-sampled distance correlation estimator works for both one-sample and two-sample cases. The effectiveness of using biased samples to estimate the distance correlation compared to using unbiased samples seems to depend on the weight functions. For instance, $\widehat{R}_{n,\{5\}}$ and $\widehat{R}_{n}^u$ have similar biases across different $\rho$ values, while $\widehat{R}_{n,\{6\}}$ shows larger biases. Notably, $\widehat{R}_{2n,\{2,4\}}$, despite not using any unbiased samples, exhibits similar biases with slightly larger standard deviations compared to $\widehat{R}_{2n}^u$. Comparing $\widehat{R}_{2n,\{1,2\}}$ or $\widehat{R}_{2n,\{1,3\}}$ with $\widehat{R}_n^u$, we see that the inclusion of biased samples can substantially improve the estimation accuracy.

\begin{table}
\centering
\footnotesize
\begin{tabular}{rlllllll}
  &\multicolumn{7}{c}{$\rho = 0.2$  ($\mathcal{R}^2(X, Y) = 0.18$) }  \\ \hline \rule{0pt}{12pt}
 $n$ & $\widehat{\mathcal{R}}_{2n,\{1,2\}}$ &$\widehat{\mathcal{R}}_{2n,\{1,3\}}$& $\widehat{\mathcal{R}}_{2n,\{2, 4\}}$& $\widehat{\mathcal{R}}_{n,\{5\}}$& $\widehat{\mathcal{R}}_{n,\{6\}}$& $\widehat{\mathcal{R}}_{n}^u$ &$\widehat{\mathcal{R}}_{2n}^u$  \\
  \hline
$50$ & 7.65 (7.1) & 6.76 (6.73) & 7.52 (7.04) & 10.21 (6.79) & 16.54 (8.19) & 11.15 (7.57) & 6.03 (6.41) \\ 
  $100$ & 3.96 (5.86) & 3.46 (5.52) & 3.89 (5.81) & 5.75 (5.72) & 10.99 (6.69) & 6.16 (6.44) & 3.21 (5.3) \\ 
  $250$ & 1.61 (4.26) & 1.45 (3.98) & 1.63 (4.28) & 2.4 (4.12) & 6.1 (4.96) & 2.54 (4.92) & 1.22 (3.74) \\ 
  $500$ & 0.79 (3.17) & 0.72 (2.91) & 0.73 (3.14) & 1.2 (3.12) & 3.79 (3.78) & 1.28 (3.76) & 0.63 (2.75) \\ 
   \hline \\
     & \multicolumn{7}{c}{$\rho = 0.4$  ($\mathcal{R}^2(X, Y) = 0.36$) }\\ \hline \rule{0pt}{12pt}
      $n$ & $\widehat{\mathcal{R}}_{2n,\{1,2\}}$ &$\widehat{\mathcal{R}}_{2n,\{1,3\}}$& $\widehat{\mathcal{R}}_{2n,\{2, 4\}}$& $\widehat{\mathcal{R}}_{n,\{5\}}$& $\widehat{\mathcal{R}}_{n,\{6\}}$& $\widehat{\mathcal{R}}_{n}^u$ &$\widehat{\mathcal{R}}_{2n}^u$\\ \hline
  $50$ & 3.38 (8.37) & 3.02 (7.87) & 3.09 (8.3) & 5.24 (8.44) & 11.17 (9.23) & 5.33 (9.42) & 2.49 (7.43) \\ 
 $100$ & 1.76 (6.35) & 1.61 (5.94) & 1.64 (6.39) & 2.73 (6.46) & 7 (7.16) & 2.64 (7.41) & 1.28 (5.56) \\ 
 $250$ & 0.65 (4.19) & 0.63 (3.82) & 0.57 (4.11) & 1.17 (4.25) & 3.67 (4.92) & 1.08 (5.05) & 0.52 (3.55) \\ 
 $500$ & 0.37 (2.99) & 0.37 (2.72) & 0.31 (2.97) & 0.56 (3.07) & 2.18 (3.63) & 0.45 (3.6) & 0.27 (2.58) \\ 
   \hline \\
 & \multicolumn{7}{c}{$\rho = 0.6$  ($\mathcal{R}^2(X, Y) = 0.55$) }\\ \hline \rule{0pt}{12pt}
      $n$ & $\widehat{\mathcal{R}}_{2n,\{1,2\}}$ &$\widehat{\mathcal{R}}_{2n,\{1,3\}}$& $\widehat{\mathcal{R}}_{2n,\{2, 4\}}$& $\widehat{\mathcal{R}}_{n,\{5\}}$& $\widehat{\mathcal{R}}_{n,\{6\}}$& $\widehat{\mathcal{R}}_{n}^u$ &$\widehat{\mathcal{R}}_{2n}^u$\\
     \hline
$50$ & 1.61 (7.34) & 1.49 (6.81) & 1.47 (7.13) & 3.06 (7.86) & 8.42 (9.48) & 2.63 (8.72) & 1.4 (6.38) \\ 
 $100$ & 0.82 (5.35) & 0.79 (4.87) & 0.75 (5.17) & 1.52 (5.82) & 5.33 (7.38) & 1.41 (6.39) & 0.7 (4.67) \\ 
  $250$ & 0.33 (3.37) & 0.29 (3.13) & 0.32 (3.35) & 0.66 (3.74) & 2.8 (5.08) & 0.51 (4.18) & 0.27 (2.93) \\ 
  $500$ & 0.19 (2.38) & 0.17 (2.2) & 0.16 (2.36) & 0.31 (2.67) & 1.61 (3.77) & 0.23 (2.95) & 0.13 (2.1) \\ 
   \hline \\
&  \multicolumn{7}{c}{$\rho = 0.8$  ($\mathcal{R}^2(X, Y) = 0.76$) }\\ \hline \rule{0pt}{12pt}
      $n$ & $\widehat{\mathcal{R}}_{2n,\{1,2\}}$ &$\widehat{\mathcal{R}}_{2n,\{1,3\}}$& $\widehat{\mathcal{R}}_{2n,\{2, 4\}}$& $\widehat{\mathcal{R}}_{n,\{5\}}$& $\widehat{\mathcal{R}}_{n,\{6\}}$& $\widehat{\mathcal{R}}_{n}^u$ &$\widehat{\mathcal{R}}_{2n}^u$\\
     \hline
$50$ & 0.64 (4.66) & 0.58 (4.35) & 0.54 (4.62) & 1.59 (5.83) & 6.05 (8.25) & 1.12 (5.83) & 0.61 (4.2) \\ 
  $100$ & 0.36 (3.37) & 0.34 (3.14) & 0.29 (3.32) & 0.83 (4.2) & 3.81 (6.63) & 0.62 (4.2) & 0.27 (3) \\ 
  $250$ & 0.15 (2.13) & 0.13 (1.97) & 0.11 (2.1) & 0.35 (2.69) & 1.99 (4.66) & 0.22 (2.67) & 0.13 (1.91) \\ 
  $500$ & 0.06 (1.52) & 0.08 (1.4) & 0.06 (1.51) & 0.18 (1.91) & 1.17 (3.62) & 0.1 (1.91) & 0.06 (1.34) \\ 
   \hline
\end{tabular}
\caption{Bias of distance correlation estimate in a bivariate normal model ($\times 10^{-2}$). The standard deviation is in parentheses ($\times 10^{-2}$).}
\label{table:bivariate_bias_sd}
\end{table}

Next, we illustrate testing independence based on multiple biased samples using the proposed bias-corrected sampled distance correlation. Suppose that $Y = \rho (X_1^2 + X_2^2) + \varepsilon$, where $X_1, X_2 \stackrel{i.i.d.}{\sim} U(0, 1)$ and $\varepsilon \sim N(0, 0.5^2)$. Thus, when $\rho \neq 0$, $Y$ is non-linearly dependent on $X_1$ and $X_2$. When $\rho = 0$, $(X_1,X_2)$ and $Y$ are independent. The weight functions considered are listed below.
\begin{table}[H]
  \centering
    \begin{tabular}{l|lllllll}
    \hline
          & $\mathcal{D}_{1,n}$    & $\mathcal{D}_{2,n}$   & $\mathcal{D}_{3,n}$    & $\mathcal{D}_{4,n}$    & $\mathcal{D}_{5,n}$ & 
           \\ \hline
    $w(x, y)$ & $1$  & $\mathbbm{1}(y < x_1)$ & $\mathbbm{1}(x_2 > 0.2)$ & $\mathbbm{1}(x_2 > 0.8)$ & $\mathbbm{1}(y < 1)$  \\
    \hline
    \end{tabular}%
\end{table}%
As before, $\mathcal{D}_{i,n}$ denotes the biased samples of size $n$ according to the corresponding weight function.
We set $n = 100$. Denote $T_{N,A}$, where $A \subset \{1,\ldots,5\}$, as the test based on the bias-corrected sample distance covariance using $\cup_{a \in A} \mathcal{D}_{a,n}$ with total sample size being $N$. Here, we consider $T_{2n,\{1,2\}}$, $T_{2n, \{1,3\}}$, $T_{2n,\{1,4\}}$, $T_{2n,\{1,5\}}$, $T_{2n,\{3,5\}}$, and $T_{2n,\{4,5\}}$. Note that except for $T_{2n,\{1,2\}}$, which involves weight functions that are non-separable in the sense that it cannot be written as the form in \eqref{eq:separable_weight}, we can perform the basic unweighted permutation test to determine the $p$-value because all permutations will be valid and equally likely, and we do not need to apply the MH algorithm. For $T_{2n,\{1,2\}}$, we apply the MH algorithm to draw the permutations. We set the number of permutations to be $500$ in each test.
For comparison, we benchmark against the independence test based on the usual sample distance correlation using unbiased data of sample sizes $n$ and $2n$. We denote the corresponding tests as $T_n^u$ and $T_{2n}^u$, respectively. The empirical proportions of rejection based on $1,000$ replications are given in Table \ref{table:parabola_test}. The significance level of all the tests considered in this article is $\alpha = 0.05$.

From Table \ref{table:parabola_test}, we observe that the permutation test controls the type I error well, as expected. Although all the tests with biased samples have the same total sample size, their powers differ. Intuitively, if the weight function is closer to 1, the sample should more closely resemble the unbiased one. For example, comparing $T_{2n,\{1,3\}}$ and $T_{2n,\{1,4\}}$, we see that $T_{2n,\{1,3\}}$ has higher power, possibly because $\mathcal{D}_{3,n}$ contains more information than $\mathcal{D}_{4,n}$. We also see that adding biased samples to the unbiased samples ($T_{2n,\{1,2\}}$, $T_{2n, \{1,3\}}$, $T_{2n,\{1,4\}}$, $T_{2n,\{1,5\}}$) increases the power of the test compared to using only the unbiased data ($T_n^u$). Depending on the weight function, using a combination of $n$ unbiased data an $n$ biased data can result in a test with comparable power to using $2n$ unbiased data (compare $T_{2n,\{1,5\}}$ and $T_{2n}^u$). Notably, $T_{2n,\{3,5\}}$, which does not use any unbiased samples, performs only slightly worse than using $2n$ unbiased samples.

\begin{table}[H]
\centering
\begin{tabular}{r|rrrrrrrrr}
  \hline
 & $\rho$ & $T_{2n,\{1,2\}}$ & $T_{2n, \{1,3\}}$ & $T_{2n,\{1,4\}}$ & $T_{2n,\{1,5\}}$ & $T_{2n,\{3,5\}}$ & $T_{2n,\{4,5\}}$ & $T_n^u$ & $T_{2n}^u$ \\ 
  \hline
$H_0$ &$0$ & 0.057 & 0.057 & 0.063 & 0.049 & 0.047 & 0.051 & 0.053 & 0.051 \\ \hline
$H_1$ & $0.2$& 0.389 & 0.385 & 0.282 & 0.423 & 0.409 & 0.254 & 0.248 & 0.443 \\ 
  &$0.3$ & 0.700 & 0.767 & 0.584 & 0.771 & 0.777 & 0.503 & 0.546 & 0.834 \\ 
  &$-0.2$ & 0.405 & 0.388 & 0.262 & 0.451 & 0.368 & 0.253 & 0.247 & 0.484 \\ 
  &$-0.3$ & 0.780 & 0.759 & 0.539 & 0.840 & 0.752 & 0.512 & 0.524 & 0.834 \\ 
   \hline
\end{tabular}
\caption{Empirical proportion of rejections under $H_0$ and $H_1$, where $Y = \rho(X_1^2 + X_2^2) + \varepsilon$, for different values of $\rho$ when $n = 100$.}
\label{table:parabola_test}
\end{table}

Now, we consider two other settings where the underlying random vector follows a Clayton copula or a Gumbel copula. For the Clayton copula, the distribution function is $G_\theta(x,y) = (x^{-\theta} + y^{-\theta}-1)^{-1/\theta}$ for $x,y\in[0,1]$ when $\theta > 0$. When $\theta = 0$, $X$ and $Y$ are independent uniform random variables. For the Gumbel copula, the distribution function is $G_\theta(x, y) = \exp[ -((-\log x)^\theta + (-\log y)^\theta)^{1/\theta}]$ for $x,y\in [0, 1]$, $\theta \geq 1$. When $\theta = 1$, $X$ and $Y$ are independent uniform random variables. The weight functions considered for both settings are listed below.
\begin{table}[H]
  \centering
    \begin{tabular}{l|lllllll}
    \hline
          & $\mathcal{D}_{1,n}$    & $\mathcal{D}_{2,n}$   & $\mathcal{D}_{3,n}$  & $\mathcal{D}_{4,n}$ 
           \\ \hline
    $w(x, y)$ & $x+y$  & $xy$ & $x$ & $1$ \\
    \hline
    \end{tabular}%
\end{table}%
We use the same notation $\mathcal{D}_{i,n}$ and $T_{N,A}$, $A \subset \{1,\ldots,4\}$, to denote biased samples and the test based on the bias-corrected sample distance covariance using $\cup_{a \in A} \mathcal{D}_{a,n}$. The tests using the usual sample distance correlation are denoted by $T_{n,\{1\}}^u, \ldots, T_{n,\{4\}}^u$, where $T_{n,\{i\}}^u$ is the test using $\mathcal{D}_{i,n}$. Tables \ref{table:clayton} -- \ref{table:clayton2} and Tables \ref{table:gumbel} -- \ref{table:gumbel2} show the empirical proportion of rejections based on $1,000$ replications for the Clayton and Gumbel copula cases, respectively. For these weight functions, we apply the MH algorithm to draw the permutations. 
Under $H_0$, the tests based on different biased samples have type I error close to $0.05$. When the weight function is $w(x, y) = x + y$, the $X$ and $Y$ in the biased samples become dependent. Thus, $T_{n,\{1\}}^u$ is not a valid test and rejects the null more often than it should. For the weight functions $w(x,y)= xy$ and $w(x,y) = x$, the $X$ and $Y$ in the biased samples are independent if and only if those in the unbiased samples are. Thus, $T_{n,\{2\}}^u$ and $T_{n,\{3\}}^u$ are also valid tests. For the Clayton copula, $T_{n,\{2\}}$ and $T_{n,\{3\}}$ have similar power as $T_{n,\{2\}}^u$ and $T_{n,\{3\}}^u$; while for the Gumbel copula, the test without adjusting the weights $w(x, y) =xy$ and $w(x, y) = x$ are more powerful. In Tables \ref{table:clayton} -- \ref{table:gumbel2}, we see that combining more biased samples results in a more powerful test, as expected.

\begin{table}[H]
\centering
\begin{tabular}{r|rrrrrrrr}
  \hline
 &$\theta$ & $T_{n, \{1\}}$ & $T_{n, \{2\}}$ & $T_{n, \{3\}}$ & $T_{2n, \{1, 2\}}$ & $T_{2n, \{2, 3\}}$ & $T_{2n, \{1, 3\}}$ & $T_{3n, \{1,2,3\}}$ \\  
  \hline
$H_0$ &0 & 0.042 & 0.056 & 0.049 & 0.058 & 0.050 & 0.055 & 0.056 \\  \hline
 $H_1$ &0.2 & 0.174 & 0.090 & 0.119 & 0.196 & 0.158 & 0.246 & 0.263  \\ 
  &0.4 & 0.399 & 0.147 & 0.282 & 0.449 & 0.392 & 0.596 & 0.642 \\ 
  &0.6 & 0.656 & 0.228 & 0.504 & 0.725 & 0.629 & 0.852 & 0.890 \\ 
  &0.8 & 0.840 & 0.342 & 0.694 & 0.897 & 0.818 & 0.967 & 0.983\\ 
  &1 & 0.930 & 0.487 & 0.845 & 0.970 & 0.940 & 0.995 & 0.998  \\ 
   \hline
\end{tabular}
\caption{Empirical proportion of rejections under $H_0$ and $H_1$, where $(X,Y)$ follows the Clayton copula, for different values of $\theta$ when $n = 100$ (Part 1).}
\label{table:clayton}
\end{table}

\begin{table}[H]
\centering
\begin{tabular}{r|rrrrr}
  \hline
 &$\theta$ &  $T_{n,\{1\}}^u$ & $T_{n,\{2\}}^u$ & $T_{n,\{3\}}^u$ & $T_{n,\{4\}}^u$ \\  
  \hline
$H_0$ &0 & 0.127 & 0.055 & 0.060 & 0.039 \\  \hline
 $H_1$ &0.2 & 0.047 & 0.069 & 0.102 & 0.246 \\ 
  &0.4 & 0.080 & 0.151 & 0.269 & 0.696 \\ 
  &0.6 & 0.268 & 0.233 & 0.467 & 0.939 \\ 
  &0.8 & 0.521 & 0.365 & 0.695 & 0.990 \\ 
  &1 & 0.712 & 0.519 & 0.831 & 0.997 \\ 
   \hline
\end{tabular}
\caption{Empirical proportion of rejections under $H_0$ and $H_1$, where $(X,Y)$ follows the Clayton copula, for different values of $\theta$ when $n = 100$ (Part 2).}
\label{table:clayton2}
\end{table}

\begin{table}[H]
\centering
\begin{tabular}{r|rrrrrrrr}
  \hline
 &$\theta$ & $T_{n, \{1\}}$ & $T_{n, \{2\}}$ & $T_{n, \{3\}}$ & $T_{2n, \{1, 2\}}$ & $T_{2n, \{2, 3\}}$ & $T_{2n, \{1, 3\}}$ & $T_{3n, \{1,2,3\}}$  \\  \hline
$H_0$& 1 & 0.053 & 0.046 & 0.049 & 0.057 & 0.048 & 0.053 & 0.053\\  \hline
 $H_1$ &1.1 & 0.138 & 0.078 & 0.138 & 0.144 & 0.148 & 0.212 & 0.231 \\ 
  &1.2 & 0.385 & 0.137 & 0.320 & 0.435 & 0.390 & 0.604 & 0.645 \\ 
  &1.3 & 0.673 & 0.244 & 0.555 & 0.752 & 0.676 & 0.916 & 0.935 \\ 
  &1.4 & 0.860 & 0.357 & 0.719 & 0.924 & 0.835 & 0.979 & 0.987 \\ 
  &1.5 & 0.960 & 0.521 & 0.881 & 0.992 & 0.920 & 0.997 & 0.999 \\ 
   \hline
\end{tabular}
\caption{Empirical proportion of rejections under $H_0$ and $H_1$, where $(X,Y)$ follows the Gumbel copula, for different values of $\theta$ when $n = 100$ (Part 1).}
\label{table:gumbel}
\end{table}

\begin{table}[H]
\centering
\begin{tabular}{r|rrrrr}
  \hline
 &$\theta$ & $T_{n,\{1\}}^u$ & $T_{n,\{2\}}^u$ & $T_{n,\{3\}}^u$ & $T_{n,\{4\}}^u$ \\  \hline
$H_0$& 1 & 0.115 & 0.048 & 0.054 & 0.052 \\  \hline
 $H_1$ & 1.1 & 0.107 & 0.318 & 0.269 & 0.237 \\ 
  &1.2 & 0.429 & 0.733 & 0.663 & 0.649 \\ 
  &1.3 & 0.785 & 0.918 & 0.911 & 0.901 \\ 
  &1.4 & 0.952 & 0.987 & 0.977 & 0.977 \\ 
  &1.5 & 0.993 & 0.999 & 0.997 & 0.998 \\ 
   \hline
\end{tabular}
\caption{Empirical proportion of rejections under $H_0$ and $H_1$, where $(X,Y)$ follows the Gumbel copula, for different values of $\theta$ when $n = 100$ (Part 2).}
\label{table:gumbel2}
\end{table}

\section{Real data applications}\label{sect:real}
In this section, we illustrate the use of the independence test based on our proposed bias-sampled distance covariance estimator on two real datasets, where we perform outcome-dependent sampling, treating the original dataset as the whole population. The first dataset is the Boston housing dataset, which contains housing data for $506$ census tracts of Boston from the $1970$ census. We use the dataframe \texttt{BostonHousing2} from the \texttt{mlbench} package in \texttt{R}. The dataset consists of $19$ attributes including the median value of owner-occupied homes and the crime rate by town, among other variables.
\cite{clemenccon2022statistical} also considered this dataset to illustrate the effectiveness of empirical risk minimization when training observations are generated from biased sampling models.  They sampled $100$ data points uniformly from the whole sample, so that $w_1(z) = 1$, and $200$ data points among the cheapest houses, with values lower than $22$, so that $w_2(z) = \mathbbm{1}(\text{value} \leq 22)$. We follow the same outcome-dependent sampling scheme and consider testing the independence between the crime rate and housing value using our proposed method with both samples.
The $p$-value of our test is $0.002$, indicating that the two variables are not independent.

The second dataset is air quality data consisting of daily measurements of ozone concentration, wind speed, temperature and solar radiation in New York City from May to September of $1973$, with a total of $111$ observations. The dataset is available as the dataframe \texttt{environmental} in the \texttt{R} package \texttt{lattice}. This dataset was also used by \cite{yan2017statistical} to illustrate the application of outcome-dependent sampling designs to estimate the regression coefficients under generalized linear models. Yan et al. concluded that the performance of the outcome-dependent sampling design is very close to the analysis of full data. We follow their design by randomly sampling $40$ observations from the full dataset, an additional $10$ observations from the subset where the ozone level is below the $30$th percentile, and another $10$ observations from the subset where the ozone level is above the $70$th percentile. This sampling scheme corresponds to the weight functions $w_1(z) = 1$, $w_2(z) = \mathbbm{1}(\text{ozone} < 20)$, and $w_3(z) = \mathbbm{1}(\text{ozone} > 49)$. We test the independence of ozone concentration with wind speed, temperature, and solar radiation in three separate tests using our proposed method with all three samples. All the $p$-values are below $0.05$, indicating these variables are not independent of ozone concentration.

These examples illustrate the usage of our approach to combine data to perform tests of independence based on distance covariance in the presence of multiple biased samples.

\section{Discussion}\label{sec:conclusion}
In this article, we study the estimation of distance covariance and distance correlation under multiple biased sampling models, with application to independence testing. Our approach accounts for selection bias and combines multiple data sources. 

In the following, we discuss some potential generalizations and related dependence measures under our framework. First, we have assumed that the weight functions are known. An extension is to consider the semi-parametric multiple biased sampling models studied in \cite{gilbert1999maximum} and \cite{gilbert2000large}, which allows the weight functions to depend on an unknown finite-dimensional parameter. In such cases, we can still follow the approach used here to compute estimates of the distance covariance and correlation after we obtain the estimates of the finite-dimensional parameters for the weight functions and the corresponding estimate of the unbiased distribution.

Next, we can generalize our estimator to the class of $\alpha$-distance dependence measures as described in \cite{szekely2007measuring}, defined as
\begin{equation*}
	\mathcal{V}^{2(\alpha)}(X, Y) := \int_{\mathbb{R}^{p+q}} \frac{|\varphi_{X,Y}(s, t) - \varphi_X(s)\varphi_Y(t)|^2}{c_{p,\alpha} c_{q,\alpha} \|s\|^{\alpha+p} \|t\|^{\alpha+q}}\,ds\,dt,
\end{equation*}
where $c_{p,\alpha}$ and $c_{q,\alpha}$ are some constants. This measure is well-defined when $\mathbb{E}(\|X\|^\alpha + \|Y\|^\alpha) < \infty$. When $\alpha = 1$, it reduces to the distance covariance. Our estimator can be readily adapted to estimate $\alpha$-distance dependence by replacing $\|\cdot\|$ with $\|\cdot\|^\alpha$ in Proposition \ref{prop:computation}. The theoretical properties of the resulting estimator can be established analogously.

The consideration of distance covariance or $\alpha$-distance dependence measure here can be further generalized to the Hilbert–Schmidt independence criterion (HSIC) \cite{gretton2007kernel}, as they are special cases of HSIC with specific choices of kernel \cite{sejdinovic2013equivalence}. Let $k_X$ and $k_Y$ be two kernels, and let $(X,Y), (X', Y'), (X'', Y''), (X''',Y''')$ be independent and identically distributed. The HSIC can be computed as
\begin{equation*}
	\mathbb{E}(k_X(X,X')k_Y(Y,Y')) + \mathbb{E}(k_X(X,X')k_Y(Y'',Y''')) - 2\mathbb{E}(k_X(X,X')k_Y(Y,Y'')).
\end{equation*}
For multiple biased samples, we can estimate HSIC by replacing the above expectations with integrals with respect to the NPMLE $\mathbb{G}_n$:
\begin{align*}
	&\iint k_X(x,x')k_Y(y,y')\, d \mathbb{G}_n(x,y)\, d \mathbb{G}_n(x',y')\\
 & \quad + 
 \iint k_X(x,x') d \mathbb{G}_n(x,y)\, d \mathbb{G}_n(x',y') \iint k_Y(y,y')) d \mathbb{G}_n(x,y)\, d \mathbb{G}_n(x',y')\\ 
 &\quad - 2\ \iiint k_X(x,x')k_Y(y,y'')
 \, d \mathbb{G}_n(x,y) \, d \mathbb{G}_n(x',y') \, d \mathbb{G}_n(x'',y'').
\end{align*}
This will result in a similar form as in Proposition \ref{prop:computation} with kernels replacing the Euclidean distances.
However, in Theorem  \ref{thrm:null_distribution}, we use extensively the properties of characteristic functions in establishing the asymptotic null distribution of our test statistic. It requires different techniques for general kernels, which is left to future research.




\appendix

\section{Proofs for Sections \ref{sect:est_def} and \ref{sect:testing_indep_permutation}}\label{appendix:23}

\begin{proof}[Proof of Proposition \ref{prop:computation}]
From the result in \cite{szekely2007measuring} (by treating $\mathbb{G}_n$ as a probability measure), we obtain
\begin{align*}
	\widehat{\mathcal{V}}^2_n & = \iint \|x_1-x_2\|\cdot\|y_1-y_2\| \,d\mathbb{G}_n(x_1,y_1) \,d\mathbb{G}_n(x_2,y_2) \\
	& \quad - 2 \iiint    \|x_1-x_2\|\cdot\|y_1-y_3\| \, d\mathbb{G}_n(x_1,y_1) \,d\mathbb{G}_n(x_2,y_2) \,d \mathbb{G}_n(x_3, y_3)  \\
	& \quad + \iiiint  \|x_1-x_2\|\cdot \|y_3-y_4\| \,d\mathbb{G}_n(x_1,y_1)\, d\mathbb{G}_n(x_2,y_2) \,d \mathbb{G}_n(x_3, y_3) \,d \mathbb{G}_n(x_4,y_4).
\end{align*}	
The computational form in the proposition then follows as $\mathbb{G}_n(x,y) = \sum^n_{i=1} \widehat{p}_i \mathbbm{1}(\widetilde{X}_i \leq x, \widetilde{Y}_i \leq y)$.
\end{proof}

\begin{proof}[Proof of Lemma \ref{lemma:joint_dist_same}]
Using (\ref{eq:permut_prob}) and the law of total probability,  we have
\begin{align*}
    &f_{\pi(\mathcal{D}_n)}((x_{11},y_{11}),\ldots,(x_{K,n_K}, y_{K,n_K})) \\
    &= \sum_{\pi \in \mathcal{S}_n} \left\{  \prod^K_{i=1}   \frac{\prod_{j=1}^{n_i} f_{i}(x_{ij}, y_{ij})}{  \sum_{\pi_i' \in S_{n_i}} \prod_{j'=1}^{n_i} f_{i}(x_{ij'},y_{i,\pi_i'(j')}) } \prod_{i=1}^K \prod_{j=1}^{n_i} f_{i}(x_{ij}, y_{i,\pi_i(j)}) \right\}\\
    &=  \prod^K_{i=1} \left\{ \sum_{\pi_i \in S_{n_i}} \frac{\prod_{j=1}^{n_i} f_{i}(x_{ij}, y_{ij})}{  \sum_{\pi_i' \in S_{n_i}} \prod_{j'=1}^{n_i} f_{i}(x_{ij'},y_{i,\pi_i'(j')}) }  \prod_{j=1}^{n_i} f_{i}(x_{ij}, y_{i,\pi_i(j)}) \right\}\\
    &= \prod_{i=1}^K \prod_{j=1}^{n_i} f_{i}(x_{ij}, y_{ij}).
\end{align*}
\end{proof}

\begin{proof}[Proof of Proposition \ref{prop:permu_level}]
    First, note that for any $\pi_k \in \mathcal{S}_n$,
\begin{equation*}
\{ \pi(d_n): \pi \in \mathcal{S}_n\} = \{\pi(\pi_k(d_n)) : \pi \in \mathcal{S}_n\}. 
\end{equation*}
Thus,
\begin{align}
    \widehat{R}_{T_n}(\pi_{k}(d_n); t) = \sum_{\pi \in \mathcal{S}_n} \mathbb{P}_0(\pi|\pi_k(d_n)) \mathbbm{1}(T_n(\pi(\pi_k(d_n))) \leq t) 
    = \widehat{R}_{T_n}(d_n;t) .  \label{eq:Rhat_Tn}
\end{align}
By Lemma \ref{lemma:joint_dist_same}, we have $\pi(\mathcal{D}_n) \stackrel{d}{=} \mathcal{D}_n$. Therefore, 
\begin{align}
    \mathbb{E}(\phi_n(\mathcal{D}_n)) &= \mathbb{E}(\phi_n(\pi(\mathcal{D}_n))) \nonumber \\
    &= \mathbb{E} \left[  \mathbb{E}\left( \phi_n(\pi(\mathcal{D}_n)) \bigg|\mathcal{D}_n  \right) \right] \nonumber  \\
    &= \mathbb{E} \left[   \sum_{\pi\in\mathcal{S}_n} \mathbb{P}_0( \pi(\mathcal{D}_n)| \mathcal{D}_n ) 
    \mathbbm{1}(T_n(\pi(\mathcal{D}_n) > (\widehat{R}_{T_n}(\pi(\mathcal{D}_n;\cdot))^{-1}(1-\alpha )\right] \nonumber  \\
    &= \mathbb{E} \left[   \sum_{\pi\in\mathcal{S}_n} \mathbb{P}_0( \pi(\mathcal{D}_n)| \mathcal{D}_n ) 
    \mathbbm{1}(T_n(\pi(\mathcal{D}_n) > (\widehat{R}_{T_n}(\mathcal{D}_n;\cdot))^{-1}(1-\alpha )\right] \nonumber  \\
    &= \mathbb{E} \left[ \mathbb{E} (\mathbbm{1}(T_n(\pi(\mathcal{D}_n))) > (\widehat{R}_{T_n}(\mathcal{D}_n;\cdot))^{-1}(1-\alpha) | \mathcal{D}_n ) \right], \label{eq:permut1}
\end{align}
where the second last equality is due to (\ref{eq:Rhat_Tn}). Note that $T_n(\pi(d_n))$ has distribution function $\widehat{R}_{T_n}(d_n;\cdot)$. Thus, by the property of generalized inverse, 
\begin{equation}\label{eq:permut2}
    \mathbb{E} (\mathbbm{1}(T_n(\pi(\mathcal{D}_n))) > (\widehat{R}_{T_n}(\mathcal{D}_n))^{-1}(1-\alpha) | \mathcal{D}_n ) \leq \alpha.
\end{equation}
The required result follows in view of (\ref{eq:permut1}) and (\ref{eq:permut2}).
\end{proof}

\section{Consistency of bias-sampled distance covariance estimators} \label{app:consistency_dcov}
For an integer $p \geq 1$, define a real-valued function $H_{p}(\cdot; \cdot)$ on $[0,\infty) \times [0,1]$ as follows: $H_p(0;\delta) = H_p(r; 0) = 0$ for any $\delta \in [0,1]$ and $r \in [0,\infty)$, and for any $r > 0$ and $\delta \in (0,1]$, define
\begin{equation}
    \label{def:H_function}
    H_{p}(r; \delta) := \int_{\|z\| \leq \delta \ \cup \ \|z\| \geq 1/\delta} \frac{1-\cos(r z_1)}{\|z\|^{1+p}} dz,
\end{equation}
where $z \in \bR^p$ and $z_1$ is its first coordinate. For the sense of the above integral,  we refer to \cite[Lemma 1]{szekely2007measuring}. 

The following lemma summarizes some elementary facts about the function $H_p(\cdot)$.

\begin{lemma}\label{lemma:H_p}
Let $p \geq 1$ be an integer. 
\begin{enumerate}[(i)]
    \item For each fixed $r \geq 0$,  $H_p(r;\cdot)$ is a continuous, increasing function on $[0,1]$ with   $H_p(r;0) = 0$ and $H_p(r;1) = r c_p$, where $c_p$ is defined in \eqref{def:cp_const}.
\item For any $\theta \in \mathbb{R}^p$ and $\delta \geq 0$,
\begin{align*}
    \int_{\|z\| \leq\delta \ \cup \ \|z\| \geq 1/\delta} \frac{1-\cos(\theta^\top z)}{\|z\|^{1+p}} dz 
    = H_{p}(\|\theta\|;\delta).
\end{align*}
\end{enumerate}
\end{lemma}
\begin{proof}
Part (i) follows from \cite[Lemma 1]{szekely2007measuring} with $\alpha=1$. Part (ii) is due to a change of variable and Part (i).
\end{proof}
Further, for $s \in \bR^{p} \backslash \{0\}$ and $t \in \bR^{q} \backslash \{0\}$, define
\begin{equation}
    \label{def:weight_function}
    \cI(s,t) := \frac{1}{c_p c_q \|s\|^{1+p} \|t\|^{1+q}}.
\end{equation}
Recall that $\varphi_{X,Y}, \varphi_X, \varphi_Y$ are the joint and marginal characteristic functions of the random vector $(X,Y)$, and that $\varphi^{(n)}_{X,Y}, \varphi^{(n)}_X, \varphi^{(n)}_Y$ are the estimated versions under the NPMLE $\bG_n$. For each $(s,t), (x,y) \in \bR^{p+q}$,  define
\begin{equation}\label{def:uv_functions}
    u(s, x) := \exp(\sqrt{-1} s^\top x) - \varphi_X(s) \quad \text{ and } \quad v(t, y) := \exp(\sqrt{-1} t^\top y) - \varphi_Y(t).
\end{equation}

\begin{lemma}\label{lemma:bound_u}
Let $x \in \bR^{p}$ and $\delta \in (0,1]$. Then
\begin{align*}
	\int_{\|s\|  \leq \delta \; \cup \; \|s\| \geq 1/\delta} \frac{|u(s, x)|^2}{ |s|^{1+p}_p}ds   \leq 2  \mathbb{E}\left(H_p(\|x - X\|; \delta) \right).
\end{align*}
Further, 
\begin{align*}
	\int_{\bR^p} \frac{|u(s, x)|^2}{ |s|^{1+p}_p}ds   \leq 2 c_p  \mathbb{E}(\|x - X\|).
\end{align*}
\end{lemma}
\begin{proof}
By definition,
\begin{align*}
|u(s, x)|^2 = 1 + |\varphi_X(s)|^2 - e^{\sqrt{-1} s^\top x} \overline{\varphi_X(s)} - e^{-\sqrt{-1} s^\top x} \varphi_X(s).
\end{align*}
Let $X'$ be a random variable that is independent from $X$ and has the same distribution as $X$. Then 
\begin{align*}
   |u(s, x)|^2  =1 + \mathbb{E}(\cos(s^\top (X-X'))) - 2 \mathbb{E}(\cos(s^\top (x - X))).
\end{align*}
By Lemma \ref{lemma:H_p} Part (ii), we have
\begin{align}\label{u_int_aux1}
	\int_{\|s\|  \leq \delta \ \cup \ \|s\| \geq 1/\delta} \frac{|u(s, x)|^2}{ \|s\|^{1+p}}ds 
 = \Exp\left[2H_p(\|x-X\|;\delta) - H_p(\|X-X'\|;\delta)\right].
\end{align}

Then the first result follows since $H_p(\cdot)$ is a non-negative function. Further, by setting $\delta = 1$, we prove the second result due to Lemma \ref{lemma:H_p} Part (i).
\end{proof}

\begin{lemma}
\label{lemma:simple_cal1}
For any $(s,t) \in \bR^{p+q}$,  we have
\begin{align*}
& |\varphi^{(n)}_{X,Y}(s, t) - \varphi^{(n)}_X(s)\varphi^{(n)}_Y(t)|^2 \\
& \leq 4 \int_{\bR^{p+q}} 
 |u(s, x)|^2 \, d\bG_n(x,y) \int_{\bR^{p+q}} 
 |v(t, y)|^2 \, d\bG_n(x,y).
\end{align*}
\end{lemma}
\begin{proof}
Fix $(s,t) \in \bR^{p+q}$. 
By definition, we have
\begin{align*}
    &\varphi^{(n)}_{X,Y}(s, t) - \varphi^{(n)}_X(s)\varphi^{(n)}_Y(t) \\
     =& \int (u(s,x) +  \varphi_X(s)) (v(t, y) + \varphi_Y(t)) d\mathbb{G}_n(x, y)   \\
    &\quad - \int (u(s,x) +  \varphi_X(s))   d\mathbb{G}_n(x, y)   \int (v(t, y) + \varphi_Y(t)) d\mathbb{G}_n(x, y)   \\
    =&\int u(s, x)v(t, y) d\mathbb{G}_n(x, y) - \int u(s, x) d\mathbb{G}_n(x, y) \int v(t, y) d\mathbb{G}_n(x, y).
\end{align*}
Thus, using   the inequality $|x+y|^2 \leq 2|x|^2 + 2|y|^2$ and due to the Cauchy–Schwarz inequality, we have
\begin{align*}
&	|\varphi^{(n)}_{X,Y}(s, t) - \varphi^{(n)}_X(s)\varphi^{(n)}_Y(t)|^2  \\
 \leq &2 \left| \int u(s, x)v(t, y) d \mathbb{G}_n(x, y) \right|^2 + 2 \left| \int u(s, x) d\mathbb{G}_n(x, y) \int v(t, y) d\mathbb{G}_n(x, y) \right|^2\\
	\leq &4 \int |u(s, x)|^2 d \mathbb{G}_n(x, y)  \int |v(t, y)|^2 d\mathbb{G}_n(x, y) .
\end{align*}
\end{proof}

We now prove Theorem \ref{thm:consistency_dcov}, which follows the argument of the proof of Theorem 2 in \cite{szekely2007measuring}.
\begin{proof}[Proof of Theorem \ref{thm:consistency_dcov}] Clearly, it suffices to prove the first statement regarding the distance covariance estimator.  
For each $\delta \in (0,1)$, 
recall the definition of $D(\delta)$ in \eqref{def:D_delta}
 and define
\begin{align*}
&	\mathcal{V}^2_{\delta}:=  \int_{D(\delta)} {|\varphi_{X,Y}(s, t) - \varphi_X(s)\varphi_Y(t)|^2}{ \cI(s,t)}\,ds\,dt, \quad \text{ and } \quad\\
&	\widehat{\mathcal{V}}^2_{n,\delta}:=  \int_{D(\delta)} {|\varphi^{(n)}_{X,Y}(s, t) - \varphi^{(n)}_X(s)\varphi^{(n)}_Y(t)|^2}{ \cI(s,t)}\,ds\,dt.
\end{align*}
By Lemma \ref{lemma:consistency_G}, for each $(s, t) \in \bR^{p+q}$, $\varphi^{(n)}_{X,Y}(s, t)$ converges to $\varphi_{X,Y}(s, t)$ almost surely. Thus, by the bounded convergence theorem, we have, with probability $1$,
\begin{equation*}
	\lim_{n \rightarrow \infty}\widehat{\mathcal{V}}^2_{n,\delta} =\mathcal{V}^2_{\delta}.
\end{equation*}
By the triangle inequality,
\begin{equation*}
	|\widehat{\mathcal{V}}^2_n - \mathcal{V}^2| \leq |\widehat{\mathcal{V}}_n^2 - \widehat{\mathcal{V}}_{n,\delta}^2| + |\widehat{\mathcal{V}}_{n,\delta}^2 - \mathcal{V}_{\delta}^2| + |{\mathcal{V}}_{\delta}^2 - \mathcal{V}^2|.
\end{equation*}
Clearly, $\lim_{\delta \downarrow 0} \mathcal{V}_\delta^2 = \mathcal{V}^2$. Thus, it remains to show that
\begin{equation}\label{aux1}
	\limsup_{\delta \downarrow 0} \limsup_{n \rightarrow \infty}|\widehat{\mathcal{V}}_n^2 - \widehat{\mathcal{V}}_{n,\delta}^2| = 0.
\end{equation}
For each $\delta \in (0,1)$, we have
\begin{align}
\begin{split}
    |\widehat{\mathcal{V}}_n^2 - \widehat{\mathcal{V}}_{n,\delta}^2| & \leq \int_{\|s\| \leq \delta \ \cup \ \|s\| \geq \delta^{-1}} | \varphi^{(n)}_{X,Y}(s, t) - \varphi^{(n)}_X(s)\varphi^{(n)}_Y(t)|^2 \cI(s,t) \,ds\,dt.  \\
	& \quad + \int_{\|t\| \leq \delta \ \cup \ \|t\| \geq \delta^{-1}}  | \varphi^{(n)}_{X,Y}(s, t) - \varphi^{(n)}_X(s)\varphi^{(n)}_Y(t)|^2 \cI(s,t) \,ds\,dt.  
 \end{split}
 \label{eq:4terms}
\end{align}
By Lemmas \ref{lemma:simple_cal1} and \ref{lemma:bound_u},
\begin{align*}
&	\int_{\|s\| \leq \delta \ \cup \ \|s\| \geq \delta^{-1}} | \varphi^{(n)}_{X,Y}(s, t) - \varphi^{(n)}_X(s)\varphi^{(n)}_Y(t)|^2 \cI(s,t) \,ds\,dt  \\
\leq & 4 \int \left( \int_{\|s\| \leq \delta \ \cup \ \|s\| \geq \delta^{-1}} \frac{|u(s, x)|^2}{c_p |s|^{1+p}_p}ds\right) d \mathbb{G}_n(x, y) \int \left( \int_{\mathbb{R}^q} \frac{|v(t, y)|^2}{c_q |t|^{1+q}_q}dt \right) d \mathbb{G}_n(x ,y) \\
\leq & \frac{16}{c_p } \int \mathbb{E}\|y - Y\| d \mathbb{G}_n(x, y)\int \mathbb{E}\left(H_p(\|x-X\|;\delta) \right) d \mathbb{G}_n(x, y).
\end{align*}

By the weak convergence of $\bG_n$ to $G$ in Lemma \ref{lemma:consistency_G} and the Portmanteau lemma \cite[Theorem 1.3.4]{vanderVaart1996}, since  $\Exp(\|X\| + \|Y\|) < \infty$, we have that almost surely
\begin{align*}
&	\limsup_{n \rightarrow \infty} \int_{\|s\| \leq \delta \ \cup \ \|s\| \geq \delta^{-1}} | \varphi^{(n)}_{X,Y}(s, t) - \varphi^{(n)}_X(s)\varphi^{(n)}_Y(t)|^2 \cI(s,t) \,ds\,dt  \\
\leq &\frac{16}{c_p} \int  \mathbb{E}\|y - Y\| d G(x, y)\int \mathbb{E}  \left(H_p(\|x-X\|;\delta) \right)  d G(x, y)\\
\leq &\frac{32}{c_p}   \mathbb{E}\|Y\|   \ \mathbb{E} \left(H_p(\|X'-X\|;\delta) \right),
\end{align*}
where $X'$ is independent from $X$ and has the same distribution as $X$. 

Since  $\Exp(\|X\| + \|Y\|) < \infty$ and $0 \leq H_p(\|X'-X\|;\delta) \leq c_p \|X'-X\|$ (Part (i) in Lemma \ref{lemma:H_p}), by the dominated convergence theorem,
\begin{equation*}
\limsup_{\delta \downarrow 0}	\limsup_{n \rightarrow \infty} \int_{\|s\| \leq \delta \ \cup \ \|s\| \geq \delta^{-1}} | \varphi^{(n)}_{X,Y}(s, t) - \varphi^{(n)}_X(s)\varphi^{(n)}_Y(t)|^2 \cI(s,t) \,ds\,dt   = 0.
\end{equation*}
By the same argument, the second term on the right-hand side of \eqref{eq:4terms} also goes to zero as $n \to \infty$ and $\delta \downarrow 0$. Then the proof is complete.
\end{proof}

\section{Proofs regarding the limiting Gaussian process under the null}
First, we introduce some notations that shall be frequently used.  Recall the NPMLE in \eqref{def:NPMLE} and denote by
$$
\mathbb{Z}_n := \sqrt{n}\left( \mathbb{G}_n - G \right).
$$
Let $(X,Y)$, $(X',Y')$, $(X'',Y'')$ and $(X''',Y''')$ be independent and identically distributed random vectors with the unbiased distribution $G$, i.e.,
\begin{equation}\label{def:independ_copies}
    (X,Y), \, (X',Y'), \, (X'',Y'')\, (X''',Y''')\, \quad \overset{i.i.d.}{\sim} \quad G.
\end{equation}
For a function $f:\bR^{p+q} \to \bR^{d}$, for some integer $d \geq 1$, we denote by $G(f) := \int f(x,y) dG(x,y)$ whenever the integral is defined, where the integration is component-wise when $d > 1$.

\subsection{Proof of Theorem \ref{thrm:convergence_of_characterists}}\label{app:cf_distribution}
Note that for each $(s,t) \in \mathbb{R}^{p} \times \mathbb{R}^q$,
\begin{align*}
  &\sqrt{n}  \left(\varphi_{X,Y}^{(n)}(s,t) - \varphi^{(n)}_X(s) \varphi^{(n)}_Y(t) \right)  
  =    \int e^{\sqrt{-1}(s^T x + t^T y)} \, d\mathbb{Z}_n(x, y) + \sqrt{n} \varphi_{X,Y}(s,t) \\
  &\qquad - \frac{1}{\sqrt{n}}\left(\int e^{\sqrt{-1}(s^T x )} \, d\mathbb{Z}_n(x, y) + \sqrt{n} \varphi_{X}(s) \right)
  \left(\int e^{\sqrt{-1}(t^T y )} \, d\mathbb{Z}_n(x, y) + \sqrt{n} \varphi_{Y}(t) \right) \\
  = &\int \left(e^{\sqrt{-1}(s^T x + t^T y)} -  e^{\sqrt{-1}(s^T x )} \varphi_{Y}(t)
  - \varphi_{X}(s) e^{\sqrt{-1}(t^T y )} \right)
   \, d\mathbb{Z}_n(x, y) \\
   &- \frac{1}{\sqrt{n}}\left(\int e^{\sqrt{-1}(s^T x )} \, d\mathbb{Z}_n(x, y) \right)
  \left(\int e^{\sqrt{-1}(t^T y )} \, d\mathbb{Z}_n(x, y)\right) 
 + \sqrt{n} \left(\varphi_{X,Y}(s,t)  - \varphi_{X}(s)\varphi_{Y}(t)\right).
\end{align*}
Under the null, the last term is zero. Further, since $\int \, d\mathbb{Z}_n(x, y) = 0$, we have
\begin{align}\label{aux:cf_comp}
\begin{split}
          \sqrt{n}  \left(\varphi_{X,Y}^{(n)}(s,t) - \varphi^{(n)}_X(s) \varphi^{(n)}_Y(t) \right)  &= \int \left(u(s, x) v(t,y)\right)
   \, d\mathbb{Z}_n(x, y)  \\
   &\qquad - \frac{1}{\sqrt{n}}\left(\int u(s, x) \, d\mathbb{Z}_n(x, y) \right) \left(\int v(t,y) \, d\mathbb{Z}_n(x, y)\right).
   \end{split}
\end{align}
where we recall that $u(\cdot)$ and $v(\cdot)$ are defined in \eqref{def:uv_functions}.

Denote by $\Bar{F} := \sum_{i=1}^{K} \lambda_i F_i$. By the definition of $r(\cdot)$   in  \eqref{def:r_underline_omega}, $dG/ d\Bar{F} = r$.

\begin{proof}[Proof of Theorem \ref{thrm:convergence_of_characterists}]
Recall the definitions of $h_{s,t}(\cdot)$, $r(\cdot),  \underline{\omega}(\cdot)$, $D(\delta)$, and the covariance function $C(\cdot)$ in \eqref{def:hst_xy}, \eqref{def:r_underline_omega}, \eqref{def:D_delta} and \eqref{def:cov_function}. Define a constant function $h_e(x,y) := 4$ for $(x,y) \in \bR^{p+q}$, and note that for any $(s,t) \in \bR^{p+q}$, $|h_{s,t}(x,y)| \leq h_e(x,y)$ for any $(x,y) \in \bR^{p+q}$. Further,the condition \eqref{def:moment_cond1} implies that 
$\partial_s \varphi_X(s) = \mathbb{E}(\sqrt{-1}Xe^{\sqrt{-1}s^\top X})$, and thus
\begin{align*}
\partial_s h_{s,t}(x,y)  = \sqrt{-1} x h_{s,t}(x,y) + \sqrt{-1}\left(x \varphi_X(s) - \mathbb{E}(X e^{\sqrt{-1}s^\top X})\right)\left(e^{\sqrt{-1}(t^\top y )} -\varphi_Y(t) \right), \\
\partial_t h_{s,t}(x,y)  = \sqrt{-1} y h_{s,t}(x,y) + 
\sqrt{-1}\left(e^{\sqrt{-1}(s^\top x)} - \varphi_X(s)\right) \left(y \varphi_Y(t) - \mathbb{E}(Y e^{\sqrt{-1} t^\top Y})\right),
\end{align*}
which implies that
\begin{align*}
  \|\partial_s h_{s,t}(x,y)\| \leq 4 \|x\| + 2\left(\|x\| + \sqrt{p} \Exp\|X\|\right),\quad
  \|\partial_t h_{s,t}(x,y)\| \leq 4 \|y\| + 2\left(\|y\| + \sqrt{q} \Exp\|Y\|\right).
\end{align*}
Note that $r(x,y) \underline{\omega}_i(x,y) \leq 1/\lambda_i$ for $1 \leq i \leq K$, and that the condition \eqref{def:moment_cond1} implies that $\Exp[(1+r(X,Y))(1+\|X\| + \|Y\|)] < \infty$. Thus,
$$
\sup_{(s,t),(s',t') \in \bR^{p+q}} \left\|
\left( (\partial_s , \partial_t, \partial_{s'}, \partial_{t'})^\top C((s,t),(s',t')) \right)
\right\| < \infty.
$$
Then by Kolmogorov-Chentsov Theorem \cite[Theorem 2.8]{alma998444753405158} (see Problem 2.9 therein for the multivariate case), for any $\delta \in (0,1)$, there exists a Gaussian process $\bZ^{(\delta)}$ on $D(\delta)$ with the covariance function $C(\cdot,\cdot)$ such that it has almost surely continuous sample paths. Then by the same argument as in \cite[Corollary 2.11]{alma998444753405158}, we can extend it to $\bR^{p+q}$. This shows the existence of $\bZ$.

For each $\delta \in (0,1)$, define a family of functions on $\bR^{p+q}$ indexed by $(s,t) \in D(\delta)$ as follows:
$$
\cH_{\delta} := \{h_{s,t} r: (s,t) \in D(\delta)\}.
$$
By the same argument as before,
\begin{align*}
    &\|\partial_s (h_{s,t}(x,y) r(x,y))\| \leq r(x,y) (6\|x\| + 2\sqrt{p} \Exp\|X\|), \\
    &\|\partial_t (h_{s,t}(x,y) r(x,y))\|  \leq r(x,y) (6\|y\| + 2\sqrt{q} \Exp\|Y\|).
\end{align*}
Thus, we have for $(s,t), (x,y) \in \bR^{p+q}$,
$$
\left\|\left(\partial_s h_{s,t}(x,y),\ \partial_t h_{s,t}(x,y) \right)^\top \right\| \leq 6r(x,y) (\|x\| + \|y\| + \sqrt{p} \Exp\|X\| +\sqrt{q} \Exp\|Y\|).
$$
 The condition \eqref{def:moment_cond1} implies that $\Exp[r(X,Y)(1+\|X\|^2 + \|Y\|^2)] < \infty$, which implies that for each $ 1 \leq i \leq K$,
$$
\int r(x,y)^2(1+\|x\| + \|y\|)^2\  dF_i(x,y) < \infty.
$$
Thus by \cite{van2000asymptotic} (see Example 19.7),  $\cH_{\delta}$ is $F_i$-Donsker for each $ 1 \leq i \leq K$ \cite{van2000asymptotic}, which implies that the condition in Equation (2.8) of \cite{gill1988large} is fulfilled due to \cite[Theorem 1.3]{dudley1983invariance}. Further, $G(h_e r) = 4G(r) < \infty$ by condition \eqref{def:moment_cond1}. As a result, by \cite[Theorem 2.2]{gill1988large}, there exists a zero mean Gaussian process $\mathbb{\xi}^{(\delta)} := \{\mathbb{\xi}^{(\delta)}(s,t): (s,t) \in D(\delta) \}$ with the covariance function in \eqref{def:cov_function} such that
$$
\sup_{(s,t) \in D(\delta)}\left|\int h_{s,t}(x,y) d \bZ_n(x,y) - \mathbb{\xi}^{(\delta)}(s,t) \right| \to 0 \quad \text{ in probability.}
$$
Since $\left\{\bZ(s,t): (s,t) \in D(\delta)\right\}$ and $\mathbb{\xi}^{(\delta)}$ have the same covariance function, they have the same distribution as random elements in $\mathcal{C}(D_{\delta})$. This implies that
$$
\left\{\int h_{s,t}(x,y) d \bZ_n(x,y): \ (s,t) \in D(\delta) \right\} \quad \overset{d}{\longrightarrow}
\quad \left\{\bZ(s,t): (s,t) \in D(\delta)\right\}
$$
in the space $\mathcal{C}(D_{\delta})$. Finally, by the same argument, we can show that
$$
\sup_{(s,t) \in D(\delta)} \left|\int u(s, x) \, d\mathbb{Z}_n(x, y)\right| + \left|\int v(t,y) \, d\mathbb{Z}_n(x, y)\right|
$$
is bounded in probability. As a result, due to \eqref{aux:cf_comp},
$$
\left\{\sqrt{n}  \left(\varphi_{X,Y}^{(n)}(s,t) - \varphi^{(n)}_X(s)\varphi^{(n)}_Y(t)\right) \ : (s,t) \in D(\delta) \right\} \quad \overset{d}{\longrightarrow} \quad \left\{\bZ(s,t): (s,t) \in D(\delta)\right\}
$$
in the space $\mathcal{C}(D_{\delta})$. The proof is complete.
\end{proof}

\subsection{Bounding the integral of the limiting Gaussian process}\label{app:GP_integral}

\begin{lemma}\label{lemma:GP_integral}
Assume the conditions in Theorem \ref{thrm:convergence_of_characterists} hold. Then $\bP(\bQ < \infty) = 1$, where $\bQ$ is defined in Theorem \ref{thrm:null_distribution}.
\end{lemma}

\begin{proof}
It suffices to show that $\Exp[\bQ] < \infty$. Recall the definition of the covariance function in \eqref{def:cov_function}. For $(s,t) \in \bR^{p+q}$,     $\Exp\left[ |\bZ(s,t)|^2\right] = C((s,t), (s,t))$. Since by definition $0 \leq r(x,y) \underline{w}_i \leq 1/\lambda_i$
for $i = 1,\ldots, K$, there exits some constant $L > 0$ such that
\begin{equation*}
    \Exp\left[ |\bZ(s,t)|^2\right] \leq L \Exp\left[(1+r(X_,Y)) |h_{s,t}(X,Y)|^2 \right]
\end{equation*}
Recall that $(X',Y')$ and $(X'',Y'')$ in \eqref{def:independ_copies} are independent copies of $(X,Y)$. Note that 
\begin{align*}
    \left| e^{\sqrt{-1}(s^\top X)} - \varphi_X(s)  \right|^2
 = \mathbb{E} \left[ 1 - 2 \cos(s^T(X - X')) + \cos(s^\top (X' - X^{''})) \ \vert \ X,Y
 \right].
\end{align*}
As a result, by the law of iterated expectation,
\begin{align*}
     \Exp\left[ |\bZ(s,t)|^2\right] \leq L \Exp & \left[
    (1+r(X, Y)) \left(1 - 2 \cos(s^\top (X - X^{'})) + \cos(s^\top (X^{'} - X^{''})) \right) \right. \\
    &\quad \left.  \left(1 - 2 \cos(t^\top (Y - Y^{'})) + \cos(t^\top (Y^{'} - Y^{''}))\right)
 \right]
\end{align*}
Note that 
\begin{equation*}
    1 - 2 \cos(s^\top(X - X^{'}) + \cos(s^\top (X^{'} - X^{''})
= 2[1- \cos(s^\top (X - X^{'}))] - [1- \cos(s^\top (X^{'} - X^{''}))],
\end{equation*}
which, by \cite[Lemma 1]{szekely2007measuring}, implies that 
\begin{align*}
\mathbb{E}(Q) &= \mathbb{E} \left[   \int \frac{|\bZ(s,t)|^2}{c_p c_q \|s\|^{1+p} \|t\|^{1+q}} ds dt\right]  \\
&\leq   L \mathbb{E} \left[ (1+r(X,Y)) \left(2\|X -X^{'}\| - \|X^{'} - X^{''}\|\right)
 \left(2\|Y -Y^{'}\|  - \|Y^{'} - Y^{''}\|\right)
\right] \\
&\leq  4L \mathbb{E}  \left[(1+r(X,Y))(\|X-X'\|\|Y-Y'\|+\|X'-X''\|\|Y'-Y''\|) \right],
\end{align*}
which is finite under Condition (\ref{def:moment_cond1}). The proof is complete.
\end{proof}

\section{A Lemma on independence}

\begin{lemma}\label{lemma:independence lemma}
Let $(X,Y)$, $(X',Y')$, $(X'',Y'')$ and $(X''',Y''')$ be independent and identically distributed random vectors from the unbiased distribution $G$.  Let $f: \bR^{2p} \to [0,\infty)$, $g:\bR^{2q} \to [0,\infty)$ and $h: \bR^{p+q} \to [0,\infty)$ be three deterministic nonnegative functions.
Assume further that $f$ and $g$ are symmetric, that is, 
$$
f(x,x') = f(x',x) \text{ for } x, x' \in \bR^p,\quad
\text{ and } \quad
g(y,y') = g(y',y) \text{ for } y, y' \in \bR^q.
$$
Under the null that $X$
 and $Y$ are independent, 
 \begin{align*}
&\Exp\left[f(X,X') (g(Y,Y') + g(Y'', Y'''))(h(X,Y) +h(X',Y')+h(X'',Y'')+h(X''',Y''') )\right]\\
&=2\Exp\left[ f(X,X') g(Y,Y'') (h(X,Y) +h(X',Y')+h(X'',Y'')+h(X''',Y''') ) \right].
\end{align*}
 \end{lemma}
 
 \begin{proof}
Since $(X,Y)$, $(X',Y')$, $(X'',Y'')$ and $(X''',Y''')$ are i.i.d., due to the symmetry of $f,g$, we have that
\begin{align*}
 &\Exp\left[f(X,X') g(Y,Y') (h(X,Y) +h(X',Y')+h(X'',Y'')+h(X''',Y''') )\right]\\
= &  2\Exp\left[f(X,X') g(Y,Y') h(X,Y)\right] +2\Exp\left[f(X,X') g(Y,Y')\right] \Exp\left[ h(X,Y)\right],
\end{align*}
and that
\begin{align*}
 &\Exp\left[f(X,X') g(Y'',Y''') (h(X,Y) +h(X',Y')+h(X'',Y'')+h(X''',Y''') )\right]\\
= &  2\Exp\left[f(X,X') h(X,Y)\right] \Exp\left[g(Y,Y')\right] +2\Exp\left[f(X,X')\right]  \Exp\left[ g(Y,Y') h(X,Y)\right],
\end{align*}
and that 
\begin{align*}
 &2\Exp\left[f(X,X') g(Y,Y'') (h(X,Y) +h(X',Y')+h(X'',Y'')+h(X''',Y''') )\right]\\
= &  2\Exp\left[f(X,X') g(Y,Y'') h(X,Y)\right] +2\Exp\left[f(X,X') g(Y,Y'') h(X',Y')\right] \\
&+ 2\Exp\left[f(X,X') g(Y,Y'') h(X'',Y'')\right]
+ 2\Exp\left[f(X,X') g(Y,Y'')\right] \Exp\left[ h(X,Y)\right]. 
\end{align*}

Now under the null that $X$ and $Y$ are independent, we have that $(X,Y), (X', Y')$ have the same distribution as $(X,Y),(X',Y'')$, and thus
\begin{align*}
&\Exp\left[f(X,X') g(Y,Y'') h(X,Y)\right] =
\Exp\left[f(X,X') g(Y,Y') h(X,Y)\right].\\
&\Exp\left[f(X,X') g(Y,Y'')\right] \Exp\left[ h(X,Y)\right] =
    \Exp\left[f(X,X') g(Y,Y')\right] \Exp\left[ h(X,Y)\right].
\end{align*}
Again due to independence and the symmetry of the function $f$, we have
\begin{align*}
 \Exp\left[f(X,X') g(Y,Y'') h(X',Y')\right]
 =& \Exp\left[f(X,X')  h(X',Y')\right] \Exp\left[ g(Y,Y'')\right] \\
 =& \Exp\left[f(X,X')  h(X,Y)\right] \Exp\left[ g(Y,Y')\right]
\end{align*}
Similarly, due to independence and the symmetry of the function $g$, we have
\begin{align*}
    \Exp\left[f(X,X') g(Y,Y'') h(X'',Y'')\right]
    =  &   \Exp\left[f(X,X')\right] \Exp\left[ g(Y,Y'') h(X'',Y'')\right] \\
    = &\Exp\left[f(X,X')\right] \Exp\left[ g(Y,Y') h(X,Y)\right].
\end{align*}
Then the proof is complete by combining the above equalities.

\end{proof}

\section{Proofs regarding the limiting distribution under the null}\label{app:null_distribution}

\begin{proof}[Proof of Theorem \ref{thrm:null_distribution}]
 By Theorem \ref{thrm:convergence_of_characterists} and the continuous mapping theorem \cite[Theorem 2.7]{billingsley2013convergence}, for any $\delta \in (0,1)$,  
\begin{align*}
    \int_{D(\delta)} \frac{\left|\sqrt{n}  \left(\varphi_{X,Y}^{(n)}(s,t) - \varphi^{(n)}_X(s) \varphi^{(n)}_Y(t) \right)\right|^2}{|s|_p^{1+p}|t|_q^{1+q}} \,ds\, dt 
\quad \overset{d}{\longrightarrow} \quad
 \int_{D(\delta)} \frac{|\bZ(s,t)|^2}{c_p c_q \|s\|^{1+p} \|t\|^{1+q}}\,ds\,dt.
\end{align*}

In Lemma \ref{lemma:GP_integral}, Appendix \ref{app:GP_integral}, we show that $\bP(\bQ < \infty) = 1$. Thus, by the monotone convergence theorem, as $\delta \downarrow 0$,
$$
\int_{D(\delta)} \frac{|\bZ(s,t)|^2}{c_p c_q \|s\|^{1+p} \|t\|^{1+q}}\,ds\,dt \quad \overset{d}{\longrightarrow} \quad \bQ.
$$
Furthermore,  Lemma \ref{lemma:bound_outside_compacta} in Appendix \ref{app:null_distribution} (ahead) implies that for any $\epsilon > 0$ 
\begin{align*} 
\lim_{\delta \downarrow 0} \limsup_{n \to \infty}   \bP\left(\int_{ D(\delta)^c} \frac{\left|\sqrt{n}  \left(\varphi_{X,Y}^{(n)}(s,t) - \varphi^{(n)}_X(s) \varphi^{(n)}_Y(t) \right)\right|^2}{c_p c_q\|s\|^{1+p}\|t\|^{1+q}} ds dt \geq \epsilon \right) = 0,
\end{align*}
Then, the proof is complete due to \cite[Theorem 3.2]{billingsley2013convergence}
\end{proof}

We introduce additional notations. Define an index set as follows:
$$
\cI := \left\{(i,j): 1 \leq i \leq K, 1 \leq j \leq n_i\right\}.
$$
For $(i,j),(i',j') \in \cI$ and $\delta > 0$, define
\begin{align}\label{def:cross_prod}
A_{ij, i'j'}(\delta) := c_p^{-1} H_p(\|X_{ij} - X_{i'j'}\|;\delta), \quad B_{ij,i'j'}:=  \|Y_{ij} - Y_{i'j'}\|,
\end{align}
where $H_p(\cdot)$ is defined in \eqref{def:H_function}. Furthermore, for $\theta \in \bR^{K}$ and $\delta \in (0,1]$, define
\begin{align}
    \label{def:Delta_n}
    \begin{split}
        \Delta_n(\delta, \theta) := \frac{1}{n^3} &\sum_{(i,j),(i',j'),(i'',j''), (i''',j''') \in \mathcal{I}} 
    A_{ij,i'j'}(\delta) 
\left(B_{ij,i'j'} + B_{i''j'',i'''j'''}
- 2 B_{ij, i''j''}
\right)\\
&\widetilde{\omega}_n^{-1}(X_{ij},Y_{ij}; \theta)
\widetilde{\omega}_n^{-1}(X_{i'j'},Y_{i'j'}; \theta)
\widetilde{\omega}_n^{-1}(X_{i''j''},Y_{i''j''}; \theta)
\widetilde{\omega}_n^{-1}(X_{i'''j'''},Y_{i'''j'''}; \theta),
    \end{split}
\end{align}
where $\widetilde{\omega}_n(\cdot)$ is defined in  \eqref{def:avg_weight_funcs}.
For $(i,j) \in \cI$ and $n \geq 1$, define
$$
\widetilde{w}_{n,ij} := \sum_{i=1}^{K} \frac{\lambda_{n,i} \omega_i(X_{ij},Y_{ij})}{\bW_{n,i}} = \widetilde{\omega}_n(X_{ij},Y_{ij}; \bW_n),
$$
 By the definition of NPMLE in \eqref{def:NPMLE}, due to the second equation, we have 
\begin{equation}
    \label{obs:normalization}
1 = \bG_n(\bR^{p+q}) = \frac{1}{n}\sum_{(i,j) \in \cI} \widetilde{w}_{n,ij}^{-1}.
\end{equation}
We will denote by $V(s,t)$ a term such that
$$
\int_{A} \int_{B} V(s,t) ds dt = 0,
$$
for any set $A \subset  \mathbb{R}^{q}$ and $B \subset  \mathbb{R}^{p}$ that are rotationally invariant. Its specific form may vary from line to line.

\begin{lemma}
    \label{lemma:bound_outside_compacta}
Assume that the conditions in Theorem \ref{thrm:null_distribution} hold. For any $\epsilon > 0$, we have
\begin{align*}
&\lim_{\delta \downarrow 0} \limsup_{n \to \infty}   \bP\left(\int_{\|s\| \leq \delta \ \cup \ \|s\| \geq 1/\delta} \int_{t \in \bR^q} \frac{\left|\sqrt{n}  \left(\varphi_{X,Y}^{(n)}(s,t) - \varphi^{(n)}_X(s) \varphi^{(n)}_Y(t) \right)\right|^2}{c_p c_q\|s\|^{1+p}\|t\|^{1+q}} \,ds\,dt \geq \epsilon \right) = 0,\\
&\lim_{\delta \downarrow 0} \limsup_{n \to \infty}   \bP\left(\int_{s \in \bR^p}\int_{\|t\| \leq \delta \ \cup \ \|t\| \geq 1/\delta}  \frac{\left|\sqrt{n}  \left(\varphi_{X,Y}^{(n)}(s,t) - \varphi^{(n)}_X(s) \varphi^{(n)}_Y(t) \right)\right|^2}{c_p c_q\|s\|^{1+p}\|t\|^{1+q}}\, ds\, dt \geq \epsilon \right) = 0.
\end{align*}
\end{lemma}
\begin{proof}
We will focus on the first statement since the same argument would work for the second statement. 

For a fixed pair $(s,t) \in \mathbb{R}^{p+q}$, we first  compute
$|\varphi_{X,Y}^{(n)}(s,t) - \varphi^{(n)}_X(s) \varphi^{(n)}_Y(t)|^2$. Note that 
\begin{align*}
    &\varphi_{X,Y}^{(n)}(s,t) \overline{\varphi_{X,Y}^{(n)}(s,t)} = \frac{1}{n^2}  \sum_{(i,j), (i',j') \in \cI} e^{\sqrt{-1} (s^\top  X_{ij} + t^\top  Y_{ij})} e^{-\sqrt{-1} (s^\top  X_{i'j'} + t^\top  Y_{i'j'})} \widetilde{\omega}_{n,ij}^{-1}\widetilde{\omega}_{n,i'j'}^{-1} \\
    = &\frac{1}{n^2}  \sum_{(i,j), (i',j') \in \cI} \cos(s^\top (X_{ij} - X_{i'j'}))\cos(t^\top (Y_{ij} - Y_{i'j'}))    \widetilde{\omega}_{n,ij}^{-1}\widetilde{\omega}_{n,i'j'}^{-1} + V(s,t).
\end{align*}
Similarly, we have
\begin{align*}
&\varphi_{X,Y}^{(n)}(s,t) \overline{
\varphi_{X}^{(n)}(s)\varphi_{Y}^{(n)}(t)
}  + \overline{\varphi_{X,Y}^{(n)}(s,t)} 
\varphi_{X}^{(n)}(s)\varphi_{Y}^{(n)}(t) \\
=&\frac{2}{n^3} \sum_{(i,j),(i',j'),(i'',j'') \in \cI} \cos(s^\top (X_{ij} - X_{i'j'})) \cos(t^\top (Y_{ij}-Y_{i''j''}) )\widetilde{\omega}_{n,ij}^{-1}\widetilde{\omega}_{n,i'j'}^{-1}\widetilde{\omega}_{n,i''j''}^{-1} + V(s,t).
\end{align*}
and 
\begin{align*}
&\varphi_{X}^{(n)}(s)\varphi_{Y}^{(n)}(t) \overline{    \varphi_{X}^{(n)}(s)\varphi_{Y}^{(n)}(t)} 
= V(s,t) + \\
&\frac{1}{n^4} \sum_{(i,j),(i',j'),(i'',j''), (i''',j''') \in \cI}   \cos(s^\top (X_{ij} - X_{i'j'})) \cos(t^\top (Y_{i''j''} - Y_{i'''j'''})) \widetilde{\omega}_{n,ij}^{-1}\widetilde{\omega}_{n,i'j'}^{-1} \widetilde{\omega}_{n,i''j''}^{-1}\widetilde{\omega}_{n,i'''j'''}^{-1}.
\end{align*}

Note that for each $u,v \in \bR$,
\begin{align*}
    \cos(u)\cos(v) = 1 - (1-\cos(u)) - (1-\cos(v)) + (1-\cos(u)) (1-\cos(v)).
\end{align*}
If we apply this identity to the above three equalities, and  due to \eqref{obs:normalization}, we have
\begin{align*}
&|\varphi_{X,Y}^{(n)}(s,t) - \varphi^{(n)}_X(s) \varphi^{(n)}_Y(t)|^2 = V(s,t) 
+ 
\frac{1}{n^4} \sum_{(i,j),(i',j'),(i'',j''), (i''',j''') \in \cI}   \\
&\qquad
\left(\  (1-\cos(s^\top (X_{ij} - X_{i'j'})))(1-\cos(t^\top (Y_{ij} - Y_{i'j'}))) \right.\\
&\qquad
-2 (1-\cos(s^\top (X_{ij} - X_{i'j'})))(1-\cos(t^\top (Y_{ij} - Y_{i''j''})))\\
&\qquad 
\left.+(1-\cos(s^\top (X_{ij} - X_{i'j'})))(1-\cos(t^\top (Y_{i''j''} - Y_{i'''j'''}))) \, \right)\widetilde{\omega}_{n,ij}^{-1}\widetilde{\omega}_{n,i'j'}^{-1} \widetilde{\omega}_{n,i''j''}^{-1}\widetilde{\omega}_{n,i'''j'''}^{-1}.
\end{align*}

Due to the definition of $H_p(\cdot)$ and $A_{ij,i'j'}(\delta), B_{ij,i'j'}$ in \eqref{def:H_function} and \eqref{def:cross_prod}, and by Lemma \ref{lemma:H_p}, for any $\delta \in (0,1]$,
\begin{align}\label{null_cal}
&n\widehat{\cV}_n^2(\delta) := \int_{\|s\| \leq \delta \ \cup \ \|s\| \geq 1/\delta} \int_{t \in \bR^q} \frac{\left|\sqrt{n}  \left(\varphi_{X,Y}^{(n)}(s,t) - \varphi^{(n)}_X(s) \varphi^{(n)}_Y(t) \right)\right|^2}{c_p c_q\|s\|^{1+p}\|t\|^{1+q}} ds dt   \nonumber \\
= &\frac{1}{n^3} \sum_{(i,j),(i',j'),(i'',j''), (i''',j''') \in \cI} 
\left(\, A_{ij,i'j'}(\delta)B_{ij,i'j'} 
- 2 A_{ij,i'j'}(\delta)B_{ij,i''j''}
+ A_{ij,i'j'}(\delta)B_{i''j'',i'''j'''} 
\, \right)\nonumber \\
&\qquad \qquad \qquad \qquad \qquad  \qquad  \widetilde{\omega}_{n,ij}^{-1}\widetilde{\omega}_{n,i'j'}^{-1} \widetilde{\omega}_{n,i''j''}^{-1}\widetilde{\omega}_{n,i'''j'''}^{-1}. \nonumber \\
=& \Delta_n(\delta, \bW_n),
\end{align}
where $\Delta_n(\cdot)$ is defined in \eqref{def:Delta_n}. Further, define
\begin{align}\label{def:remainder_term}
\mathcal{R}_n(\delta) := \Delta_n(\delta, \mathbb{W}_n)  - \Delta_n(\delta, W) - \left(\frac{1}{\sqrt{n}}\nabla \Delta_n(\delta,W)\right)^T \sqrt{n}(\mathbb{W}_n-W),
\end{align}
where the gradient is with respect to $W$-coordinates. 
By \cite[Proposition 2.2]{gill1988large}, $\sqrt{n}(\mathbb{W}_n-W)$ is bounded in probability. Then the proof is complete due to Lemmas \ref{lemma:Delta_n_W}, \ref{lemma:bound_grad}, and \ref{lemma:remainder}.
\end{proof}

\subsection{Preliminary calulation}\label{sub:prelim}
In this subsection, we assume $\delta \in (0,1]$ is fixed and drop the dependence on $\delta$ to lighten the notations. Define
\begin{equation}
    \label{def:L_upper}
L_{n} := \max\left\{ \frac{1}{\lambda_{n,1}}, \ldots,  \frac{1}{\lambda_{n,K}}, \; 
\frac{1}{W_1}, \ldots,  \frac{1}{W_K}
\right\}.
\end{equation}
Then by the definitions in  \eqref{def:avg_weight_funcs} and \eqref{def:r_underline_omega}, since $\lambda_i \in (0,1)$ for $1 \leq i \leq K$, we have for any $(x,y) \in \bR^{p+q}$, 
\begin{align}\label{obs:omega_r}
    \max_{1 \leq i \leq K} \widetilde{\omega}_n^{-1}(x,y; W) \frac{\omega_i(x,y)}{W_i} \leq L_{n}, \quad \text{ and } \quad
    \widetilde{\omega}_n^{-1}(x,y; W) \leq L_{n} r(x,y).
\end{align}
We will denote a pair $(i,j) \in \cI$ by $\iota$ and for $(\iota_1,\iota_2,\iota_3,\iota_4) \in \cI^4$, define
\begin{align}\label{def:LT_iota}
\begin{split}
    \cT_{\iota_1,\iota_2,\iota_3,\iota_4} := &A_{\iota_1,\iota_2}(\delta) 
\left(B_{\iota_1,\iota_2} + B_{\iota_3,\iota_4}
- 2 B_{\iota_1, \iota_3}
\right)\\
&\widetilde{\omega}_n^{-1}(X_{\iota_1},Y_{\iota_1}; W)
\widetilde{\omega}_n^{-1}(X_{\iota_2},Y_{\iota_2};  W)
\widetilde{\omega}_n^{-1}(X_{\iota_3},Y_{\iota_3};  W)
\widetilde{\omega}_n^{-1}(X_{\iota_4},Y_{\iota_4};  W), 
\end{split}
\end{align}
where $A_{\iota_1,\iota_2}(\delta), B_{\iota_1,\iota_2}$ are defined in \eqref{def:cross_prod}.
Then by the definition of $\Delta_n$ in \eqref{def:Delta_n}, we have
\begin{align}\label{comp:Delta_n_W}
     \Delta_n(\delta, W) =  \frac{1}{n^3} \sum_{(\iota_1,\iota_2,\iota_3,\iota_4) \in \cI^4} \cT_{\iota_1,\iota_2,\iota_3,\iota_4}.
\end{align}       
We first partition $\cI^4$ into sub-classes and denote by $\cI^{(k)}$ the subset of $\cI^4$ that contains  $k$-distinct elements for $k = 1,\ldots,4$. Specifically, 
\begin{align*}
& \cI^{(1)} := \{ (\iota_1,\iota_2,\iota_3,\iota_4) \in \cI^4: \iota_1 = \iota_2 = \iota_3 = \iota_4 \}, \\
&\cI^{(2)} :=  \left\{ (\iota_1,\iota_2,\iota_3,\iota_4) \in \cI^4: \iota_1 = \iota_2 \neq  \iota_3 = \iota_4, \; \text{ or } \; \iota_1 = \iota_3 \neq  \iota_2 = \iota_4 \; \text{ or } \; \iota_1 = \iota_4 \neq  \iota_2 = \iota_3 \right.\\
&\left. \qquad \iota_1 = \iota_2 =  \iota_3 \neq \iota_4   \; \text{ or } \; \iota_1 = \iota_2 =  \iota_4 \neq \iota_3 \; \text{ or } \; \iota_1 = \iota_3 =  \iota_4 \neq \iota_2 \; \text{ or } \;  \iota_2 = \iota_3 = \iota_4 \neq \iota_1   \right\},\\
& \cI^{(3)} :=  \left\{ (\iota_1,\iota_2,\iota_3,\iota_4) \in \cI^4: \iota_1 = \iota_2 \neq  \iota_3 \neq \iota_4, \; \text{ or } \; \iota_1 = \iota_3 \neq  \iota_2 \neq \iota_4 \; \text{ or } \; \iota_1 = \iota_4 \neq  \iota_2 \neq \iota_3 \right.\\
&\left. \qquad \qquad \qquad   \; \text{ or } \; \iota_2 = \iota_3 \neq  \iota_1 \neq \iota_4 \; \text{ or } \; \iota_2 = \iota_4 \neq  \iota_1 \neq \iota_3 \; \text{ or } \;  \iota_3 = \iota_4 \neq \iota_1 \neq \iota_2    \right\}, \\
& \cI^{(4)} := \{ (\iota_1,\iota_2,\iota_3,\iota_4) \in \cI^4: \iota_1 \neq \iota_2 \neq \iota_3 \neq \iota_4 \}.
\end{align*}
We note the following elementary bounds on the size of each subset:
\begin{align}
\begin{split}
        &|\cI^{(1)}| = n, \qquad 
    |\cI^{(2)}| = 7n(n-1) \leq 7n^2, \\
    &|\cI^{(3)}| = 6n(n-1)(n-2) \leq 6n^3, \qquad 
    |\cI^{(4)}| =  n(n-1)(n-2)(n-3) \leq n^4.
    \end{split}
\label{eq:size_bound}
\end{align}

Next, we deal with $\cT_{\iota_1,\iota_2,\iota_3,\iota_4}$   
within each sub-class separately.\\

For $(\iota_1,\iota_2,\iota_3,\iota_4) \in \cI^{(1)}$, by definition in \eqref{def:cross_prod}, $\cT_{\iota_1,\iota_2,\iota_3,\iota_4} = 0$.\\

For $(\iota_1,\iota_2,\iota_3,\iota_4) \in \cI^{(2)}$, $\cT_{\iota_1,\iota_2,\iota_3,\iota_4} = 0$ 
if $\iota_1 = \iota_2$ or $\iota_1 = \iota_4 \neq \iota_2 = \iota_3$. We consider the remaining cases. 
    \begin{itemize}
      \setlength{\itemindent}{0em}
        \item Case 1: $\iota_1 = \iota_3 \neq \iota_2 = \iota_4$: 
        \begin{align*}
 &           \Exp\left|\cT_{\iota_1,\iota_2,\iota_3,\iota_4} \right| =2 c_p^{-1}  \Exp\left(H_p(\|X_{\iota_1} - X_{\iota_2}\|;\delta) \|Y_{\iota_1} - Y_{\iota_2}\| \widetilde{\omega}_n^{-2}(X_{\iota_1},Y_{\iota_1}; W)
\widetilde{\omega}_n^{-2}(X_{\iota_2},Y_{\iota_2};  W)\right)\\
&= 2c_p^{-1}  \int H_p(\|x - x'\|;\delta) \|y - y'\|\\
&\qquad \qquad  \widetilde{\omega}_n^{-2}(x,y; W)
 \widetilde{\omega}_n^{-2}(x',y';  W)   \frac{\omega_i(x,y)}{W_i}
\frac{\omega_{i'}(x',y')}{W_{i'}} dG(x,y)dG(x',y'),
    \end{align*}
    where $i$ and $i'$ are the first components of $\iota_1$ and $\iota_2$, respectively. Due to \eqref{obs:omega_r} and by Lemma \ref{lemma:H_p},
    \begin{align}
      \Exp\left|\cT_{\iota_1,\iota_2,\iota_3,\iota_4} \right| 
      \leq 2  L_n^4 \Exp\left[\|X-X'\|\|Y-Y'\| r(X,Y) r(X',Y')  \right]. \label{case:22}
    \end{align}
        \item Case 2: $\iota_1 = \iota_3 = \iota_4 \neq \iota_2$ or 
        $\iota_2 = \iota_3 = \iota_4 \neq \iota_1$. By a similar calculation, we have
    \begin{align} \label{case:13}
      \Exp\left|\cT_{\iota_1,\iota_2,\iota_3,\iota_4} \right| 
      \leq  L_n^4 \Exp\left[\|X-X'\|\|Y-Y'\|(r^{2}(X,Y) + 
 r^{2}(X',Y'))  \right].
    \end{align}       
\end{itemize}

\medskip

For $(\iota_1,\iota_2,\iota_3,\iota_4) \in \cI^{(3)}$, by a similar argument as above, we have
   \begin{align}\label{case:I3}
      \Exp\left|\cT_{\iota_1,\iota_2,\iota_3,\iota_4} \right| 
      \leq  2 c_p^{-1} L_n^4 \Exp[ & H_p(\|X-X'\|;\delta)(\|Y-Y'\| + \|Y'-Y''\| + \|Y-Y''\|) \\
      &(r(X,Y) + 
 r(X'',Y''))  ]. \nonumber
    \end{align}   

\medskip

Finally, for $(\iota_1,\iota_2,\iota_3,\iota_4) \in \cI^{(4)}$,  
\begin{align}
\begin{split}
          \Exp\left(\cT_{\iota_1,\iota_2,\iota_3,\iota_4} \right)  =
      c_p^{-1} \int \; &  H_p(\|x-x'\|; \delta) \left( \|y-y'\| + \|y''-y'''\| - 2\|y - y''\| \right)\\
    &\widetilde{\omega}_n^{-1}(x,y; W)
\widetilde{\omega}_n^{-1}(x',y'; W)
\widetilde{\omega}_n^{-1}(x'',y''; W)
\widetilde{\omega}_n^{-1}(x''',y'''; W)\\
& \frac{\omega_i(x,y)}{W_i}
\frac{\omega_{i'}(x',y')}{W_{i'}}
\frac{\omega_{i''}(x'',y'')}{W_{i''}}
\frac{\omega_{i'''}(x''',y''')}{W_{i'''}} \\
&dG(x,y)dG(x',y')dG(x'',y'')dG(x''',y'''),
\end{split}
\label{def:order_four_computation}
\end{align}
where $i,i',i'',i'''$ are the first components of $\iota_1,\ldots,\iota_4$, respectively.

\subsection{Bounding the term $\Delta_n(\delta, W)$}\label{bound_exp_value}

In this subsection, we focus on the term $\Delta_n(\delta, W)$ defined in \eqref{def:Delta_n} with $\theta = W$ and prove the following lemma.

\begin{lemma}
    \label{lemma:Delta_n_W}
Assume that the conditions in Theorem \ref{thrm:null_distribution} hold. We have that $\Delta_n(\delta, W) \geq 0$ and 
$$
\lim_{\delta \downarrow 0} \limsup_{n \to \infty} \Exp\left[\Delta_n(\delta, W) \right] = 0.
$$
\end{lemma}

\begin{proof}
We first define a probability measure as follows: define 
$$
\cN_{n} := \int_{\bR^{p+q}} \left[ \widetilde{\omega}_n(x,y; W) \right]^{-1} \,d \mathbb{F}_n(x,y),
$$
and for any Borel $A \subset \bR^{p+q}$,
$$
\bG_{n}^{(0)}(A) := \frac{1}{\cN_{n} } \int_{A} \left[ \widetilde{\omega}_n(x,y; W) \right]^{-1}\, d \mathbb{F}_n(x,y).
$$
Note that compared to the NPMLE in \eqref{def:NPMLE}, $\bW_n$ is replaced by $W$. Then by the same calculation as in the proof of Lemma \ref{lemma:bound_outside_compacta},  for $\delta \in (0,1]$,
\begin{align*}
    \int_{\|s\| \leq \delta \ \cup \ \|s\| \geq 1/\delta} \int_{t \in \bR^q} \frac{\left|\sqrt{n}  \left(\varphi_{X,Y}^{(n,0)}(s,t) - \varphi^{(n,0)}_X(s) \varphi^{(n,0)}_Y(t) \right)\right|^2}{c_p c_q\|s\|^{1+p}\|t\|^{1+q}} ds dt 
    = \frac{1}{\cN_{n}^4} \Delta(\delta, W),
\end{align*}
where $\varphi_{X,Y}^{(n,0)}$, $\varphi_{X}^{(n,0)}$ and $\varphi_{Y}^{(n,0)}$ are respectively the characteristic functions of $(X,Y)$, $X$ and $Y$ under $\bG_n^{(0)}$. This shows that $\Delta_n(\delta, W) \geq 0$.

Next, we compute the expectation of $\Delta_n(\delta, W)$. Recall the notations and computation in the previous Subsection \ref{sub:prelim}. We have that
\begin{align*}
    \Exp\left[\Delta_n(\delta, W) \right] =& \frac{1}{n^3}\sum_{(\iota_1,\iota_2,\iota_3,\iota_4) \in  \cI^{(2)}} \Exp\left[\cT_{\iota_1,\iota_2,\iota_3,\iota_4} \right]+
 \frac{1}{n^3}\sum_{(\iota_1,\iota_2,\iota_3,\iota_4) \in \cI^{(3)}} \Exp\left[\cT_{\iota_1,\iota_2,\iota_3,\iota_4}\right]\\
 &\quad +
 \frac{1}{n^3}\sum_{(\iota_1,\iota_2,\iota_3,\iota_4) \in \cI^{(4)}} \Exp\left[\cT_{\iota_1,\iota_2,\iota_3,\iota_4}\right].
\end{align*}
By \eqref{case:22} and \eqref{case:13}, and due to \eqref{eq:size_bound}, we have
\begin{align*}
     \frac{1}{n^3}\sum_{(\iota_1,\iota_2,\iota_3,\iota_4) \in  \cI^{(2)}} \Exp\left|\cT_{\iota_1,\iota_2,\iota_3,\iota_4} \right|  \leq \frac{7n^2}{n^3}  L_n^4 \Exp\left[\|X-X'\|\|Y-Y'\|(r^{2}(X,Y) + 
 r^{2}(X',Y'))  \right],
\end{align*}
where we recall the definition of $L_n$ in \eqref{def:L_upper}. 
Then due to assumption \eqref{def:moment_cond2}, we have
\begin{align}\label{aux:I2}
   \limsup_{n \to \infty} \frac{1}{n^3}\sum_{(\iota_1,\iota_2,\iota_3,\iota_4) \in  \cI^{(2)}} \Exp\left|\cT_{\iota_1,\iota_2,\iota_3,\iota_4} \right| = 0. 
\end{align}
Furthermore, by \eqref{case:I3} and \eqref{eq:size_bound}, $ n^{-3} \sum_{(\iota_1,\iota_2,\iota_3,\iota_4) \in  \cI^{(3)}} \Exp\left|cT_{\iota_1,\iota_2,\iota_3,\iota_4} \right| $ is upper bounded by
\begin{align*}
 \frac{6n^3}{n^3} 2 c_p^{-1} L_n^4  \Exp[ H_p(\|X-X'\|;\delta)(\|Y-Y'\| + \|Y'-Y''\| + \|Y-Y''\|) 
      (r(X,Y) + 
 r(X'',Y''))].
\end{align*}
Then due to assumption \eqref{def:moment_cond2} and Lemma \ref{lemma:H_p}, and by the dominated convergence theorem, 
\begin{align}\label{aux:I3}
 \lim_{\delta \downarrow 0}  \limsup_{n \to \infty}\frac{1}{n^3}\sum_{(\iota_1,\iota_2,\iota_3,\iota_4) \in  \cI^{(3)}} \Exp\left|\cT_{\iota_1,\iota_2,\iota_3,\iota_4} \right| = 0. 
\end{align}

It remains to show that $ \lim_{\delta \downarrow 0}  \limsup_{n \to \infty}  n^{-3}\sum_{(\iota_1,\iota_2,\iota_3,\iota_4) \in  \cI^{(4)}} \Exp\left[\cT_{\iota_1,\iota_2,\iota_3,\iota_4} \right]  = 0$. By \eqref{def:order_four_computation},
\begin{align*}
    &n^{-3}\sum_{(\iota_1,\iota_2,\iota_3,\iota_4) \in  \cI^{(4)}} \Exp\left[\cT_{\iota_1,\iota_2,\iota_3,\iota_4} \right] = \\
   & c_p^{-1}\int \; H_p(\|x-x'\|; \delta) \left( \|y-y'\| + \|y''-y'''\| - 2\|y - y''\| \right) \\
&    \widetilde{\omega}_n^{-1}(x,y; W)
\widetilde{\omega}_n^{-1}(x',y'; W)
\widetilde{\omega}_n^{-1}(x'',y''; W)
\widetilde{\omega}_n^{-1}(x''',y'''; W)
\\
&\left( n^{-3}\sum_{(\iota_1,\iota_2,\iota_3,\iota_4) \in  \cI^{(4)}}  \frac{\omega_i(x,y)}{W_i}
\frac{\omega_{i'}(x',y')}{W_{i'}}
\frac{\omega_{i''}(x'',y'')}{W_{i''}}
\frac{\omega_{i'''}(x''',y''')}{W_{i'''}}\right) \\
&dG(x,y)dG(x',y')dG(x'',y'')dG(x''',y'''),
\end{align*}
Note that $\cI^{(4)} = \cI^4 \setminus (\cI^{(1)} \cup \cI^{(2)} \cup \cI^{(3)})$, and that, since $\lambda_{n,i} = n_i/n$ for $1 \leq i \leq K$, we have
\begin{align*}
 &n^{-3}\sum_{(\iota_1,\iota_2,\iota_3,\iota_4) \in  \cI^4}  \frac{\omega_i(x,y)}{W_i}
\frac{\omega_{i'}(x',y')}{W_{i'}}
\frac{\omega_{i''}(x'',y'')}{W_{i''}}
\frac{\omega_{i'''}(x''',y''')}{W_{i'''}} \\
=&  n \left(\sum_{i=1}^{K} \frac{\lambda_{n,i}\omega_i(x,y)}{W_i}\right)
\left(\sum_{i=1}^{K} \frac{\lambda_{n,i}\omega_i(x',y')}{W_i}\right)
\left(\sum_{i=1}^{K} \frac{\lambda_{n,i}\omega_i(x'',y'')}{W_i}\right)
\left(\sum_{i=1}^{K} \frac{\lambda_{n,i}\omega_i(x''',y''')}{W_i}\right)
\\
=& 
n \widetilde{\omega}_n(x,y; W)\widetilde{\omega}_n(x',y'; W)\widetilde{\omega}_n(x'',y''; W)\widetilde{\omega}_n(x''',y'''; W),
\end{align*}
where the last equality is due to the definition of $\widetilde{w}_n$ in  \eqref{def:avg_weight_funcs}. Thus,
\begin{align*}
    n^{-3}\sum_{(\iota_1,\iota_2,\iota_3,\iota_4) \in  \cI^{(4)}} \Exp\left[\cT_{\iota_1,\iota_2,\iota_3,\iota_4} \right] =   c_p^{-1}(J_{1,n,\delta} - J_{2,n,\delta}),
\end{align*}
where 
\begin{align*}
J_{1,n,\delta} &:=   n \int  H_p(\|x-x'\|; \delta) \left( \|y-y'\| + \|y''-y'''\| - 2\|y - y''\| \right) \\
&\qquad\qquad  dG(x,y)dG(x',y')dG(x'',y'')dG(x''',y''') \\
&=n \Exp\left[ H_p(\|X-X'\|; \delta) \left( \|Y-Y'\| + \|Y''-Y'''\| - 2\|Y - Y''\| \right)\right],
\end{align*}
and 
\begin{align*}
J_{2,n,\delta} :=  & \int \;  H_p(\|x-x'\|; \delta) \left( \|y-y'\| + \|y''-y'''\| - 2\|y - y''\| \right) \\
&  \quad  \widetilde{\omega}_n^{-1}(x,y; W)
\widetilde{\omega}_n^{-1}(x',y'; W)
\widetilde{\omega}_n^{-1}(x'',y''; W)
\widetilde{\omega}_n^{-1}(x''',y'''; W)
\\
&\quad \left( n^{-3}\sum_{(\iota_1,\iota_2,\iota_3,\iota_4) \in  \cI^{(1)} \cup \cI^{(2)} \cup \cI^{(3)}}  \frac{\omega_i(x,y)}{W_i}
\frac{\omega_{i'}(x',y')}{W_{i'}}
\frac{\omega_{i''}(x'',y'')}{W_{i''}}
\frac{\omega_{i'''}(x''',y''')}{W_{i'''}}\right) \\
& \quad dG(x,y)dG(x',y')dG(x'',y'')dG(x''',y''').
\end{align*}
By Lemma \ref{lemma:independence lemma} with $h(\cdot)$ being a constant function, we have that $J_{1,n,\delta} = 0$ for any $n \geq 1$ and $\delta \in (0,1)$. Further, due to \eqref{obs:omega_r} and \eqref{eq:size_bound},
\begin{align*}
    &    \widetilde{\omega}_n^{-1}(x,y; W)
\widetilde{\omega}_n^{-1}(x',y'; W)
\widetilde{\omega}_n^{-1}(x'',y''; W)
\widetilde{\omega}_n^{-1}(x''',y'''; W)
\\
&\left( n^{-3}\sum_{(\iota_1,\iota_2,\iota_3,\iota_4) \in  \cI^{(1)} \cup \cI^{(2)} \cup \cI^{(3)}}  \frac{\omega_i(x,y)}{W_i}
\frac{\omega_{i'}(x',y')}{W_{i'}}
\frac{\omega_{i''}(x'',y'')}{W_{i''}}
\frac{\omega_{i'''}(x''',y''')}{W_{i'''}}\right) \\
\leq & \frac{6(n+1)^3}{n^3} L_n^4.
\end{align*}
Thus, under assumption \eqref{def:moment_cond2}, by the dominated convergence theorem, we have
\begin{align*}
    \lim_{\delta \downarrow 0}  \limsup_{n \to \infty}  |J_{2,n,\delta}| = 0.
\end{align*}
Thus the proof is complete.
\end{proof}

\subsection{Bounding the gradient $\nabla \Delta_n(\delta,W)$}
For $(\iota_1,\iota_2,\iota_3,\iota_4) \in \cI^4$, define
\begin{align*}
\nabla \cT_{\iota_1,\iota_2,\iota_3,\iota_4} := &A_{\iota_1,\iota_2}(\delta) \left(B_{\iota_1,\iota_2} + B_{\iota_3,\iota_4}- 2 B_{\iota_1, \iota_3} \right) \\
&\nabla \left(\widetilde{\omega}_n^{-1}(X_{\iota_1},Y_{\iota_1}; W)\widetilde{\omega}_n^{-1}(X_{\iota_2},Y_{\iota_2};  W)\widetilde{\omega}_n^{-1}(X_{\iota_3},Y_{\iota_3};  W)\widetilde{\omega}_n^{-1}(X_{\iota_4},Y_{\iota_4};  W)\right), 
\end{align*}
where $A_{\iota_1,\iota_2}(\delta), B_{\iota_1,\iota_2}$ are defined in \eqref{def:cross_prod} and $\nabla$ denotes the gradient with respect to $W$. Then
$$
\nabla \left(\frac{1}{\sqrt{n}} \Delta_n(\delta, W)\right)  =  \frac{1}{n^{7/2}} \sum_{(\iota_1,\iota_2,\iota_3,\iota_4) \in \cI^4} \nabla \cT_{\iota_1,\iota_2,\iota_3,\iota_4}.
$$


Recall the definition of $\widetilde{\omega}_n(\cdot)$ in \eqref{def:avg_weight_funcs} and note that for $(x,y) \in \bR^{p+q}$, $\theta \in (0,\infty)^K$, and $\delta \in (0,1)$,
\begin{align*}
    \nabla \widetilde{\omega}_n(x,y; \theta)  = -\begin{bmatrix}
    \lambda_{n1} w_1(x,y)/\theta_1^2 \\
    \vdots\\
     \lambda_{nK} w_K(x,y)/\theta_K^2
    \end{bmatrix}.
\end{align*}
Thus due to \eqref{obs:omega_r} and by the definition of $L_n$ in \eqref{def:L_upper}, 
\begin{align}\label{obs:grad}
\begin{split}
        &\|\nabla \widetilde{\omega}_n(x,y; W)\|_{\infty} \leq L_n \widetilde{\omega}_n(x,y; W), \\
    & 
    \|\nabla \left(\widetilde{\omega}_n^{-1}(x,y; W)\widetilde{\omega}_n^{-1}(x',y';  W)\widetilde{\omega}_n^{-1}(x'',y'';  W)\widetilde{\omega}_n^{-1}(x''',y''';  W)\right) \|_{\infty}
     \\
    & \leq  4L_n \widetilde{\omega}_n^{-1}(x,y; W)\widetilde{\omega}_n^{-1}(x',y';  W)\widetilde{\omega}_n^{-1}(x'',y'';  W)\widetilde{\omega}_n^{-1}(x''',y''';  W).
\end{split}
\end{align}
where for a vector or a matrix, $\|\cdot\|_{\infty}$ denotes the $\ell_{\infty}$-norm.

\begin{lemma}\label{lemma:bound_grad}
 Assume that the conditions in Theorem \ref{thrm:null_distribution} hold. For any $\epsilon > 0$, we have
\begin{align*}
&\lim_{\delta \downarrow 0} \limsup_{n \to \infty}   \bP\left( \left\|\nabla \left(\frac{1}{\sqrt{n}} \Delta_n(\delta, W)\right) \right\| \geq \epsilon \right) = 0.
\end{align*}
\end{lemma}
\begin{proof}
Recall the notations in Subsection \ref{sub:prelim}. As $T_{\iota_1,\iota_2,\iota_3,\iota_4} = 0$ for $(\iota_1, \iota_2, \iota_3, \iota_4) \in \mathcal{I}^{(1)}$, we have  $\nabla \left(\frac{1}{\sqrt{n}} \Delta_n(\delta, W)\right) = V_{1,n,\delta} + V_{2,n,\delta}$, where we define
\begin{align*}
V_{1,n,\delta} :=  \frac{1}{n^{7/2}} \sum_{(\iota_1,\iota_2,\iota_3,\iota_4) \in \cI^{(4)}} \nabla \cT_{\iota_1,\iota_2,\iota_3,\iota_4}, \qquad
V_{2,n,\delta}  := 
\frac{1}{n^{7/2}} \sum_{(\iota_1,\iota_2,\iota_3,\iota_4) \in \cI^{(2)} \cup \cI^{(3)}} \nabla \cT_{\iota_1,\iota_2,\iota_3,\iota_4}.
\end{align*}
Due to the discussion in \eqref{obs:grad}, we have
\begin{align*}
    \frac{1}{n^{7/2}} \sum_{(\iota_1,\iota_2,\iota_3,\iota_4) \in \cI^{(2)} \cup \cI^{(3)}} \|\nabla \cT_{\iota_1,\iota_2,\iota_3,\iota_4} \|_{\infty} 
    \leq  \frac{4L_n}{\sqrt{n}} \frac{1}{n^3} \sum_{(\iota_1,\iota_2,\iota_3,\iota_4) \in \cI^{(2)}\; \cup \; \cI^{(3)}} \left| \cT_{\iota_1,\iota_2,\iota_3,\iota_4} \right|,
\end{align*}
where recall that $\cT_{\iota_1,\iota_2,\iota_3,\iota_4}$ is defined in \eqref{def:LT_iota}. Then due to \eqref{aux:I2} and \eqref{aux:I3}, we have that  for any $\epsilon > 0$,  
\begin{align*}
&\lim_{\delta \downarrow 0} \limsup_{n \to \infty}   \bP\left( \left\|V_{2,n,\delta}\right\| \geq \epsilon \right) = 0.
\end{align*}

Next, we focus on $V_{1,n,\delta}$ and consider its mean and variance. Note that
\begin{align*}
    \Exp\left[V_{1,n,\delta}\right] =   
   & c_p^{-1} \int \;  H_p(\|x-x'\|; \delta) \left( \|y-y'\| + \|y''-y'''\| - 2\|y - y''\| \right) \\
& \quad  \nabla\left(  \widetilde{\omega}_n^{-1}(x,y; W)
\widetilde{\omega}_n^{-1}(x',y'; W)
\widetilde{\omega}_n^{-1}(x'',y''; W)
\widetilde{\omega}_n^{-1}(x''',y'''; W)\right)
\\
&\quad  \left( n^{-7/2}\sum_{(\iota_1,\iota_2,\iota_3,\iota_4) \in  \cI^{(4)}}  \frac{\omega_i(x,y)}{W_i}
\frac{\omega_{i'}(x',y')}{W_{i'}}
\frac{\omega_{i''}(x'',y'')}{W_{i''}}
\frac{\omega_{i'''}(x''',y''')}{W_{i'''}}\right) \\
& \quad  dG(x,y)dG(x',y')dG(x'',y'')dG(x''',y''').
\end{align*}
By the same argument as in the proof of Lemma \ref{lemma:Delta_n_W}, we have
\begin{align*}
\lim_{\delta \downarrow 0}  \limsup_{n \to \infty}\| \Exp\left[V_{1,n,\delta}\right] - c_p^{-1}\widetilde{J}_{1,n,\delta}\|_\infty = 0,
\end{align*}
where 
\begin{align*}
    \widetilde{J}_{1,n,\delta}  := &  n^{1/2}  \int \; H_p(\|x-x'\|; \delta) \left( \|y-y'\| + \|y''-y'''\| - 2\|y - y''\| \right) \\
&  \nabla\left(  \widetilde{\omega}_n^{-1}(x,y; W)
\widetilde{\omega}_n^{-1}(x',y'; W)
\widetilde{\omega}_n^{-1}(x'',y''; W)
\widetilde{\omega}_n^{-1}(x''',y'''; W)\right)
\\
&\left(\widetilde{\omega}_n(x,y; W)
\widetilde{\omega}_n(x',y'; W)
\widetilde{\omega}_n(x'',y''; W)
\widetilde{\omega}_n(x''',y'''; W)\right) \\
&dG(x,y)dG(x',y')dG(x'',y'')dG(x''',y''').
\end{align*}
By Lemma \ref{lemma:independence lemma},   $\widetilde{J}_{1,n,\delta} = 0$ for any $n \geq 1$ and $\delta \in (0,1)$. As a result,
\begin{align}\label{aux:mean_to_zero}
\lim_{\delta \downarrow 0}  \limsup_{n \to \infty}\| \Exp\left[V_{1,n,\delta}\right]\|_\infty = 0,
\end{align}
Now we fixed $1 \leq i \leq K$ and compute the variance of $V_{1,n,\delta,i}$, the $i$-th coordinate of $V_{1,n,\delta}$. We denote by $\partial_i$ the partial derivative with respect to the $i$-th coordinate of $W$.  Note that
\begin{align*}
    \Var(V_{1,n,\delta,i}) = \frac{1}{n^{7}} \sum_{(\iota_1,\iota_2,\iota_3,\iota_4), (\iota_1',\iota_2',\iota_3',\iota_4') \in \cI^{(4)}} \text{Cov}\left(\partial_i \cT_{\iota_1,\iota_2,\iota_3,\iota_4}, \partial_i \cT_{\iota_1',\iota_2',\iota_3',\iota_4'}\right).
\end{align*}
If $(\iota_1,\iota_2,\iota_3,\iota_4)$ and $(\iota_1',\iota_2',\iota_3',\iota_4')$ do not have any common element, then the corresponding covariance is zero. Further, the number of pairs that do share at least one element is at most $14n^7$. Thus, by Cauchy-Schwarz inequality, for $1 \leq i \leq K$,
\begin{align*}
    \Var(V_{1,n,\delta,i}) \leq 14 \max_{(\iota_1,\iota_2,\iota_3,\iota_4) \in \cI^{(4)}} \Exp\left(\partial_i \cT_{\iota_1,\iota_2,\iota_3,\iota_4} \right)^2.
\end{align*}
Note that by definition
\begin{align*}
    \Exp\left(\partial_i \cT_{\iota_1,\iota_2,\iota_3,\iota_4} \right)^2 = & c_p^{-2}\int \;  H_p^2(\|x-x'\|; \delta) \left( \|y-y'\| + \|y''-y'''\| - 2\|y - y''\| \right)^2 \\
&  \left(\partial_i\left(  \widetilde{\omega}_n^{-1}(x,y; W)
\widetilde{\omega}_n^{-1}(x',y'; W)
\widetilde{\omega}_n^{-1}(x'',y''; W)
\widetilde{\omega}_n^{-1}(x''',y'''; W)\right)\right)^2
\\
&\frac{\omega_i(x,y)}{W_i}
\frac{\omega_{i'}(x',y')}{W_{i'}}
\frac{\omega_{i''}(x'',y'')}{W_{i''}}
\frac{\omega_{i'''}(x''',y''')}{W_{i'''}} \\
&dG(x,y)dG(x',y')dG(x'',y'')dG(x''',y'''),
\end{align*}
where $i',i'',i''',i'''$ are respectively the first component of $\iota_1,\ldots,\iota_4$. By \eqref{obs:grad} and \eqref{obs:omega_r}
 \begin{align*}
 &\left(\partial_i\left(  \widetilde{\omega}_n^{-1}(x,y; W)
\widetilde{\omega}_n^{-1}(x',y'; W)
\widetilde{\omega}_n^{-1}(x'',y''; W)
\widetilde{\omega}_n^{-1}(x''',y'''; W)\right)\right)^2
\\
&\frac{\omega_i(x,y)}{W_i}
\frac{\omega_{i'}(x',y')}{W_{i'}}
\frac{\omega_{i''}(x'',y'')}{W_{i''}}
\frac{\omega_{i'''}(x''',y''')}{W_{i'''}} \\
\leq & 16 L_n^{10} r(x,y)r(x',y')r(x'',y'')r(x''',y''').
 \end{align*}
 As a result,
 \begin{align*}
         \Exp\left(\partial_i \cT_{\iota_1,\iota_2,\iota_3,\iota_4} \right)^2 \leq 16 c_p^{-2} L_n^{10} &\Exp[H_p^2(\|X-X'\|;\delta)\left( \|Y-Y'\| + \|Y''-Y'''\| - 2\|Y - Y''\| \right)^2    \\
         &\qquad \left. r(X,Y)r(X',Y')r(X'',Y'')r(X''',Y''')\right].
 \end{align*}
 Thus, under assumption \ref{def:moment_cond2} and due to Lemma \ref{lemma:H_p}, by dominated convergence theorem,
 \begin{align*}
\lim_{\delta \downarrow 0}  \limsup_{n \to \infty} \max_{1 \leq i \leq K}     \max_{(\iota_1,\iota_2,\iota_3,\iota_4) \in \cI^{(4)}} \Exp\left(\partial_i \cT_{\iota_1,\iota_2,\iota_3,\iota_4} \right)^2  = 0,
 \end{align*}
 which implies that
 $$
 \lim_{\delta \downarrow 0}  \limsup_{n \to \infty} \max_{1 \leq i \leq K}       \Var(V_{1,n,\delta,i}) = 0.
 $$
 Together with \eqref{aux:mean_to_zero}, we have
 that  for any $\epsilon > 0$,  
\begin{align*}
&\lim_{\delta \downarrow 0} \limsup_{n \to \infty}   \bP\left( \left\|V_{1,n,\delta}\right\| \geq \epsilon \right) = 0.
\end{align*}
The proof is complete.
\end{proof}

\subsection{Bounding the remainder term $\cR_{n}(\delta)$}
In this subsection, we bound the term $\cR_{n}(\delta)$ defined in \eqref{def:remainder_term}.  Recall the definition of  $\widetilde{\omega}_n(x,y; \theta)$ in \eqref{def:avg_weight_funcs}.
For $(x_i,y_i) \in \mathbb{R}^p \times \mathbb{R}^q$ and $\theta \in (0,\infty)^{K}$,
\begin{align*}
    \nabla \left( \prod_{i=1}^{4}\widetilde{\omega}_n^{-1}(x_i,y_i;\theta) \right)
    = \left(\prod_{i=1}^{4}\widetilde{\omega}_n^{-1}(x_i,y_i;\theta) \right)\left( \sum_{i=1}^{4}
    \frac{\nabla \widetilde{\omega}_n(x_i,y_i; \theta)}{\widetilde{\omega}_n(x_i,y_i; \theta)}
    \right),
\end{align*}
and further for $1 \leq k, \ell \leq K$, denote by $ \partial_{kl}$ the partial derivative with respect to first $\theta_k$ and then $\theta_\ell$, and we have
\begin{align*}
    \partial_{kl} \left( \prod_{i=1}^{4}\widetilde{\omega}_n^{-1}(x_i,y_i;\theta) \right) 
    = &\left(\prod_{i=1}^{4}\widetilde{\omega}_n^{-1}(x_i,y_i;\theta) \right)\left( \sum_{i=1}^{4}
    \frac{\partial_{k} \widetilde{\omega}_n(x_i,y_i; \theta)}{\widetilde{\omega}_n(x_i,y_i; \theta)}
    \right)\left( \sum_{i=1}^{4}
    \frac{\partial_{\ell} \widetilde{\omega}_n(x_i,y_i; \theta)}{\widetilde{\omega}_n(x_i,y_i; \theta)}
    \right) \\
    +& \left(\prod_{i=1}^{4}\widetilde{\omega}_n^{-1}(x_i,y_i;\theta) \right)\left( \sum_{i=1}^{4}
    \frac{\partial_{k\ell} \widetilde{\omega}_n(x_i,y_i; \theta)}{\widetilde{\omega}_n(x_i,y_i; \theta)}
    \right) \\
    -&\left(\prod_{i=1}^{4}\widetilde{\omega}_n^{-1}(x_i,y_i;\theta) \right)\left( \sum_{i=1}^{4}
    \frac{\partial_{k} \widetilde{\omega}_n(x_i,y_i; \theta)\partial_{\ell} \widetilde{\omega}_n(x_i,y_i; \theta)}{\widetilde{\omega}_n^2(x_i,y_i; \theta)}
    \right).
\end{align*}
Next, we define 
\begin{align}
    \label{def:Q_*}
    \mathcal{Q}^*:= \{\theta \in \mathbb{R}^K: W_i/2 < \theta_i < 2W_i \;\; \text{ for } 1 \leq i \leq K \}.
\end{align}
For $\theta \in \mathcal{Q}^*$, and for any $1 \leq k,\ell \leq K$,
\begin{align*}
 \widetilde{\omega}_n^{-1}(x,y;\theta) \leq   2  \widetilde{\omega}_n^{-1}(x,y;W), \quad
 \frac{\left|\partial_{k} \widetilde{\omega}_n^{-1}(x,y;\theta)\right|}{\widetilde{\omega}_n^{-1}(x,y;\theta)} \leq  2L_n, \quad
   \frac{\left|\partial_{k\ell} \widetilde{\omega}_n(x_i,y_i; \theta)\right|}{\widetilde{\omega}_n(x_i,y_i; \theta)} \leq 8 L_n^2,
\end{align*}
where we recall the definition of $L_n$ in \eqref{def:L_upper}.
Thus, due to \eqref{obs:omega_r}, we have for $\theta \in \mathcal{Q}^*$, 
\begin{align}\label{equ:sec_deriv_bounds}
\max_{1 \leq k, \ell, \leq K} \left| \partial_{kl} \left( \prod_{i=1}^{4}\widetilde{\omega}_n^{-1}(x_i,y_i;\theta) \right) \right|    \leq 224 L_n^6 \prod_{i=1}^{4} r(x_i,y_i).
\end{align}

\begin{lemma}
   \label{lemma:remainder}
 Assume that the conditions in Theorem \ref{thrm:null_distribution} hold. For any $\epsilon > 0$, we have
\begin{align*}
&\lim_{\delta \downarrow 0} \limsup_{n \to \infty}   \bP\left( \left| \cR_{n}(\delta) \right| \geq \epsilon \right) = 0.
\end{align*}
\end{lemma}
\begin{proof}
 By the definition  of $\cR_{n}(\delta)$  in \eqref{def:remainder_term} and the mean value theorem, there exits a random vector $\widetilde{\bW}_n$, on the segment between $\bW_n$ and $W$, such that
 \begin{align*}
  \cR_n(\delta)   = \frac{1}{2}\left(\sqrt{n}(\mathbb{W}_n-W)\right)^\top
\widetilde{H}_{n,\delta}
  \left(\sqrt{n}(\mathbb{W}_n-W)\right),
 \end{align*}
where 
\begin{align*}
 \widetilde{H}_{n,\delta} :=    \frac{1}{n^4} &\sum_{(i,j),(i',j'),(i'',j''), (i''',j''') \in \mathcal{I}} 
    A_{ij,i'j'}(\delta) 
\left(B_{ij,i'j'} + B_{i''j'',i'''j'''}
- 2 B_{ij, i''j''}
\right)\\
&\nabla^2 \left(\widetilde{\omega}_n^{-1}(X_{ij},Y_{ij}; \widetilde{\bW}_n)
\widetilde{\omega}_n^{-1}(X_{i'j'},Y_{i'j'}; \widetilde{\bW}_n)
\widetilde{\omega}_n^{-1}(X_{i''j''},Y_{i''j''}; \widetilde{\bW}_n)
\widetilde{\omega}_n^{-1}(X_{i'''j'''},Y_{i'''j'''}; \widetilde{\bW}_n)\right).
\end{align*}
Note that $\nabla^2$ denotes the Hessian matrix with respective to the $W$-coordinates, and that  $A_{\iota_1,\iota_2}(\delta), B_{\iota_1,\iota_2}$ are defined in \eqref{def:cross_prod}.

 On the event $\{\bW_n \in \mathcal{Q}^{*}\}$, due to 
 \eqref{equ:sec_deriv_bounds}, we have
 $   \|\widetilde{H}_{n,\delta}\|_{\infty} \leq 224 L_n^6\widehat{H}_{n,\delta}$,  where   $\|\cdot\|_{\infty}$ denotes the supremum norm of a matrix and 
 \begin{align*}
 \widehat{H}_{n,\delta} := \frac{1}{n^4} &\sum_{(i,j),(i',j'),(i'',j''), (i''',j''') \in \mathcal{I}} 
    A_{ij,i'j'}(\delta) 
\left(B_{ij,i'j'} + B_{i''j'',i'''j'''}
+ 2 B_{ij, i''j''}
\right)\\
&r(X_{ij},Y_{ij})r(X_{i'j'},Y_{i'j'})
r(X_{i''j''},Y_{i''j''})
r(X_{i'''j'''},Y_{i'''j'''}).
 \end{align*}
Note that $\widehat{H}_{n,\delta} \geq 0$, and 
\begin{align*}
    \Exp\left[ \widehat{H}_{n,\delta} \right] \leq  \max_{(i,j),(i',j'),(i'',j''), (i''',j''') \in \mathcal{I}} 
    \Exp[& A_{ij,i'j'}(\delta) 
\left(B_{ij,i'j'} + B_{i''j'',i'''j'''}
+ 2 B_{ij, i''j''}
\right) \\
&r(X_{ij},Y_{ij})r(X_{i'j'},Y_{i'j'})
r(X_{i''j''},Y_{i''j''})
r(X_{i'''j'''},Y_{i'''j'''})].
\end{align*}
The upper bound above does not depend on $n$. Then 
due to assumption \eqref{def:moment_cond2} and by the dominated convergence theorem,
$$
\lim_{\delta \downarrow 0} \limsup_{n\to \infty}\Exp\left[ \widehat{H}_{n,\delta} \right] = 0.
$$
Finally, we note that
\begin{align*}
  \bP\left( \left| \cR_{n}(\delta) \right| \geq \epsilon \right)
  &\leq   \bP\left(\bW_n \in  (\mathcal{Q}^*)^c \right) + \bP\left( \left| \cR_{n}(\delta) \right| \geq \epsilon, \bW_n \in  \mathcal{Q}^* \right) \\
  & \leq  \bP\left(\bW_n \in  (\mathcal{Q}^*)^c \right) + \bP\left( 224L_n^6 K \|\sqrt{n}(\mathbb{W}_n-W)\|^2 \widehat{H}_{n,\delta} \geq \epsilon \right).
\end{align*}
By \cite[Proposition 2.2]{gill1988large}, $\sqrt{n}(\mathbb{W}_n-W)$ is bounded in probability, which completes the proof.
\end{proof}

\section{Proof of Lemma \ref{lemma:alternative_derivatives}}
\label{app:alter_V_derivatives}

For $\rho \in [0,1]$, consider
\begin{align}
    \label{def:alter_iid}
    (X_{\rho},Y_{\rho}), (X_{\rho}',Y_{\rho}'), (X_{\rho}'',Y_{\rho}''), (X_{\rho}''',Y_{\rho}''') \;\; \overset{i.i.d.}{\sim} \;\; \text{ density } g_{\rho}. 
\end{align}
\begin{proof}[Proof of Lemma \ref{lemma:alternative_derivatives}]
For $\rho \in [0,1]$, by \cite[Equation (2.7)]{sejdinovic2013equivalence},
\begin{align*}
    \cV^2(\rho) = \Exp\left[\|X_{\rho} - X_{\rho}'\|\left(\|Y_{\rho} - Y_{\rho}'\| + \|Y_{\rho}'' - Y_{\rho}'''\| - 2\|Y_{\rho} - Y_{\rho}''\| \right) \right].
\end{align*}
As a result, 
\begin{align*}
    \cV^2(\rho) =  \int \; & \|x-x'\| \left( \|y-y'\| + \|y''-y'''\| - 2\|y - y''\| \right) \\
& g_{\rho}(x,y)
g_{\rho}(x',y')
g_{\rho}(x'',y'')
g_{\rho}(x''',y''') d\nu(x,y)d\nu(x',y')d\nu(x'',y'')d\nu(x''',y''').
\end{align*}
We note that for $\rho \in [0,1)$,
\begin{align*}
&   \frac{d}{d\rho}\left(g_{\rho}(x,y)
g_{\rho}(x',y')
g_{\rho}(x'',y'')
g_{\rho}(x''',y''')\right)=g_{\rho}(x,y)
g_{\rho}(x',y')
g_{\rho}(x'',y'')
g_{\rho}(x''',y''')\\
&\qquad \qquad \times \left( \frac{d g_{\rho}(x,y)/d\rho}{g_{\rho}(x,y)} 
+ \frac{d g_{\rho}(x',y')/d\rho}{g_{\rho}(x',y')} 
+ \frac{d g_{\rho}(x'',y'')/d\rho}{g_{\rho}(x'',y'')}
+ \frac{d g_{\rho}(x''',y''')/d\rho}{g_{\rho}(x''',y''')}
\right),
\end{align*}
and 
\begin{align*}
&\frac{d^2}{d\rho^2}\left(g_{\rho}(x,y)
g_{\rho}(x',y')
g_{\rho}(x'',y'')
g_{\rho}(x''',y''')\right)=g_{\rho}(x,y)
g_{\rho}(x',y')
g_{\rho}(x'',y'')
g_{\rho}(x''',y''') \times \\
&\left(\left( \frac{d g_{\rho}(x,y)/d\rho}{g_{\rho}(x,y)} 
+ \frac{d g_{\rho}(x',y')/d\rho}{g_{\rho}(x',y')} 
+ \frac{d g_{\rho}(x'',y'')/d\rho}{g_{\rho}(x'',y'')}
+ \frac{d g_{\rho}(x''',y''')/d\rho}{g_{\rho}(x''',y''')}
\right)^2\right. \\
&\quad + \left(\frac{d^2 g_{\rho}(x,y)/d\rho^2}{g_{\rho}(x,y)} 
+ \frac{d^2 g_{\rho}(x',y')/d\rho^2}{g_{\rho}(x',y')} 
+ \frac{d^2 g_{\rho}(x'',y'')/d\rho^2}{g_{\rho}(x'',y'')}
+ \frac{d^2 g_{\rho}(x''',y''')/d\rho^2}{g_{\rho}(x''',y''')}\right) \\
&\quad \left.- \left( 
\left(\frac{d g_{\rho}(x,y)}{d\rho} \right)^2
+ \left(\frac{d g_{\rho}(x',y')}{d\rho}  \right)^2
+ \left(\frac{d g_{\rho}(x'',y'')}{d\rho} \right)^2
+ \left(\frac{d g_{\rho}(x''',y''')}{d\rho} \right)^2
\right)\right).
\end{align*}
By the dominated convergence theorem \cite[Theorem 2.27]{folland1999real}, under Assumption \ref{assumption:dominated}, we can exchange the order of differentiation ($d/d\rho$) and integration, and thus $\cV^2(\rho)$ is twice differentiable on $[0,\epsilon)$ and
\begin{align*}
& \left.\frac{d (\cV^2(\rho))}{d\rho} \right\vert_{\rho = 0} =  \int \;  \|x-x'\| \left( \|y-y'\| + \|y''-y'''\| - 2\|y - y''\| \right) \\
& \qquad g_0(x,y)
g_0(x',y')
g_0(x'',y'')
g_0(x''',y''')\\
&\qquad \left(\frac{d g_0(x,y)/d\rho}{g_0(x,y)} 
+ \frac{d g_0(x',y')/d\rho}{g_0(x',y')} 
+ \frac{d g_0(x'',y'')/d\rho}{g_0(x'',y'')}
+ \frac{d g_0(x''',y''')/d\rho}{g_0(x''',y''')}
\right) \\
&\qquad d\nu(x''',y''') d\nu(x,y)d\nu(x',y')d\nu(x'',y'')d\nu(x''',y''').
\end{align*}
As a result,
\begin{align*}
 &\left.\frac{d (\cV^2(\rho))}{d\rho} \right\vert_{\rho = 0} = \Exp\bigg[\ \|X_{0} - X_{0}'\|\left(\|Y_{0} - Y_{0}'\| + \|Y_{0}'' - Y_{0}'''\| - 2\|Y_{0} - Y_{0}''\| \right) \\
 &\qquad\left(\frac{d g_0(X_{0},Y_{0})/d\rho}{g_0(X_{0},Y_{0})} 
+ \frac{d g_0(X_{0}',Y_{0}')/d\rho}{g_0(X_{0}',Y_{0}')} 
+ \frac{d g_0(X_{0}'',Y_{0}'')/d\rho}{g_0(X_{0}'',Y_{0}'')}
+ \frac{d g_0(X_{0}''',Y_{0}''')/d\rho}{g_0(X_{0}''',Y_{0}''')}
\right) \ \bigg].
\end{align*}
Then, since $X_0$ and $Y_0$ are independent, by Lemma \ref{lemma:independence lemma}, 
$$\left.\frac{d (\cV^2(\rho))}{d\rho} \right\vert_{\rho = 0} = 0,
$$
which completes the proof.
\end{proof}

\section{Limiting behaviour under alternatives} \label{app:alternative}
Recall the definition of $D(\delta)$ in \eqref{def:D_delta}. Furthermore, for $\rho \in [0,1]$, recall that $\varphi_{X,Y}(\cdot;\rho)$, $ \varphi_X(\cdot;\rho)$ and $\varphi_Y(\cdot;\rho)$ are respectively the characteristics function of $(X,Y)$, $X$ and $Y$, when $(X,Y)$ has the joint density $g_{\rho}$. For $\rho \in [0,1]$ and $\delta \in (0,1)$, define
\begin{equation}
    \label{def:cV_rho_delat}
\cV^2(\rho; \delta) :=
     \int_{D(\delta)} \frac{|\varphi_{X,Y}(s, t;\rho) - \varphi_X(s;\rho)\varphi_Y(t;\rho)|^2}{c_p c_q \|s\|^{1+p} \|t\|^{1+q}}\,ds\,dt.
\end{equation}
Clearly, $\cV^2(0;\delta) = 0$ for any $\delta \in (0,1)$.

\begin{lemma}
    \label{lemma:alter_second_derivative_delta}
    Assume that the conditions in Theorem \ref{thrm:null_distribution} hold. Then, there exists $\delta \in (0,1)$ such that
$$
\lim_{\rho \downarrow 0} \frac{1}{\rho^2} \cV^{2}(\rho; \delta) > 0.
$$
\end{lemma}
\begin{proof}
Recall the random vectors in \eqref{def:alter_iid} for $\rho \in [0,1]$. For an integer $p \geq 1$, define a real-valued function $\widetilde{H}_{p}(\cdot; \cdot)$ on $[0,\infty) \times (0,1]$ as follows: $\widetilde{H}_{p}(0;\delta) = 0$ for any $\delta \in (0,1]$, and for any $r > 0$ and $\delta \in (0,1]$, define
\begin{equation*}
    \widetilde{H}_{p}(r; \delta) := \frac{1}{c_p}\int_{ \delta \leq \|z\| \leq 1/\delta} \frac{1-\cos(r z_1)}{\|z\|^{1+p}} dz,
\end{equation*}
where $z \in \bR^p$ and $z_1$ is its first coordinate. By Lemma \ref{lemma:H_p}, we have for each $r \geq 0$, $ \widetilde{H}_{p}(r; \cdot)$ is a decreasing function on $(0,1)$ and 
\begin{align*}
    \lim_{\delta \downarrow 0} \widetilde{H}_{p}(r; \delta) = r.
\end{align*}

By a similar calculation as  \cite[Equation (2.7)]{sejdinovic2013equivalence} or as in the proof of Lemma \ref{lemma:bound_outside_compacta}, we have that $\cV^2(\rho; \delta)$ is equal to the following
\begin{align*}
\Exp\left[\widetilde{H}_p(\|X_{\rho}-X_{\rho}'\|;\delta)\left(
   \widetilde{H}_q(\|Y_{\rho}-Y_{\rho}'\|;\delta) + 
   \widetilde{H}_q(\|Y_{\rho}''-Y_{\rho}'''\|;\delta) 
   -2\widetilde{H}_q(\|Y_{\rho}-Y_{\rho}''\|;\delta)
   \right)\right].
\end{align*}

Note that $0 \leq \widetilde{H}_{p}(r; \delta) \leq r$ for any $\delta \in (0,1)$. Then due to assumption \ref{assumption:dominated}, by the dominated convergence theorem (similar to the proof of Lemma \ref{lemma:alternative_derivatives}),  we have that
for each fixed $\delta \in (0,1)$, $\cV^2(\cdot;\delta)$ is twice differentiable on $[0,\epsilon)$ and that we can interchange the order of differentiation and expectation. Thus,
\begin{align*}
&    \left.\frac{d (\cV^2(\rho;\delta))}{d\rho} \right\vert_{\rho = 0}  \\
= &\Exp\bigg[\widetilde{H}_p(\|X_{0} - X_{0}'\|;\delta)\left(\widetilde{H}_p(\|Y_{0} - Y_{0}'\|;\delta) + \widetilde{H}_p(\|Y_{0}'' - Y_{0}'''\|;\delta) - 2\widetilde{H}_p(\|Y_{0} - Y_{0}''\|;\delta) \right) \\
 &\qquad\left(\frac{d g_0(X_{0},Y_{0})/d\rho}{g_0(X_{0},Y_{0})} 
+ \frac{d g_0(X_{0}',Y_{0}')/d\rho}{g_0(X_{0}',Y_{0}')} 
+ \frac{d g_0(X_{0}'',Y_{0}'')/d\rho}{g_0(X_{0}'',Y_{0}'')}
+ \frac{d g_0(X_{0}''',Y_{0}''')/d\rho}{g_0(X_{0}''',Y_{0}''')}
\right) \ \bigg],
\end{align*}
which is equal to zero  for any $\delta \in (0,1)$ by Lemma \ref{lemma:independence lemma}. Similarly, by writing down the expression for the second derivative and by the dominated convergence theorem, we have
\begin{align*}
\lim_{\delta \downarrow 0}    \left.\frac{d^2 (\cV^2(\rho;\delta))}{d\rho^2} \right\vert_{\rho = 0} 
= \left.\frac{d^2 (\cV^2(\rho))}{d\rho^2} \right\vert_{\rho = 0}.
\end{align*}
By assumption, the limit on the left-hand side of the above equation is positive. Thus, there exists $\delta > 0$ such that
$$
 \left.\frac{d^2 (\cV^2(\rho;\delta))}{d\rho^2} \right\vert_{\rho = 0} > 0.
$$
For this $\delta$, since $\cV^{2}(0;\delta) = 0$ and $\left.\frac{d (\cV^2(\rho;\delta))}{d\rho} \right\vert_{\rho = 0} = 0$, the proof is complete by Taylor's theorem.
\end{proof}

Now, we will prove Theorem \ref{thrm:alternative}.

\begin{proof}[Proof of Theorem \ref{thrm:alternative}]
We shall show the unique existence of the NPMLE $(\bG_n,\bW_n)$  in Lemma \ref{lemma:alter_NPMLE} and focus on the second statement in this proof.

By Lemma \ref{lemma:alter_second_derivative_delta}, there exists some $\delta_0 \in (0,1)$ such that
\begin{equation}
    \label{def:construction_delta0}
\lim_{\rho \downarrow 0} \frac{1}{\rho^2} \cV^{2}(\rho; \delta_0) > 0.
\end{equation}
Recall the definition of $\cV^2(\rho;\delta)$ in \eqref{def:cV_rho_delat}. Note that by the triangle inequality, for any complex numbers $a,b$, we have that $|a|^2 \geq 2^{-1}|b|^2 - |a-b|^2$, which implies that 
\begin{align*}
n \widehat{\cV}^2_n \geq &\int_{D(\delta_0)} \frac{n|\varphi^{(n)}_{X,Y}(s, t) - \varphi^{(n)}_X(s)\varphi^{(n)}_Y(t)|^2}{c_p c_q \|s\|^{1+p} \|t\|^{1+q}}\,ds\,dt \\
\geq & \frac{1}{2} n \cV^{2}(\rho_n; \delta_0) - \int_{D(\delta_0)} \frac{|\mathbb{C}_{n}(s,t)|^2}{c_p c_q \|s\|^{1+p} \|t\|^{1+q}}\,ds\,dt 
\end{align*}
where for $n \geq 1$ and $(s,t) \in \bR^{p+q}$,
\begin{align*}
\mathbb{C}_{n}(s,t):=
\sqrt{n}\left( \varphi^{(n)}_{X,Y}(s, t) - \varphi^{(n)}_X(s)\varphi^{(n)}_Y(t)
    -(\varphi_{X,Y}(s, t;\rho_n) - \varphi_X(s;\rho_n)\varphi_Y(t;\rho_n))\right).
\end{align*}
Note that by \eqref{def:construction_delta0} and since $\sqrt{n}\rho_n \to \infty$ as $n \to \infty$, we have
\begin{align*}
  n \cV^{2}(\rho_n; \delta_0) = n \rho_n^2 \left(\frac{1}{\rho_n^2}   \cV^{2}(\rho_n; \delta_0) \right) \to \infty, \;\;\text{ as } n \to \infty. 
\end{align*}
Furthermore, note that
\begin{align*}
\mathbb{C}_{n}(s,t):=
&\sqrt{n}\left( \varphi^{(n)}_{X,Y}(s, t) 
    -\varphi_{X,Y}(s, t;\rho_n)\right)  \\
    &-\sqrt{n}\left( \varphi^{(n)}_{X,Y}(s,0) - \varphi_{X,Y}(s,0;\rho_n)\right) \left( \varphi^{(n)}_{X,Y}(0,t)- \varphi_{X,Y}(0,t;\rho_n)\right)\\
    &-\sqrt{n}\left( \varphi^{(n)}_{X,Y}(s,0) - \varphi_{X,Y}(s,0;\rho_n)\right) \varphi_{X,Y}(0,t;\rho_n)\\
    &-\sqrt{n} \varphi_{X,Y}(s,0;\rho_n) \left( \varphi^{(n)}_{X,Y}(0,t)- \varphi_{X,Y}(0,t;\rho_n)\right).
\end{align*}
By Lemma \ref{lemma:Cn}, we have
$$
\sup_{(s,t) \in D(\delta_0)} |\mathbb{C}_n(s,t)| \;\text{ is bounded in probability},
$$
which implies that
$$
\int_{D(\delta_0)} \frac{|\mathbb{C}_{n}(s,t)|^2}{c_p c_q \|s\|^{1+p} \|t\|^{1+q}}\,ds\,dt 
\;\text{ is bounded in probability}.
$$
Thus, the proof is complete.
\end{proof}

\subsection{Bounding the characteristics functions under alternatives}

Recall the definition of the NPMLE $(\bG_n,\bW_n)$ in \eqref{def:NPMLE}.

\begin{lemma}
    \label{lemma:alter_NPMLE}
    Suppose that Assumption  \ref{assumption:alter_G0},  \ref{assumption:dominated} and \ref{assumption:alter_lambda} hold. As $n \to \infty$, with probability approaching one, the NPMLE $(\bG_n,\bW_n)$ exists and is  unique.
\end{lemma}
\begin{proof}
We shall apply \cite[Proposition 1]{clemenccon2022statistical}. Specifically, our result follows as long as we show that for all large enough $n$, Assumptions $4$, $5$, and $6$ in \cite{clemenccon2022statistical} hold.

First, due to our Assumption \ref{assumption:alter_lambda}, there exists some constant $C > 0$ such that for large enough $n$,
$$
|\lambda_{n,i} - \lambda_i| \leq \frac{C}{\sqrt{n}},
$$
which verifies Assumption 4 in \cite{clemenccon2022statistical}.

Second, part (iii) in our Assumption \ref{assumption:alter_G0} is identical to Assumption 6 in \cite{clemenccon2022statistical}.

Third,  for any $1 \leq k \neq \ell \leq K$ and $\rho \in [0,1]$,  define
\begin{align*}
    \kappa_{\rho}(k,\ell) := \int \mathbbm{1}(w_{k}(x,y)  > 0 )\mathbbm{1}(w_{\ell}(x,y)  > 0 ) g_{\rho}(x,y)\,d\nu(x,y).
\end{align*}
Furthermore, define
\begin{align*}
    \tau := \frac{1}{2}\min\{\kappa_{0}(k,\ell): 1 \leq k \neq \ell \leq K, \kappa_{0}(k,\ell) > 0\}.
\end{align*}
Clearly, $\tau >0$. Further, for this $\tau > 0$, denote by $\bm{G}_{\rho}^{\tau}$ the (undirected) graph with vertices in
$\{1,\ldots,K\}$, and edge between $1 \leq k \neq  \ell \leq K$ if and only if $\kappa_{\rho}(k,\ell) \geq \tau$.

By our Assumption \ref{assumption:dominated} and the dominated convergence theorem, for any $1 \leq k \neq \ell \leq K$,
\begin{align*}
\lim_{n \to \infty} \kappa_{\rho_n}(k,\ell)  =
 \kappa_{0}(k,\ell) .
\end{align*}
Thus, due to the definition of $\tau$, we have that for all large enough $n$, the edges of $\bm{G}_{\rho_n}^{\tau}$  is a super set of the edges of $\bm{G}_{0}$. By part (ii) of our Assumption \ref{assumption:alter_G0}, $\bm{G}_{0}$ is connected. Thus, for all large enough $n$, $\bm{G}_{\rho_n}^{\tau}$ is connected, which verifies Assumption 5 in \cite{clemenccon2022statistical}. The proof is complete.
\end{proof}


Recall that $\varphi^{(n)}_{X,Y}(\cdot)$ is defined in Definition \ref{def:Vn_Rn} and $\varphi_{X,Y}(\cdot;\rho_n)$ is the characteristic function of $X,Y$ when they have the joint density $g_{\rho_n}$ for $n \geq 1$. Further, recall the definition of $D(\delta)$ in \eqref{def:D_delta}.

\begin{lemma}
    \label{lemma:Cn}
    Assume that the conditions in Theorem \ref{thrm:alternative} hold. For any $\delta \in (0,1)$,
$$
\sup_{(s,t) \in D(\delta)} 
\sqrt{n}\left|\varphi^{(n)}_{X,Y}(s,t)-\varphi_{X,Y}(s,t;\rho_n)\right|
$$
is bounded in probability as $n \to \infty$.
\end{lemma}
\begin{proof}
First, by considering the real and imaginary parts, it suffices to show that the following two terms are bounded in probability:
\begin{align*}
    \sup_{\theta \in D(\delta)} 
\sqrt{n}\left|
\int \cos(\theta^\top z) d\bG_n(z) - 
\int \cos(\theta^\top z) dG_n(z)
\right|, \\
\sup_{\theta \in D(\delta)} 
\sqrt{n}\left|
\int \sin(\theta^\top z) d\bG_n(z) - 
\int \sin(\theta^\top z) dG_n(z)
\right|,
\end{align*}
where we denote by $\theta = (s,t) \in \bR^{p+q}$ and
$z = (x,y) \in \bR^{p+q}$ in order to be consistent with the notations used in \cite{clemenccon2022statistical}. We will only prove that the former is bounded in probability, noting that the same argument applies to the latter.

For $\theta \in \Theta := D(\delta) \subset \bR^{p+q}$, define
$$
\psi(z;\theta) := \cos(\theta^\top z), \;\; \text{ for } z \in \bR^{p+q},
$$
and $\cF_{\Theta } := \{\psi(\cdot;\theta): \theta \in \Theta \}$ the class of functions indexed by $\Theta = D(\delta)$. Then $\int \cos(\theta^\top z) d\bG_n(z)$ and $\int \cos(\theta^\top z) dG_n(z)$ correspond to $\widetilde{L}_n(\theta)$ and $L(\theta)$ in \cite[Theorem 1]{clemenccon2022statistical}, whose analysis is non-asymptotic. Thus, if we can show Assumptions $4$--$8$ in \cite{clemenccon2022statistical} hold for all large enough $n$, then our result follows from their Theorem 1, that is,
$$
\sup_{\theta \in \Theta} \sqrt{n}|\widetilde{L}_n(\theta)-L(\theta)| \text{ is bounded in probability.}
$$

We have already shown in the proof of Lemma \ref{lemma:alter_NPMLE} that Assumptions $4$,$5$, and $6$ of \cite{clemenccon2022statistical} hold for all large enough $n$ under our assumptions.

\medskip

Next, we consider the assumption 7 in \cite{clemenccon2022statistical}. First, for $n \geq 1$ and $\bm{u}=(u_1,\ldots,u_K) \in \bR^{K}$, 
define a $K \times K$ matrix $\Bar{D}''_n(\bm{u})$ such that for $ 1 \leq k,k'\leq K$, its $(k,k')$-th entry $\left[\Bar{D}''_{n}(\bm{u})\right]_{k,k'}$ is given by
$$
\int \left[
\frac{e^{u_{k}} w_k(z) \mathbbm{1}(k = k')}{\sum_{\ell=1}^{K}e^{u_{\ell}}  w_{\ell}(z)}
-
\frac{e^{u_{k}} w_k(z) e^{u_{k'}} w_{k'}(z) }{\left(\sum_{\ell=1}^{K}e^{u_{\ell}}  w_{\ell}(z)\right)^2}
\right] \left(\sum_{\ell=1}^{K}  \frac{\lambda_{n,\ell} w_{\ell}(z)}{W_{n,{\ell}}}\right) g_{\rho_n}(z) d\nu(z),
$$
where we denote by $z = (x,y) \in \bR^{p+q}$. Further, for $\bm{u}=(u_1,\ldots,u_K) \in \bR^{K}$, define another $K\times K$ matrix $\Bar{D}''_0(\bm{u})$ such that for $ 1 \leq k,k'\leq K$, its $(k,k')$-th entry $\left[\Bar{D}''_{0}(\bm{u})\right]_{k,k'}$ is given by
$$
\int \left[
\frac{e^{u_{k}} w_k(z) \mathbbm{1}(k = k')}{\sum_{\ell=1}^{K}e^{u_{\ell}}  w_{\ell}(z)}
-
\frac{e^{u_{k}} w_k(z) e^{u_{k'}} w_{k'}(z) }{\left(\sum_{\ell=1}^{K}e^{u_{\ell}}  w_{\ell}(z)\right)^2}
\right] \left(\sum_{\ell=1}^{K} \frac{\lambda_{\ell} w_{\ell}(z)}{W_{0,{\ell}}}\right) g_{0}(z) d\nu(z),
$$
where $W_{0,\ell} := \int_{\bR^{p+q}} w_{\ell}(z) g_0(z) d\nu(z)$.

For a symmetric matrix $M$, denote by $\sigma_2(M)$ the second smallest eigenvalue of $M$. By the dominated convergence theorem, due to part (iii) in assumption \ref{assumption:alter_G0}, the following function is continuous:
$$
\bm{u} \in \bR^K \;\;\mapsto \;\; \Bar{D}''_0(\bm{u}) \in \bR^{K\times K}.
$$
As a result, by Weyl’s inequality (see, e.g., \cite[Theorem 4.5.3]{Vershynin_2018}), the following function is continuous:
$$
\bm{u} \in \bR^{K} \;\;\mapsto \;\; \sigma_2(\Bar{D}''_0(\bm{u})) \in  \bR.
$$
Further, by (the proof of) Proposition 1.1 in \cite{gill1988large}, since $\bm{G}_0$ is connected by Assumption \ref{assumption:alter_G0}, for each fixed $\bm{u}$, $\sigma_2(\Bar{D}''_0(\bm{u})) > 0$. Thus, for any compact set $S \subset \bR^{K}$, 
$$\tau_S := \inf_{\bm{u} \in S}  \sigma_2(\Bar{D}''_0(\bm{u})) > 0.$$

Now, for a fixed $\bm{u} \in \bR^{K}$, due to Assumption \ref{assumption:dominated},  component-wise
$$
\lim_{n \to \infty} \Bar{D}''_{n}(\bm{u}) = \Bar{D}''_{0}(\bm{u}).
$$
Furthermore, by computing the derivatives with respect to $\bm{u}$, it is elementary to show that  for any compact set $S \subset \bR^{K}$, the sequence of functions $\{\bm{u} \in S \mapsto \Bar{D}''_{n}(\bm{u}) \in \bR^{K\times K}\}$ is equicontinuous component-wise. Then by Arzel\`a-Ascoli theorem, we have that component-wise
$$
\lim_{n \to \infty} \sup_{\bm{u} \in S} \left|\Bar{D}''_{n}(\bm{u}) - \Bar{D}''_{0}(\bm{u})\right| = 0.
$$
Then, by Weyl’s inequality again,  for any compact set $S \subset \bR^{K}$ and  large enough $n$,
$$
\inf_{\bm{u} \in S}  \sigma_2(\Bar{D}''_n(\bm{u})) \geq \frac{1}{2}\tau_S,
$$
which verifies Assumption 7 of \cite{clemenccon2022statistical}.

\medskip

Finally, we consider a modified version of Assumption 8 in \cite{clemenccon2022statistical}. Clearly, $|\psi(z;\theta)| \leq 1$ for all $z \in \bR^{p+q}$ and $\theta \in \Theta$. Assumption 8 in \cite{clemenccon2022statistical} also requires the following uniform covering number condition: for some constant $C_{\Theta} > 0$ and $r > 0$,
\begin{align}
    \label{aux:uniform_covering}
    \sup_{Q} N(\zeta,\cF_{\Theta},L_2(Q)) \leq (C_{\Theta}/\zeta)^r, \text{ for all } \zeta > 0,
\end{align}
where the supremum is taken over the set of all probability measures $Q$ on $\bR^{p+q}$, and $N(\zeta,\cF_{\Theta},L_2(Q))$ is the $\zeta$-covering number under $\|\cdot\|_{L_2(Q)}$ norm, that is, the smallest number of $L_2(Q)$ balls with radius $\zeta$ that is needed to cover $\cF_{\Theta}$. Unfortunately, the condition does not hold for our $\cF_{\Theta}$. However, upon checking the proof of \cite[Theorem 1]{clemenccon2022statistical}, Assumption 8 of \cite{clemenccon2022statistical} is only needed in providing an upper bound for the right-hand side of equation (C.12)\footnote{there is a typo in (C.11) and (C.12) of \cite{clemenccon2022statistical}: $\widehat{P}_k - P$ should be $\widehat{P}_k - P_k$. Equation (3.5) in \cite{clemenccon2022statistical} has the right expression.}, where the authors use Theorem 2.14.9 in \cite{vanderVaart1996}.

Note that the conclusion of Theorem 2.14.9 in \cite{vanderVaart1996} holds under either the uniform covering number condition (Equation (2.14.6)) or the bracketing condition (Equation (2.14.7)). Assumption 8 of \cite{clemenccon2022statistical} uses the uniform covering number condition in \eqref{aux:uniform_covering}. The result in Theorem 1 of \cite{clemenccon2022statistical} continues to hold if we replace \eqref{aux:uniform_covering} with the following condition: there exist some constants $C_{\Theta} > 0$ and $r > 0$ such that for $1 \leq k \leq K$ and for all large enough $n$,
\begin{align}
    \label{our_bracketing}
    N_{[\;]}(\zeta, \cF_{\Theta},L_2(F_{n,k})) \leq (C_{\Theta}/\zeta)^r, \text{ for all } 0 < \zeta < C_{\Theta},
\end{align}
where $N_{[\; ]}(\zeta, \cF_{\Theta},L_2(F_{n,k}))$ is $\zeta$-bracketing number of $\cF_{\Theta}$ under the 
$\|\cdot\|_{L_2(Q)}$ norm (see \cite[Definition 2.1.6]{vanderVaart1996}) and $F_{n,k}$ is the biased sampling distribution for the $k$-th population as defined in Subsection \ref{sec:alternative}. Now, we show \eqref{our_bracketing} holds.

By Taylor's theorem, for any $\theta_1, \theta_2 \in \Theta$, and $z \in \bR^{p+q}$,
$$
|\psi(z;\theta_1)-\psi(z;\theta_2)| \leq m(z) \|\theta_1 - \theta_2\|,
$$
where $m(z) := \|z\|+1$. By Example 19.7 in \cite{van2000asymptotic}, there exists some constant $C_1 \geq 1$, that only depends on $\delta$ and $p+q$, such that
$$
N_{[\;]}(\zeta \|m\|_{L_2(F_{n,k})}, \cF_{\Theta},L_2(F_{n,k})) \leq \left(C_1/\zeta\right)^{p+q}, \text{ for all } 0 < \zeta < C_1,
$$
where we note that by definition, $N_{[\;]}(\zeta, \cF_{\Theta},L_2(F_{n,k})) = 1$ for any $\zeta \geq 1$.
By a change of variable, we have that
$$
N_{[\;]}(\zeta, \cF_{\Theta},L_2(F_{n,k})) \leq \left( (C_1 \|m\|_{L_2(F_{n,k})})/\zeta\right)^{p+q}, \text{ for all } 0 < \zeta < C_1\|m\|_{L_2(F_{n,k})}.
$$
By Assumptions \ref{assumption:dominated} and \ref{assumption:alter_G0} (in particular, $w_k(\cdot,\cdot) \leq 1$), there exists a constant $C_2 \geq 1$, that does not depend on $n$, such that for large enough $n$,
$$
\max_{1 \leq k \leq K} \|m\|_{L_2(F_{n,k})} \leq C_2.
$$
Then, \eqref{our_bracketing} holds for large enough $n$ with $C_{\Theta} = C_1 C_2$ and $r = p+q$. The proof is complete.
\end{proof}

\section*{Acknowledgments}
The authors would like to thank Xinrui Wang, under the supervision of the second author, for her contributions on this topic in her master project.
Yuwei Ke and Hok Kan Ling acknowledge the support by NSERC Grant RGPIN/03124-2021.
Yanglei Song acknowledges the support by NSERC Grant RGPIN-2020-04256.


\bibliographystyle{unsrt}

\bibliography{dCov_bias}

\appendix

\end{document}